\documentclass[%
 reprint,
superscriptaddress,
 amsmath,amssymb,
 aps,
]{revtex4-1}

\usepackage{soul}
\usepackage{graphicx}
\usepackage{dcolumn}
\usepackage{bm}

\usepackage{mathtools}

\usepackage[dvipsnames]{xcolor}

\usepackage[colorlinks=true,citecolor=blue]{hyperref}
\usepackage[normalem]{ulem}
\usepackage[caption=false]{subfig} 
\usepackage{amsmath}

\usepackage{multirow}
\usepackage{capt-of}
\usepackage{float}

\newcommand{\UIB}{Departament de F\'isica, Universitat de les Illes Balears, IAC3 -- IEEC, Crta. Valldemossa km 7.5, E-07122 Palma, Spain}

\newcommand{\AEI}{Max Planck Institute for Gravitational Physics (Albert Einstein Institute), Am Mühlenberg 1, D-14476 Potsdam, Germany}
\newcommand{\AEIH}{Max-Planck-Institut für Gravitationsphysik, Albert-Einstein-Institut, Callinstr. 38, D-30167 Hannover, Germany}

\definecolor{dodgerblue}{HTML}{1E90FF}
\definecolor{viennared}{HTML}{DA0A14}
\definecolor{ctorange}{HTML}{FF6C0C}
\definecolor{granadagreen}{HTML}{078931}
\definecolor{wales}{HTML}{ff0038}
\definecolor{valenciacfred}{HTML}{ee3524}
\definecolor{barcelonafcgold}{HTML}{edbb00}
\definecolor{jam}{HTML}{A50B5E}
\definecolor{austriawien}{HTML}{441678}

\AtBeginDocument{%
  \hypersetup{
    citecolor=dodgerblue,
    linkcolor=dodgerblue,   
    urlcolor=dodgerblue}}

\begin{document}


\title{Time domain phenomenological model of gravitational wave subdominant harmonics for quasi-circular non-precessing binary black hole coalescences}

\author{H\'ector Estell\'es}
\affiliation{ \UIB}

\author{Sascha Husa}
\affiliation{ \UIB}

\author{Marta Colleoni}
\affiliation{ \UIB}

\author{David Keitel}
\affiliation{ \UIB}

\author{Maite Mateu-Lucena}
\affiliation{ \UIB}

\author{Cecilio Garc\'ia-Quir\'os}
\affiliation{ \UIB}

\author{Antoni Ramos-Buades}
\affiliation{ \AEI}
\affiliation{ \UIB}

\author{Angela Borchers}
\affiliation{ \AEIH}
\affiliation{ \UIB}


\date{\today}

\begin{abstract}

In this work we present an extension of the time domain phenomenological model \texttt{IMRPhenomT} for gravitational wave signals from binary black hole coalescences to include subdominant harmonics, specifically the $(l=2, m=\pm 1)$, $(l=3, m=\pm 3)$, $(l=4, m=\pm 4)$ and $(l=5, m=\pm 5)$ spherical harmonics. We also improve our model for the dominant $(l=2, m=\pm 2)$ mode and discuss mode mixing for the $(l=3, m=\pm 2)$ mode. The model is calibrated to numerical relativity solutions of the full Einstein equations up to mass ratio 18, and to numerical solutions of the Teukolsky equations for higher mass ratios. This work complements the latest generation of traditional frequency domain phenomenological models (\texttt{IMRPhenomX}), and provides new avenues to develop computationally efficient models for gravitational wave signals from generic compact binaries.

\end{abstract}

\pacs{Valid PACS appear here}
\maketitle


\section{Introduction}\label{sec:intro}

In order to exploit the information encoded in gravitational wave (GW) signals from compact binary systems, reliable and computationally efficient models of such signals are required to perform Bayesian inference \cite{Veitch:2014wba,Ashton_2019} of the source properties.
The development of such waveform models has so far progressed in remarkable synchronicity with the challenges posed by the improvement in sensitivity of the advanced LIGO \cite{aLIGO2015} and Virgo \cite{Acernese_2014} detectors, and the discovery of new 
types of compact binaries. A decade after the numerical relativity (NR) breakthroughs to evolve black holes \cite{Pretorius2005,Zlochower2005,Koppitz2005},
the third generation of inspiral-merger-ringdown waveform models \cite{Husa:2015iqa,Khan:2015jqa,Hannam:2013oca,Bohe:PPv2,Purrer:2015tud} could be employed for parameter estimation of the first detected signal \cite{TheLIGOScientific:2016wfe}, and an extensive study of systematic errors \cite{PurrerLVC} found the models to be sufficiently accurate for this relatively loud signal, although improvements would soon be required for future events, in particular to include sub-dominant harmonics. Models that include subdominant harmonics were still not available to analyze the catalog of events from the first two observation runs \cite{Abbott_2019}, but within a year an extensive study of the particularly massive GW170729 event \cite{Chatziioannou_2019} with the first multi-mode models reported improved parameter estimation results. Several models that include both precession and subdominant harmonics were completed in time to study events in the third observation run: clear evidence for such modes has been found for two events, GW190412\cite{LIGOScientific:2020stg} and GW190814 \cite{Abbott:2020khf}, and such models have been used to analyse events for the second catalog of GW events \cite{abbott2020gwtc2}.

Further improvements in waveform models will be required to keep up with the increased sensitivity foreseen for the fourth and future observation runs of the advanced detector network \cite{Aasi:2013wya}, and to meet the goal of being able to model generic signals.
In a previous work \cite{estells2020imrphenomtp}, we presented \texttt{IMRPhenomTP}, a phenomenological model in the time domain for the dominant  $l=2, m=\pm 2$ modes in the co-precessing frame, extended to cover the $l=2$ subset of modes in the inertial frame of precessing systems through the twisting-up procedure. This model forms the cornerstone of a new family of time domain models, which are comparable in construction, accuracy and computational efficiency with the \texttt{IMRPhenomX} family of frequency domain models \cite{pratten2020setting,garcaquirs2020imrphenomxhm,phenomxphm}. The main motivation for such time domain models is twofold: First, the time domain simplifies the
separation of the inspiral, merger, and ringdown regimes, and can thus simplify tests
of general relativity, such as inspiral-merger-ringdown consistency tests \cite{Ghosh_2016, Ghosh_2017}
or parameterized tests \cite{Yunes_2009}. Second, some effects may be easier to model in the
time domain, most notably the twisting-up approximation that is often used to model
precession \cite{Hannam:2013oca,Bohe:PPv2,Ramos-Buades:2020noq, phenomxphm} and in the time domain does not require the commonly employed
stationary phase approximation \cite{Hannam:2013oca, Bohe:PPv2, phenomxphm}, or alternative
techniques which still need to be developed toward maturity \cite{Marsat:2018oam}.

Here we present an extension of the core non-precessing model \cite{estells2020imrphenomtp} that includes a subset of subdominant modes, in particular $l=2, m=\pm 1$, $l=3, m=\pm 3$, $l=4, m=\pm 4$ and $l=5, m=\pm 5$. While this work will not present results on precession, the procedures followed in \cite{estells2020imrphenomtp} carry over to the multimode extension discussed here, and in a future publication the benefits of working in the time domain for phenomenologically describing the merger-ringdown regions of precessing waveforms will be analysed.

The paper is organized as follows.
In Sec.~\ref{sec:model} we discuss the model ansatz we use, which is then calibrated  across the parameter space to a set of numerical waveforms in Sec.~\ref{sec:calib}.
In Sec.~\ref{sec:results} we test the quality of the model across the parameter space, and in Sec.~\ref{sec:PE} we test the performance of the model in Bayesian parameter estimation examples. Finally we conclude in Sec.~\ref{sec:conclusions}.

\section{Phenomenological model construction}\label{sec:model}

Gravitational radiation is encoded in two independent polarisations $h_+$ and $h_{\times}$ that can be decomposed in a suitable basis of spin-weighted spherical harmonics (SWSH) \cite{Goldberg:1966uu} (defined as in \cite{Wiaux:2005fm}):

\begin{equation}
\label{eq:polariz_decomp}
\begin{split}
    h(t;\boldsymbol{\lambda};r,\theta,\phi)&=h_+ - ih_{\times}\\
    &=\sum_{l}\sum_{-l\leq m \leq l}h_{lm}(t;\boldsymbol{\lambda};r)\ ^{-2}Y_{lm}(\theta,\phi),
\end{split}
\end{equation}
where the modes $h_{lm}(t;\boldsymbol{\lambda};r)$  express the dependence on time and the intrinsic physical properties of the source, $\boldsymbol{\lambda}$, while the relative orientation of the source with respect to the observer is encoded in the SWSH basis functions $^{-2}Y_{lm}(\theta,\phi)$. This work will be restricted to the modelling of binary black hole systems whose individual spins are aligned or anti-aligned with the orbital angular momentum direction, which is preserved 
in this situation. Then, the intrinsic degrees of freedom are $\boldsymbol{\lambda}=\{m_1,m_2,\chi_{1},\chi_{2}\}$ where $m_{1,2}$ are the component masses and $\chi_{1,2}=(\boldsymbol{S}_{1,2}/m_{1,2}
^2)\cdot\hat{\boldsymbol{L}}$ are the dimensionless spin components in the orbital angular momentum direction $\hat{\boldsymbol{L}}$. For modelling purposes, the total mass of the system $M=m_1+m_2$ is simply a scale parameter for vacuum spacetimes. We thus factor it, out working in units where time is expressed in mass units, 
and express the dependence on the masses through the 
symmetric mass ratio $\eta=m_1 m_2/M^2$.

Each mode is a complex function that can be decomposed into a real amplitude and phase as
\begin{equation}
    h_{lm}(t)=H_{lm}(t)\exp\{-i\phi_{lm}(t)\},
\end{equation}
with $\dot{\phi}_{lm}(t)\equiv\omega_{lm}(t)>0$ for $m>0$ and $\omega_{lm}(t)<0$ for $m<0$.
The model provides a phenomenological construction for the $m>0$ modes specified at the beginning, and the $m<0$ modes are computed using the symmetry relation:
\begin{equation}
\label{symm}
    h_{lm}(t)=(-1)^lh_{l,-m}^*,
\end{equation}
valid for non-precessing systems, where $^*$ indicates complex conjugation. By convention we shift the time coordinate such that $t=0$ occurs for the peak amplitude of the $(2,2)$ mode.

\subsection{Model overview}\label{sec:overview}

In a previous work \cite{estells2020imrphenomtp} a phenomenological model in the time domain for the dominant $l=2$, $m=\pm2$ modes was presented, \texttt{IMRPhenomT}. In this work we provide the extension of the model to the modes $l=2, m=\pm 1$, $l=3, m=\pm 3$, $l=4, m=\pm 4$ and $l=5, m=\pm 5$. The core structure of \texttt{IMRPhenomT} are $C^1$ analytical expressions for the amplitude $H_{22}(t)=|h_{22}(t)|$, the gravitational phase $\phi_{22}(t)$ and the wave frequency $\omega_{22}(t)=\dot{\phi}_{22}(t)$ of the $l=2$, $m=2$ mode, in terms of piece-wise functions defined on a physically motivated partition of the evolution time of the system: inspiral, merger and ringdown regions. Following the same strategy, 
the model presented here provides $C^1$ analytical expressions for the amplitudes $H_{lm}(t)=|h_{lm}(t)|$, the GW phases $\phi_{lm}(t)$ and the wave frequencies $\omega_{lm}(t)=\dot{\phi}_{lm}(t)$ for the included modes as piece-wise functions:

\begin{equation}
\label{eq:fullIMRphase}
  \phi_{lm}(t) =
  \begin{cases}
                                   \phi^{\text{insp}}_{lm}(t) & t\leq t^{\omega}_{\text{cut},lm} \\
                                   
                                   \\\phi^{\text{merger}}_{lm}(t)  & t^{\omega}_{\text{cut},lm}\leq t \leq t^\text{peak}_{22} \\
                                   \\
  								 \phi^{\text{RD}}_{lm}(t) & t \geq t^\text{peak}_{22},
  \end{cases}  
\end{equation}

\begin{equation}
\label{eq:fullIMRomega}
  \omega_{lm}(t) =
  \begin{cases}
                                   \omega^{\text{insp}}_{lm}(t) & t\leq t^{\omega}_{\text{cut},lm} \\
                                   
                                   \\\omega^{\text{merger}}_{lm}(t)  & t^{\omega}_{\text{cut},lm}\leq t \leq t^\text{peak}_{22} \\
                                   \\
  								 \omega^{\text{RD}}_{lm}(t) & t \geq t^\text{peak}_{22},
  \end{cases}  
\end{equation}

\begin{equation}
\label{eq:fullIMRamp}
       H_{lm}(t) =
 	 \begin{cases}
                                    H^{\text{insp}}_{lm}(t) & t\leq t^{H}_{\text{cut}} \\
                                   
                                   \\H^{\text{merger}}_{lm}(t)  & t^{H}_{\text{cut}}\leq t \leq t^\text{peak}_{lm} \\
                                   \\
  								 H^{\text{RD}}_{lm}(t) & t \geq t^\text{peak}_{lm},
 	 \end{cases}
\end{equation}
where $t_\text{cut}$, the separation between the inspiral and merger regimes, was selected in the previous work to correspond to the time of the minimum energy circular orbit (MECO) \cite{Cabero_2017} for the phase and frequency, and to half of this time for the amplitude. However, already in the previous work it was realised that, while this time sets a natural ending for the validity of the post-Newtonian (PN) constructions employed in the inspiral regime, it complicates the extension of the model to high mass ratios since the merger ansatz did not describe correctly the too elongated merger region for these cases, and similarly for high spins where the merger region was too short to be correctly reconstructed. For these reasons, we have now chosen a different implementation of the separation times between regions, which will be discussed in this section. We present details of the modelling of each of these regions, highlighting the main differences with the version of \texttt{IMRPhenomT} described in \cite{estells2020imrphenomtp}.

\subsection{Inspiral}\label{sec:inspiral}

For modelling the inspiral regime, we follow the approach of {\tt IMRPhenomXHM} \cite{garcaquirs2020imrphenomxhm}: We model the amplitude and the phase of the 
$\ell=\vert m \vert = 2$ mode
by adding higher order pseudo-PN terms to known PN results, and we describe the phase of the sub-dominant modes by augmenting the phase of the dominant mode with PN terms.

\subsubsection{Orbital frequency}

In the inspiral regime, where the assumptions of the PN framework and the energy balance condition are satisfied, PN analytical solutions for the dynamical evolution of the orbital frequency $\omega_\mathrm{orb}(t)\equiv\dot{\phi}_\mathrm{orb}(t)$ can be employed.
In the adiabatic approximation, we can consider the balance equations
\begin{subequations}
\begin{equation}
\label{eq:balanceeq1}
\dfrac{d\phi_\mathrm{orb}}{dt}-\dfrac{v^3}{M}=0,
\end{equation}
\begin{equation}
\label{eq:balanceeq2}
\dfrac{dv}{dt}+\dfrac{\mathcal{F}(v)}{ME'(v)}=0,
\end{equation}
\end{subequations}
where $v=\omega_\mathrm{orb}^{1/3}$ is the PN velocity, $\phi_\mathrm{orb}(t)$ is the orbital phase of the binary, $\mathcal{F}(v)$ is the gravitational wave luminosity and $E(v)$ is the binding energy of the system. Currently, the gravitational wave luminosity and the binding energy of the system are known up to 3.5PN order for quasi-circular orbits   \cite{Blanchet:1995ez,Blanchet:1997jj, PhysRevLett.93.091101}. In this work we use the expressions for the energy fluxes and binding energy at 3.5PN order, although we note that recently the 4.5PN coefficients for the energy flux have also been calculated \cite{Marchand:2016vox}. Expanding the quotient on equation \eqref{eq:balanceeq2}, different families of analytical templates \cite{buonanno2009comparison} can be found for the velocity derivative, including some analytical approximations to the orbital phase and frequency, as in the \texttt{TaylorT3} approximant:
\begin{equation}
\label{eq:taylort3eq2}
\omega_\mathrm{orb}(t)\approx\omega^{(T3)}_{n/2}(t)=\omega_{N}(t)\sum_{k=0}^{n}\hat{\omega}_k\theta^k(t),
\end{equation}
where $\theta(t)=[\eta(t_{0}-t)/(5M)]^{-1/8}$, $\omega_{N}=\theta^3/8$ and the PN coefficients $\hat{\omega}_k$ are given in Appendix A1 of \cite{estells2020imrphenomtp}.

Nevertheless, in order to improve the accuracy of the description up to the end of the inspiral region, we need to expand the expression to higher order adding 6 extra pseudo-PN orders with unknown coefficients:
\begin{equation}
\label{eq:omegainsp}
\omega^{\text{insp}}_{22}(t)=\omega^{(T3)}_{3.5}(t) + \omega_{N}(t)\sum_{k=8}^{13}\hat{c}_k\theta^k.
\end{equation}
In order to obtain the value of the coefficients across parameter space, calibration with numerical simulations is required. However, instead of performing parameter space fits of the unknown coefficients, following \cite{Husa:2015iqa,pratten2020setting} it is more convenient to fit the value of the frequency at particular times, placing collocation points, and then solve the coefficients through a linear system of equations:

\begin{equation}
\label{eq:inspsystem}
    \omega_{N,i}\sum_{k=8}^{13}\hat{c}_k\theta_i^{k}=\lambda^{\omega}_i-\omega^{(T3)}_{3.5}(\theta_i).
\end{equation}
Expression (\ref{eq:inspsystem}) corresponds to a linear system of six equations that can be solved to obtain the six unknown coefficients $\hat{c}_k$ in terms of known frequency values $\lambda^{\omega}_i$ at particular $\theta_i$ values. The frequency values $\lambda^{\omega}_i$ are fitted across parameter space from numerical simulations.

\begin{figure}
    \centering
    \includegraphics[width=0.48\textwidth]{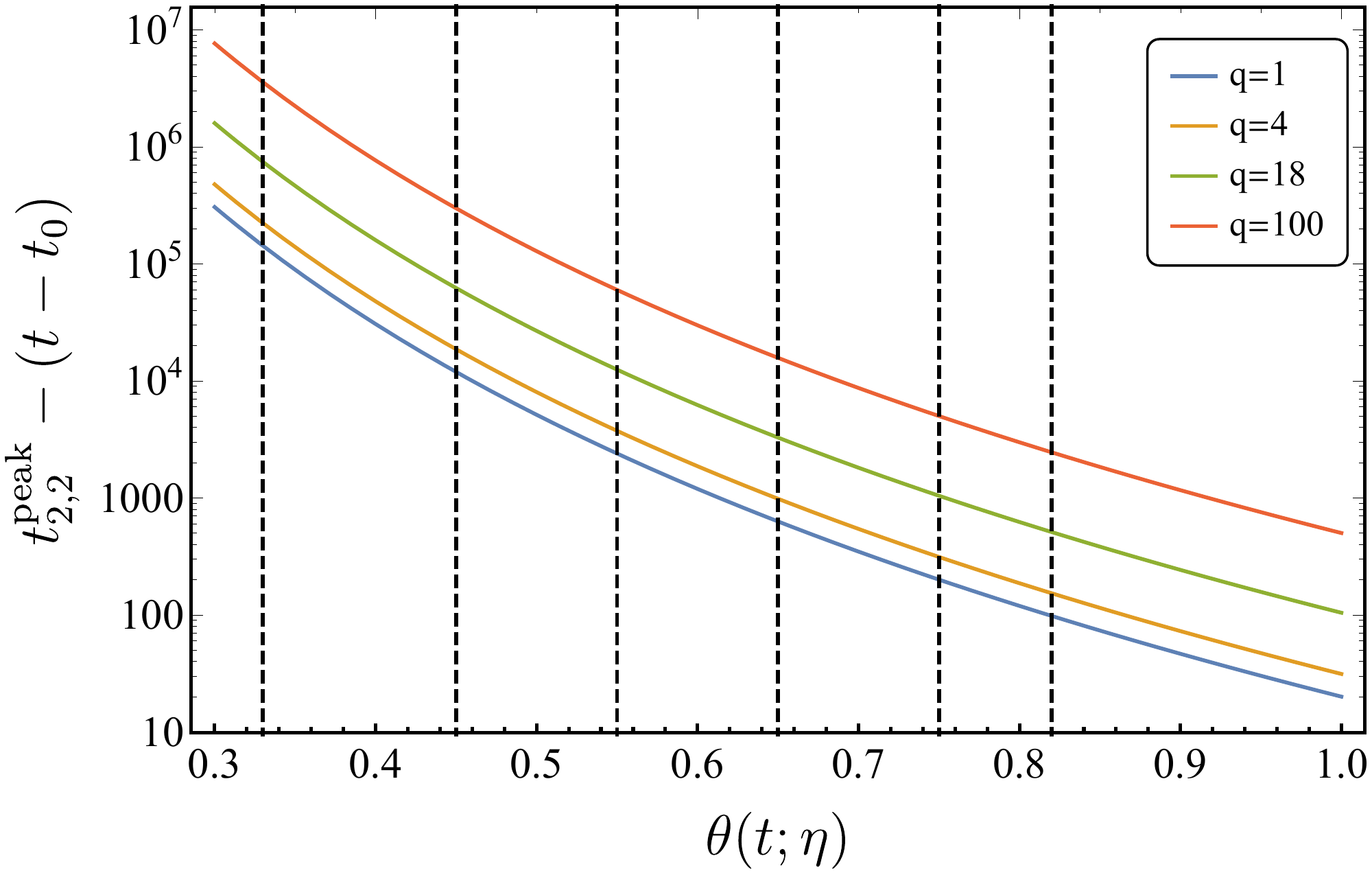}
    \caption{Placement of inspiral orbital frequency collocation points. Dashed vertical lines show the position of the different collocation points at fixed $\theta$ values, and the solid coloured lines show the relation between time and $\theta$ for different mass ratio configurations.}
    \label{fig:thetaoft}
\end{figure}

\begin{figure}
    \centering
    \includegraphics[width=0.48\textwidth]{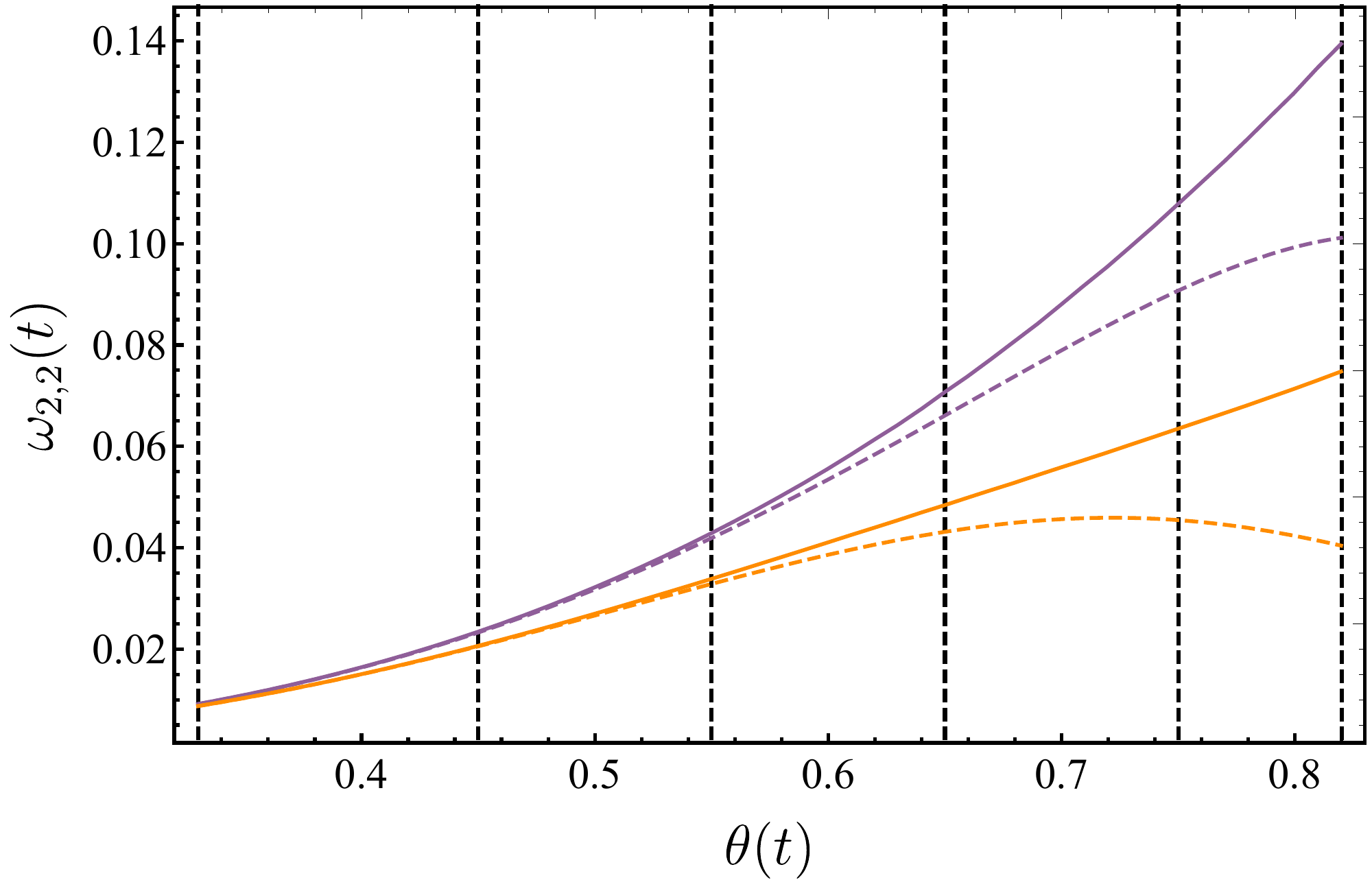}
    \caption{Frequency evolution of $l=2$, $m=2$ mode as a function of the $\theta$ parameter. Purple: $q=8,\ \chi_1=0.8$ configuration. Orange: $q=8,\ \chi_1=-0.8$ configuration. Solid: from \texttt{SEOBRNRv4} waveform. Dashed: 3.5PN \texttt{TaylorT3} implementation. Dashed black vertical lines: collocation point placement.}
    \label{fig:omegaoftheta}
\end{figure}

Both the number of coefficients and the placement of the collocation points differ from the original implementation presented in \cite{estells2020imrphenomtp}. Instead of placing the collocation points in the late inspiral at fixed times, we found that a better behaviour is achieved across parameter space if they are placed at fixed $\theta(t)$ values instead. We have selected the following placement:
\begin{equation}
\theta_i=\{0.33,\ 0.45,\ 0.55,\ 0.65,\ 0.75,\ 0.82\}
\end{equation}
and their relation with time is illustrated in Fig.~\ref{fig:thetaoft}. With this placement, the Newtonian term of \texttt{TaylorT3}, $\omega_{N}=\theta^3/8$, is the same for each collocation point across parameter space. Furthermore, the  number of extra coefficients added to the \texttt{TaylorT3} approximant has also increased with respect to the previous implementation. This was decided in order to increase the accuracy towards more unequal mass systems, especially when the primary spin magnitude is high. As an illustrative example, in Fig.~\ref{fig:omegaoftheta} we show the $l=2$, $m=2$ frequency evolution of two configurations, both at mass ratio 8, one for high positive and one for high negative primary spin magnitude, computed from the \texttt{SEOBNRv4} approximant \cite{Bohe:2016gbl} and from the \texttt{TaylorT3} approximant. As can be seen in the figure, the four later collocation points are needed for correcting the deviation of \texttt{TaylorT3} with respect to the correct solution represented by \texttt{SEOBNRv4}. The first two early collocation points are needed for regularising the behaviour at low frequency after the addition of the other extra terms, whose solution can sometimes lead to deviations in the low frequency regime.

The data to be taken for each collocation point in order to produce the needed parameter space fits is selected according to the actual time at which the collocation point is placed in each case and the region of the parameter space where the case is located. Inside the parameter space region covered by the NR simulations, if the collocation point is placed inside the time interval of the corresponding NR simulation, the value from the NR simulation is taken. If the collocation point is placed at a time earlier than the minimum available time for the corresponding NR simulation, then the result from a \texttt{SEOBNRv4} waveform is taken. For mass ratios above the NR parameter space coverage, we employ the data set of Teukolsky-EOB hybrids we have previously used in \cite{garcaquirs2020imrphenomxhm}, in particular at mass ratios 80, 200 and 1000, for different spin configurations.

\subsubsection{Mode amplitude} \label{sec:InspiralAmplitude}

For describing the inspiral amplitude of each mode, we employ the 3PN order expressions from \cite{Blanchet_2008} with 2PN spin corrections from \cite{Buonanno_2013} and 1.5PN contributions from \cite{arun2008higherorder}. For the amplitude of the $(2,2)$, we also include 3.5PN corrections from \cite{Faye_2012}. The general form of the amplitude can be written as:

\begin{equation}
\label{eq:pnamp}
    H_{lm}^{PN}(t)=2\eta\sqrt{\dfrac{16}{5}}x(t) \sum_{k=0}^{7}\hat{h}^{lm}_k x(t)^{k/2}
\end{equation}
where $x(t)=(\omega_\mathrm{orb}(t))^{2/3}=(\omega^\mathrm{insp}_{22}(t)/2)^{2/3}$ and the PN coefficients $\hat{h}^{lm}_k$ are in general complex functions of $\{\eta,\chi_1,\chi_2\}$. The original expression for the amplitude of Eq.~(\ref{eq:pnamp}) for each mode can be found in Appendix \ref{appen:pnamp}.
However, it is more convenient to rotate the complex amplitudes such that the dominant contribution is given by their real part, since in this way the absolute value of the amplitude can be approximated by the real part, while the smaller imaginary part will contribute to the phase evolution as will be explained below. The corresponding rotations applied to the expressions of Appendix \ref{appen:pnamp} are:
\begin{subequations}
\label{eq:amprot}
\begin{align}
        H_{lm}^{'PN}(t)&=\exp(i\Delta\phi_{lm})H_{lm}^{PN}(t),\\
        \Delta\phi_{22}&=0,\\
        \Delta\phi_{21}&=\pi/2,\\
        \Delta\phi_{33}&=-\pi/2,\\
        \Delta\phi_{44}&=\pi,\\
        \Delta\phi_{55}&=\pi/2.
\end{align}
\end{subequations}

Similar to what is done for the frequency, in order to improve the accuracy of the description up to the end of the inspiral, we add three extra pseudo-PN orders with unknown coefficients that will be obtained from calibration to simulations:
\begin{equation}
\label{eq:inspamp}
    H_{lm}^\mathrm{insp}(t)={H_{lm}^{'PN}}(t) +2\eta\sqrt{\dfrac{16}{5}}x(t) \sum_{k=8}^{10}\hat{d}^{lm}_k x(t)^{k/2},
\end{equation}
where now the extra coefficients $\hat{d}^{lm}_k$ are real functions of the parameter space $\{\eta,\chi_1,\chi_2\}$. As for the frequency calibration, it is more convenient to perform parameter space fits of a set of collocation points from the amplitude, and employ them to solve the value of the coefficients for each case employing a linear system of equations:
\begin{align}
&2\eta\sqrt{\dfrac{16}{5}}\begin{bmatrix}
x^5(t^H_1) & x^{11/2}(t^H_1) & x^6(t^H_1) \\
x^5(t^H_2) & x^{11/2}(t^H_2) & x^6(t^H_2) \\
x^5(t^H_3) & x^{11/2}(t^H_3) & x^6(t^H_3)  
\end{bmatrix}
\begin{bmatrix}
\hat{d}^{lm}_8 \\ \hat{d}^{lm}_9 \\ \hat{d}^{lm}_{10} 
\end{bmatrix}\\\nonumber
&=
\begin{bmatrix}
\lambda^{H}_{1,lm} - \Re{H_{lm}^{'PN}}(t^H_1)\\ \lambda^{H}_{2,lm} - \Re{H_{lm}^{'PN}}(t^H_2) \\ \lambda^{H}_{3,lm} - \Re{H_{lm}^{'PN}}(t^H_3)
\end{bmatrix},
\end{align}
where the amplitude collocation points $(t^H_i,\ \lambda^{H}_{i,lm})$ are placed at fixed time positions in the late inspiral:
\begin{equation}
    t^H_i=\{-2000M, -250M, -150M\}.
\end{equation}

\subsubsection{Phases of subdominant modes}

For modelling the subdominant harmonic phases during the inspiral, a good approximation is achieved by just rescaling the frequency (and the phase) of the $(2,2)$ mode \cite{baker2008mergers, kelly2011mergers}:
\begin{subequations}
\label{eq:inspapprox}
\begin{equation}
    \omega_{lm}^\mathrm{insp}(t)\approx \frac{m}{2} \omega_{22}^\mathrm{insp}(t),
\end{equation}
\begin{equation}  
    \phi_{lm}^\mathrm{insp}(t)\approx  \frac{m}{2} \phi_{22}^\mathrm{insp}(t).
\end{equation}
\end{subequations}
However, this approximation loses accuracy as the merger is approached, accumulating some dephasing during the late inspiral. In the modelling of the frequency domain model \texttt{IMRPhenomXHM}, it was realized that a good approximation to correct the deviation of Eq.~(\ref{eq:inspapprox}) is to employ the phase contribution from the complex mode amplitude:
\begin{equation}
\label{eq:lambda}
    \Lambda_{lm}(t)\equiv\arg{H_{lm}^\mathrm{insp}(t)}=\arctan{\Big(\dfrac{\Im{H_{lm}^\mathrm{insp}(t)}}{\Re{H_{lm}^\mathrm{insp}(t)}}\Big)}.
\end{equation}
For tracking also the correct phase relation between the modes satisfying the specific tetrad convention, we need to rotate back by the global rotation of Eq.~(\ref{eq:amprot}) applied to the amplitude. With this, the inspiral phase of the subdominant modes can be described with the approximation from Eq.~(\ref{eq:inspapprox}), corrected with the complex amplitude phase contribution of Eq.~(\ref{eq:lambda}) and rotated to the original mode orientation through the inverse global rotation of eq.~(\ref{eq:amprot}):
\begin{equation}
\label{eq:inspphaserelation}
        \phi_{lm}^\mathrm{insp}(t)=(m/2)\phi_{22}^\mathrm{insp}(t) + \Lambda_{lm}(t) - \Delta\phi_{lm}.
\end{equation}
Consequently, the mode frequency $\omega_{lm}(t)=\dot{\phi}_{lm}(t)$ is given by:
\begin{equation}
\label{eq:inspfreqrelation}
        \omega_{lm}^\mathrm{insp}(t)=(m/2)\omega_{22}^\mathrm{insp}(t) + \dfrac{d}{dt}\Lambda_{lm}(t).
\end{equation}

\begin{figure}
    \centering
    \includegraphics[width=0.48\textwidth]{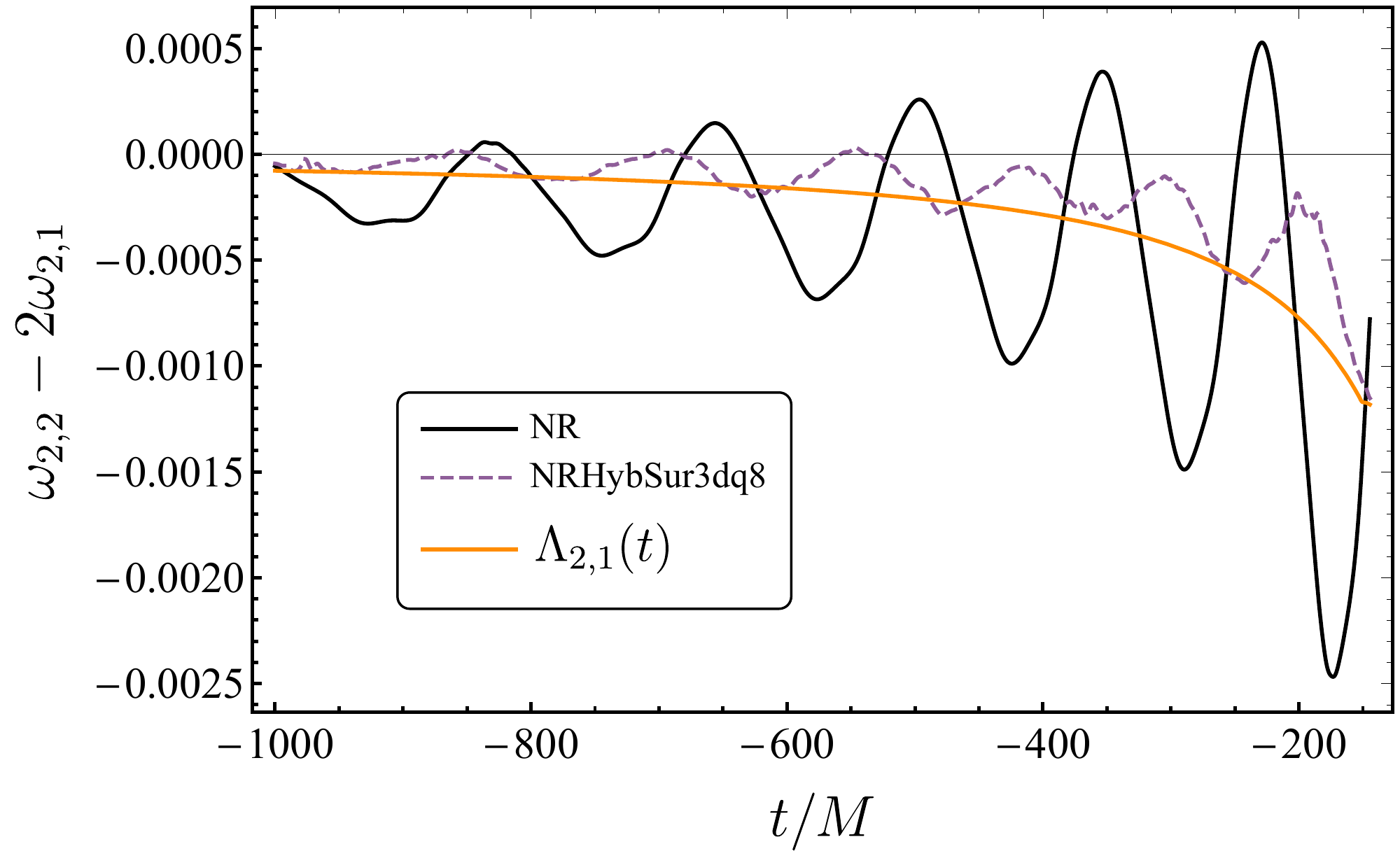}
    \includegraphics[width=0.48\textwidth]{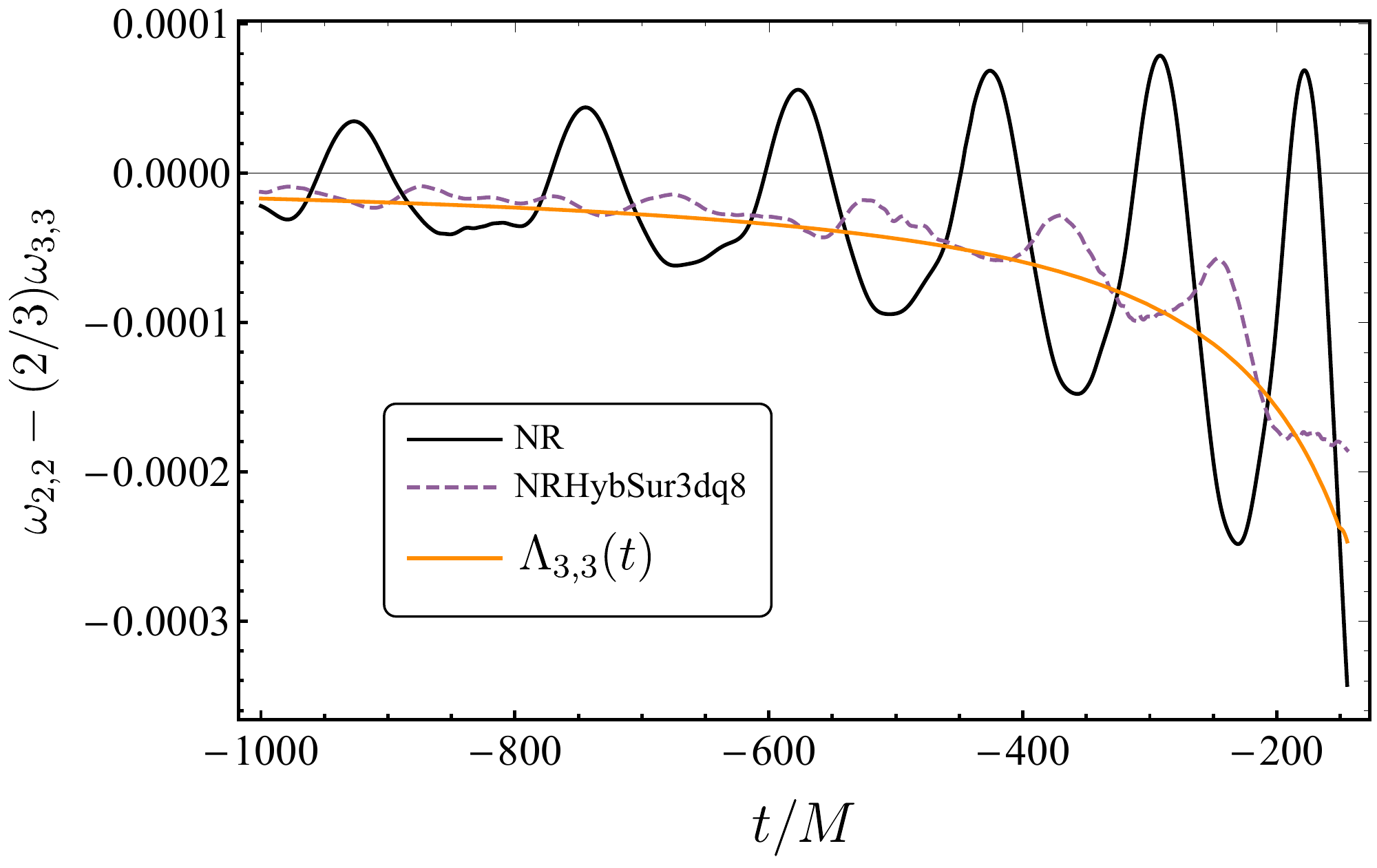}
    \caption{Frequency contribution from complex inspiral amplitude. A weighted frequency difference is shown for the modes $(l=2,m=1)$ and $(l=3,m=3)$ with respect to the $(l=2,m=2)$ frequency for the SXS NR simulation SXS:BBH:0290 with parameters $q=3$, $\chi_1=0.6$, $\chi_2=0.4$. A difference of zero would correspond to the approximation from Eq.~(\ref{eq:inspapprox}) being exact. It can be seen that the contribution $\Lambda_{lm}$ from the complex inspiral amplitude matches the behaviour of the deviation as compared with the NR simulation and to the model \texttt{NRHybSur3dq8}.}
    \label{fig:lambdapn}
\end{figure}

Since the inspiral amplitude is already a complex quantity, in the model implementation there is no need to compute the phase contribution $\Lambda_{lm}(t)$ if the modes are constructed as the complex amplitude times the complex phase exponential:
\begin{equation}
\label{eq:hlminsp}
    h^\mathrm{insp}_{lm}(t)=H_{lm}^\mathrm{insp}(t)\exp[(m/2)\phi_{22}^\mathrm{insp}(t)- \Delta\phi_{lm}].
\end{equation}
Phases and frequencies computed from the modes described by Eq.~(\ref{eq:hlminsp}) will satisfy Eq.~(\ref{eq:inspphaserelation}). As an example of the relation between $\Lambda_{lm}(t)$ and the deviation from approximation (\ref{eq:inspapprox}), in Fig. \ref{fig:lambdapn} the time derivative of $\Lambda_{lm}(t)$ and the deviation of the frequency from the scaling approximation are shown for the $l=2,m=1$ and $l=3,m=3$ modes of a NR simulation and the surrogate model \texttt{NRHybSur3dq8} for the same parameters, where it can be observed that $d\Lambda_{lm}(t)/dt$ correctly reproduces the behaviour.

\subsection{Ringdown}\label{sec:ringdown}

For describing the modes in the ringdown region, we employ the proposal described in \cite{damour2014new}, that we briefly summarise here. The damped emission of the final black hole can be well approximated in terms of linear perturbations of the Kerr solution, for which analytical approximations are known in the framework of perturbation theory in terms of a linear combination of damped quasinormal modes (QNMs) \cite{Kokkotas_1999}:

\begin{equation}\label{eq:QNM_decomp}
h^{RD}_{lm}(t)=\sum_{n=1}^{\infty}c_{nlm}\exp[i\sigma_{nlm}(t-t_0)],
\end{equation}
where indices $n$, $l$ and $m$ refer to the energy level and to the spin-weighted spheroidal mode, $c_{lnm}$ are amplitude coefficients and $\sigma_{nlm}$ is the complex frequency of the mode level, from which the asymptotic final frequency $\omega_{nlm}^{\text{RD}}$ and damping frequency $\alpha_{nlm}$ can be obtained:
\begin{subequations}
\begin{align}
    \omega_{nlm}^{\text{RD}} &= \Re (\sigma_{nlm}),\\
    \alpha_{nlm} &= \Im (\sigma_{nlm}).
\end{align}
\end{subequations}

The $n=1$ level, also called the ground state QNM, is the dominant QNM at late times, since is the most long-lived. Defining a QNM-rescaled mode as
\begin{equation}
    \bar{h}_{lm}(t)=e^{i\sigma_{1lm}(t)}h_{lm}(t),
\end{equation}
the early ringdown behaviour can be modelled with the following phenomenological expressions:

\begin{equation}
\label{eq:rdomega}
\begin{split}
    \bar{\omega}_{lm}(t)&=\omega_{lm}(t) - \omega^\mathrm{RD}_{1lm}\\
    &=c_1\dfrac{c_2(c_3e^{-c_2t} + 2c_4e^{-2c_2t})}{1+c_3e^{-c_2t} + c_4e^{-2c_2\ t}},
\end{split}
\end{equation}

\begin{equation}
\label{eq:rdamp}
\begin{split}
    |\bar{h}_{lm}(t)|&=e^{\alpha_{1lm}(t-t^{peak}_{lm})}|h|\\
    &=d_1\tanh[d_2 (t-t^\mathrm{peak}_{lm})+d_3] + d_4,
\end{split}
\end{equation}
where the following coefficients are set by imposing that the amplitude matches the value at the peak amplitude time $t^{peak}_{lm}$, the frequency matches the value at $t=0$ and that the amplitude derivative at the peak is zero:

\begin{subequations}
\label{eq:rdconstraints}
\begin{align}
    c_1 &= \dfrac{1 + c_3 + c_4}{c_2(c_3 + 2c_4)}\left(\omega_{1lm}^\mathrm{RD} - \omega_{lm}(t=t^\mathrm{peak}_{22})\right),\\
    \label{eq:c2rd}
    c_2 &=(\alpha_{2lm}-\alpha_{1lm})/2,\\
    d_1 &= H_{lm}(t=t^\mathrm{peak}_{lm})\dfrac{\alpha_{1lm}\cosh^2(d_3)}{d_2},\\
    \label{eq:d2rd}
    d_2 &=(\alpha_{2lm}-\alpha_{1lm})/2,\\
    d_4 &= H_{lm}(t=t^\mathrm{peak}_{lm}) - d_1\tanh(d_3),
\end{align}
\end{subequations}
and $c_3$, $c_4$, $d_3$ are free coefficients calibrated with NR data.

However, in the present work we choose a different treatment of the phase coefficients to improve accuracy in regions where the relation (\ref{eq:c2rd}) is not well satisfied, particularly for high unequal mass systems. In our model, $c_2$ is treated as a free coefficient to calibrate across parameter space for each mode while maintaining simplicity and avoiding complicated interactions between the different coefficients fits and $c_4$ is set to 0 (from Eq.~(\ref{eq:rdomega}) it can be seen that setting $c_4=0$ transforms the ansatz into a $\tanh$ function, as for the amplitude ansatz). While it can be argued that for the same reason the amplitude condition (\ref{eq:d2rd}) should be relaxed, as is done in other models that employ the same ringdown description as \texttt{SEOBNRv4} and \texttt{SEOBNRv4HM} \cite{Cotesta_2018}, we have found that maintaining the original amplitude treatment gives a sufficiently accurate ansatz, so in favour of simplicity and to not complicate the behaviour of the ansatz across parameter space we decided to maintain relation (\ref{eq:d2rd}).

One caveat that has been found during the construction of this model is that for the $l=5, m=5$ mode the amplitude ansatz (\ref{eq:rdamp}) does not seem to correctly reproduce the available NR data. This could be due to numerical issues, since in general the  $l=5, m=5$ is weak and difficult to extract in numerical simulations. However, this behaviour has been found in a significant number of cases, pointing to some physical explanation, that could be the interaction of higher overtones or some mode mixing effect with other modes that complicates the description in the spherical harmonic basis representation. Investigation of this issue and possible improvements of the ringdown modelling is left for future work.

In a similar spirit, we have investigated the modelling of spherical harmonic modes with mode-mixing contributions. In particular, the phenomenological model \texttt{IMRPhenomXHM} includes the $l=3$, $m=\pm2$ modes, which as discussed in \cite{garcaquirs2020imrphenomxhm} present an important contribution from the $l=2$, $m=2$ mode in the ringdown regime. We have followed the strategy of \cite{garcaquirs2020imrphenomxhm} to apply our ringdown ansatz to the $l=3$, $m=\pm2$ modes in a decomposition into spin-weighted spheroidal harmonics, and then rotate back to the spin-weighted spherical harmonic modes. As is well known, spin-weighted spheroidal spherical harmonics provide a natural basis for ringdown in the Kerr spacetime of the remnant black hole. We thus perform the phenomenological fits for the inspiral and merger in the spherical basis, and for the ringdown in the spheroidal basis, using the free coefficients from Eq.~(\ref{eq:rdconstraints}) in that basis, before rotating the resulting spheroidal modes back to the spherical basis needed for constructing the polarisations.

Therefore, it is required to calibrate $H_{lm}(t^\mathrm{peak}_{lm})$ and $\omega_{lm}(t^\mathrm{peak}_{22})$ for each mode across parameter space. The values of $\omega_{1lm}^\mathrm{RD}$, $\alpha_{1lm}$ and $\alpha_{2lm}$ can be obtained from perturbation theory as functions of the final mass and final spin of the remnant black hole. We have performed 1D fits of $\omega_{1lm}^\mathrm{RD}$, $\alpha_{1lm}$ and $\alpha_{2lm}$ as a function of the dimensionless final spin (i.e, the Kerr parameter) employing the available numerical data from \cite{Berti_2009}. In order to evaluate these fits, a calibrated function for the final mass and final spin across parameter space is needed. There are several NR fits in the literature \cite{Tichy_2008,Rezzolla_2008,Buonanno_2008,Barausse_2012,Healy_2014,Hofmann_2016,Jim_nez_Forteza_2017,Varma_2019_remnant} for the properties of the remnant black hole, and this model employs the results from \cite{Jim_nez_Forteza_2017} which are also employed in the phenomenological models \texttt{IMRPhenomXHM} and \texttt{IMRPhenomXAS}. Finally, for the three free coefficients $d_3$, $c_3$ and $c_4$ a calibration to numerical relativity is performed across parameter space.

\subsection{Merger}\label{sec:merger}

\begin{figure}
    \centering
    \includegraphics[width=0.48\textwidth]{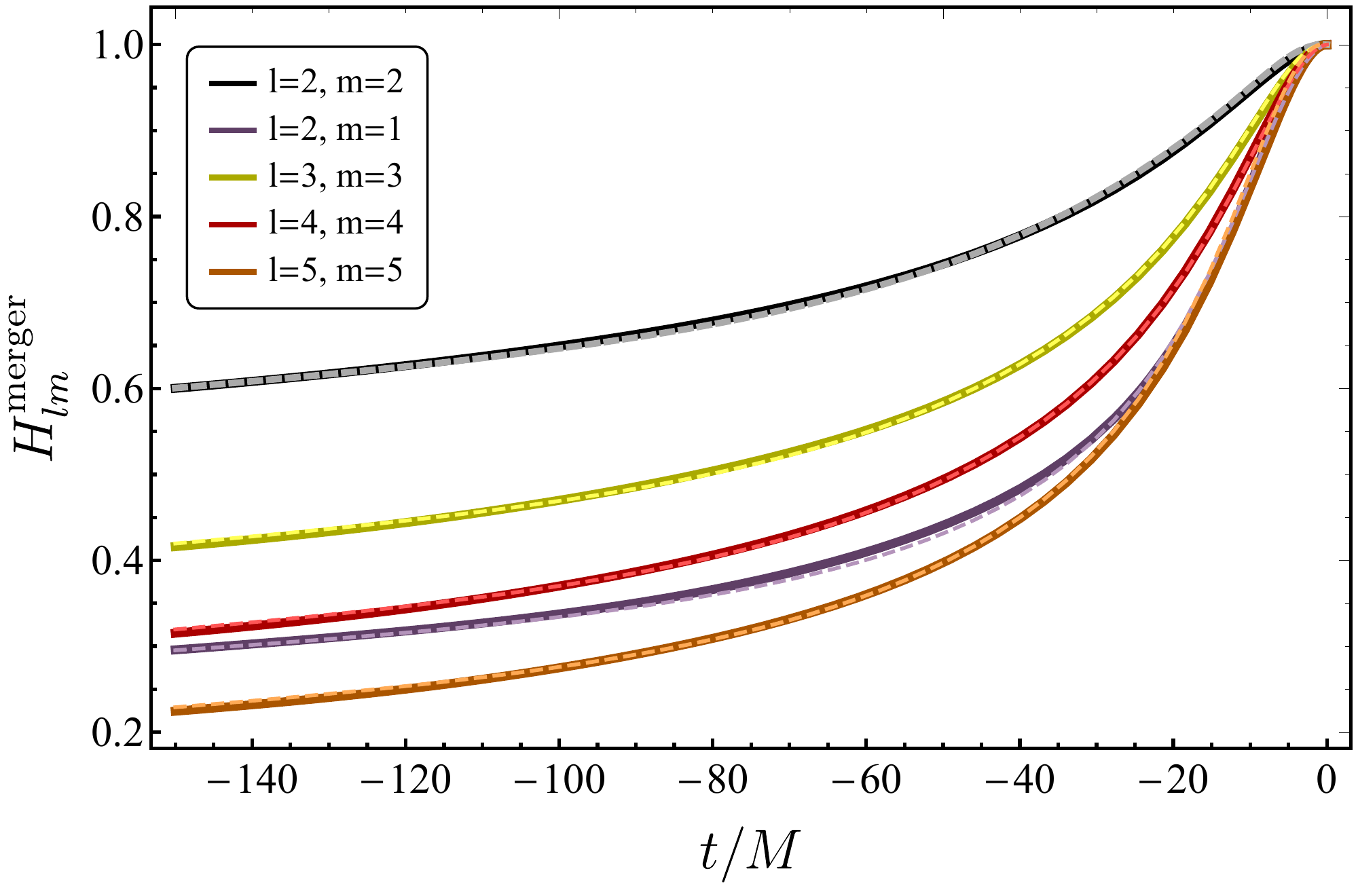}
    \caption{Amplitude of the different modes normalized to the peak amplitude and peak aligned at $t=0$, for the SXS NR simulation \texttt{SXS:BBH:1460} with parameters $q=8$, $\chi_1=0.124$, $\chi_2=0.109$. Solid: SXS NR, dashed: \texttt{IMRPhenomTHM}.} 
    \label{fig:mergeramp}
\end{figure}

To describe the behaviour in the merger regions of the amplitude and frequency of each mode, we present phenomenological expressions based on hyperbolic functions, which reproduce well the NR behaviour:
\begin{equation}
\label{eq:mergeromega}
    \omega^\mathrm{merger}_{lm}(t)=\omega_{1lm}^\mathrm{RD}\left(1 - \bar{\omega}^\mathrm{merger}_{lm}(t)\right),
\end{equation}
where
\begin{equation}
\label{eq:mergeromegarescaled}
    \bar{\omega}^\mathrm{merger}_{lm}(t)=\sum_{k=0}^{k=4}a_k\text{arcsinh}^k(\alpha_{1lm}t),
\end{equation}
\begin{equation}
\label{eq:mergeramp}
\begin{split}
    H_{lm}^\mathrm{merger}(t)= & b_0 + b_1 \tau_{lm}^2 + b_2 \text{sech}^{1/7}(2\alpha_{1lm} \tau_{lm})\\
    &+ b_3 \text{sech}(\alpha_{1lm} \tau_{lm}),
\end{split}
\end{equation}
and $\alpha_{1lm}$ is the damping frequency of the ground state QNM of the $(l,m)$ mode.
$a_i$, $b_i$ are phenomenological coefficients to be determined by imposing continuity and differentiability at the boundaries with the inspiral and ringdown regions and by imposing the ansatz to match a particular frequency or amplitude value at a specific time $t_{cp}^\mathrm{merger}=-25M$, and $\tau_{lm}=t-t^\mathrm{peak}_{lm}$ where $t^\mathrm{peak}_{lm}$ is the peak amplitude time of each mode. As an illustrative example, we show in Fig.~\ref{fig:mergeramp} the normalized amplitude of the different modes for a particular SXS NR simulation and the corresponding solution from Eq.~(\ref{eq:mergeramp}), aligning the peak of the modes at $t=0$.

In order to obtain a value for the coefficients for each case, a linear system of equations is imposed for guaranteeing continuity and differentiability at the region boundaries, and a collocation point is placed at a fixed time of both for the amplitude and the frequency:
\begin{subequations}
\begin{align}
\omega^\mathrm{merger}_{lm}(t^{\omega}_\mathrm{cut})&=\omega^\mathrm{insp}_{lm}(t^{\omega}_\mathrm{cut}),\\
\omega^\mathrm{merger}_{lm}(t^\mathrm{peak}_{22})&=\omega^\mathrm{RD}_{lm}(t^\mathrm{peak}_{22}),\\
\omega^\mathrm{merger}_{lm}(t^{\omega}_{\text{merger},lm})&=\lambda^{\omega}_{\text{merger},lm},\\
\frac{d}{dt}\omega^\mathrm{merger}_{lm}(t^{\omega}_\mathrm{cut})&=\frac{d}{dt}\omega^\mathrm{insp}_{lm}(t^{\omega}_\mathrm{cut}),\\
\frac{d}{dt}\omega^\mathrm{merger}_{lm}(t^\mathrm{peak}_{22})&=\frac{d}{dt}\omega^\mathrm{RD}_{lm}(t^\mathrm{peak}_{22}),
\end{align}
\end{subequations}
and 
\begin{subequations}
\begin{align}
H^\mathrm{merger}_{lm}(t^{H}_\mathrm{cut})&=H^\mathrm{insp}_{lm}(t^{H}_\mathrm{cut}),\\
H^\mathrm{merger}_{lm}(t^\mathrm{peak}_{lm})&=H^\mathrm{RD}_{lm}(t^\mathrm{peak}_{lm}),\\
H^\mathrm{merger}_{lm}(t^H_\mathrm{merger})&=\lambda^{H}_{\mathrm{merger},lm},\\
\frac{d}{dt}H^\mathrm{merger}_{lm}(t^{H}_\mathrm{cut})&=\frac{d}{dt}H^\mathrm{insp}_{lm}(t^H_\mathrm{cut}).
\end{align}
\end{subequations}

\begin{figure}
    \centering
    \includegraphics[width=0.48\textwidth]{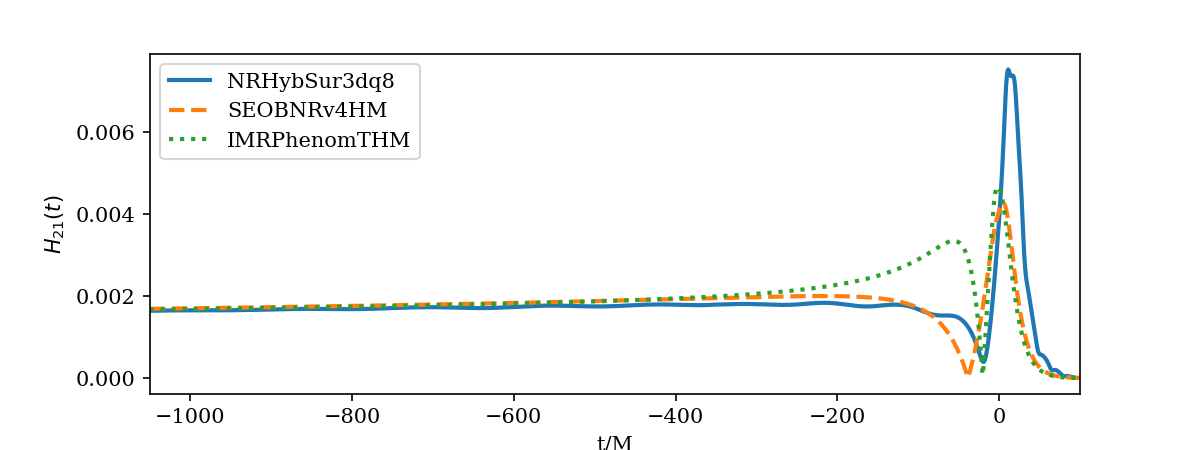}
    \includegraphics[width=0.48\textwidth]{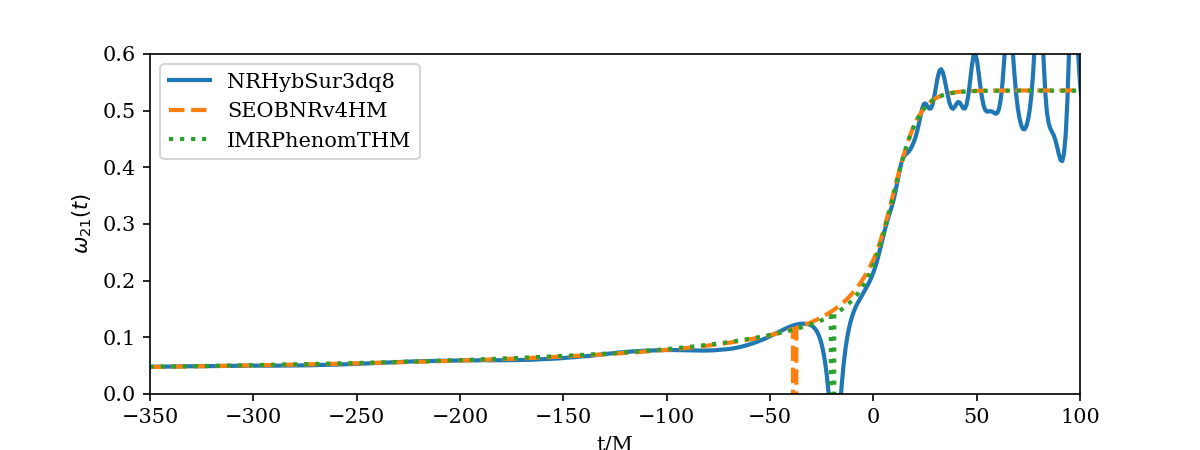}
    \includegraphics[width=0.48\textwidth]{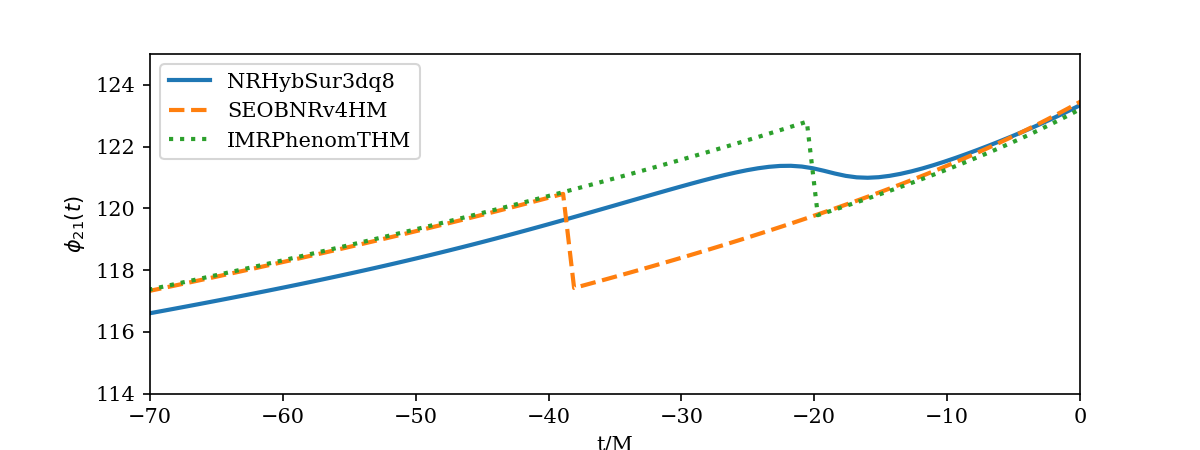}
    \caption{Example of transition to negative amplitude for the $l=2$, $m=1$ mode, for a case with parameters $q=2.894$, $\chi_1=0.871$, $\chi_2=-0.849$. Top panel: absolute value of the amplitude for the different models, where the point can be observed where the absolute amplitudes vanish. Mid panel: frequency becomes negative at the points of vanishing amplitude. Bottom panel: phase suffers a discontinuity of $\pi$ at the times of vanishing amplitude. (For \texttt{NRHybSur3dq8} this transition is softened due to the model specific construction, which does not allow for sudden discontinuities.)}
    \label{fig:mergeramp_singchange}
\end{figure}

The inspiral amplitude described in Eq.~(\ref{eq:inspamp}) is a complex quantity as we have discussed, however Eq.~(\ref{eq:mergeramp}) is a real function. In order to propagate the phase contribution from the complex inspiral amplitude to the merger, we evaluate the value of $\Lambda_{lm}(t)$ from Eq.~(\ref{eq:lambda}) at the boundary time between the inspiral and merger regions, $\Lambda_{lm}^{\text{merger}}=\Lambda_{lm}(t^H_{\text{cut}})$, and we add this contribution as a constant phase offset to the merger phase jointly with the global amplitude rotation that contains the information of the relative orientation between the modes:
\begin{equation}
    \phi_{lm}^{\text{merger}}(t)=\Lambda_{lm}^{\text{merger}} - \Delta\phi_{lm} + \int_{t'=t^{\omega}_{\text{cut},lm}}^{t}dt' \omega^{\text{merger}}_{lm}(t').
\end{equation}
This allows to keep the sign from the real part of the inspiral amplitude as well as the relative mode orientation, and since Eq.~(\ref{eq:mergeramp}) is a real quantity it also allows to track possible sign changes, i.e discontinuities of $\pi$ in the phase that could occur in the merger region. We have found during the validation of the model that this phenomenon indeed happens in some region of the parameter space for the $l=2,\ m=1$ mode, for mass ratios close to $m_1/m_2=3$, high positive primary spin $\chi_1$ and high spin difference $\chi_1-\chi_2$. Although we have not encountered this situation in NR simulations, two different models with very different modelling procedures agree in reproducing this phenomenon, the surrogate model \texttt{NRHybSur3dq8} and the Effective-One-Body model SEOBNRv4HM. A representative case of this region of parameter space is shown in Fig.~\ref{fig:mergeramp_singchange}. While it can be seen that agreement between the different models is quantitatively poor, the qualitative behaviour is shared among them. 

\section{Calibration across parameter space}\label{sec:calib}

Once the phenomenological ans{\"a}tze for describing the amplitude and frequency of each mode have been specified, there is a set of quantities that needs to be specified in order to reconstruct the modes. In summary, the following quantities have to be specified for each mode (a total of 16 for the $l=2$, $m=2$ mode and 10 for the subdominant modes):
\begin{itemize}
\item Six inspiral frequency collocation points for the $l=2$, $m=2$ mode:
\begin{equation}
\label{eq:inspwcoef}
\{\lambda^{\omega}_i\}_{i=1,...,6}  	
\end{equation}
\item Three inspiral amplitude collocation points for each mode:
\begin{equation}
\label{eq:inspampcoef}
\{\lambda^{H}_{lm,i}\}_{i=1,2,3} 	
\end{equation}
\item Two merger collocation points for each mode:
\begin{equation}
\label{eq:mergerwcoef}
\{\lambda^{\omega}_{\text{merger},lm},\ \lambda^{H}_\text{merger,lm}\}    	
\end{equation}
\item Peak amplitude and frequency at $t^{\text{peak}}_{22}$ for each mode:
\begin{equation}
\label{eq:peakquan}
\{\omega_{lm}(t^{\text{peak}}_{22}),\ H_{lm}(t^{\text{peak}}_{lm})\}    	
\end{equation}
\item Three ringdown coefficients for each mode:
\begin{equation}
\label{eq:rdampcoef}
c^{lm}_{2},\  c^{lm}_{3},\ d^{lm}_{3}.
\end{equation}
\end{itemize}

\subsection{Calibration procedure}\label{sec:calibmethod}

The above quantities are calibrated to a dataset of NR simulations, numerical Teukolsky waveforms and \texttt{SEOBNRv4} waveforms (details on this dataset are given in the next subsection) as functions of the intrinsic binary parameters $\{\eta,\chi_1,\chi_2\}$. As in the previous work \cite{estells2020imrphenomtp} and as it is done in the phenomenological models \texttt{IMRPhenomXAS} \cite{pratten2020setting} and \texttt{IMRPhenomXHM} \cite{garcaquirs2020imrphenomxhm}, we employ the hierarchical data-driven fitting approach \cite{Jim_nez_Forteza_2017} to calibrate the parameter space functions, which we briefly summarise here.

The approach consists of constructing a full parameter space ansatz through a sequence of fits to lower dimensional subspaces,  which are typically more densely populated with numerical simulations. First,  1-dimensional fits are performed for the dependence on the symmetric mass-ratio for the non-spinning subset of the dataset, and the spin dependence
of equal black holes. In order to capture in a simpler way the full two-dimensional spin-dependence, it is re-parameterized in terms of a dominant effective spin $\hat{S}$ and the spin difference $\Delta\chi$:
\begin{subequations}
    \begin{align}
    &\hat{S} = \frac{m_1^2\chi_1^2 + m_2^2\chi_2^2}{m_1^2+m_2^2},\\
    &\Delta\chi = \chi_1 - \chi_2,
    \end{align}
\end{subequations}
which were employed in the construction of the final state fits of \cite{Jim_nez_Forteza_2017} and in some of the fits constructed for \texttt{IMRPhenomXAS} and \texttt{IMRPhenomXHM} models. For the dominant effective spin effects, one performs another 1-dimensional fit for the equal spin data at a particular mass-ratio, which typically is set to $q=1$ for the fits of quantities of $m-$even modes, and to $q=2$ for quantities of the $m-$odd modes, since the odd modes vanish for $q=1$ equal spin cases due to symmetry. With both 1D fits, one constrains a 2D fit $f_{2D}(\eta,\hat{S})$ over the equal spin subset. From this fit, residuals with the unequal spin cases are computed, and these residuals are fitted as a function $h(\eta,S,\Delta\chi)$ of spin difference, effective spin and mass-ratio. The general form of a parameter space fit for a specific quantity is:
\begin{equation}
    \text{fit}(\eta,\hat{S},\Delta\chi)=g(\eta)f_{2D}(\eta,\hat{S}) + h(\eta,S,\Delta\chi),
\end{equation}
where $g(\eta)$ is a scale function of the symmetric mass ratio for taking into account known effects at the symmetric mass ratio boundaries, like vanishing amplitude when $\eta\rightarrow 0$ (then $g(\eta)=\eta$) or vanishing amplitude for the odd modes when both black holes are identical ($g(\eta)=\eta\delta m$). Further explanation and details on this procedure can be found in \cite{pratten2020setting} and \cite{garcaquirs2020imrphenomxhm}, since the procedure followed here is very similar to the procedure followed in the \texttt{PhenomX} family of waveform models.

\subsection{Calibration dataset}\label{sec:dataset}

\begin{figure}
    \centering
    \includegraphics[width=0.5\textwidth]{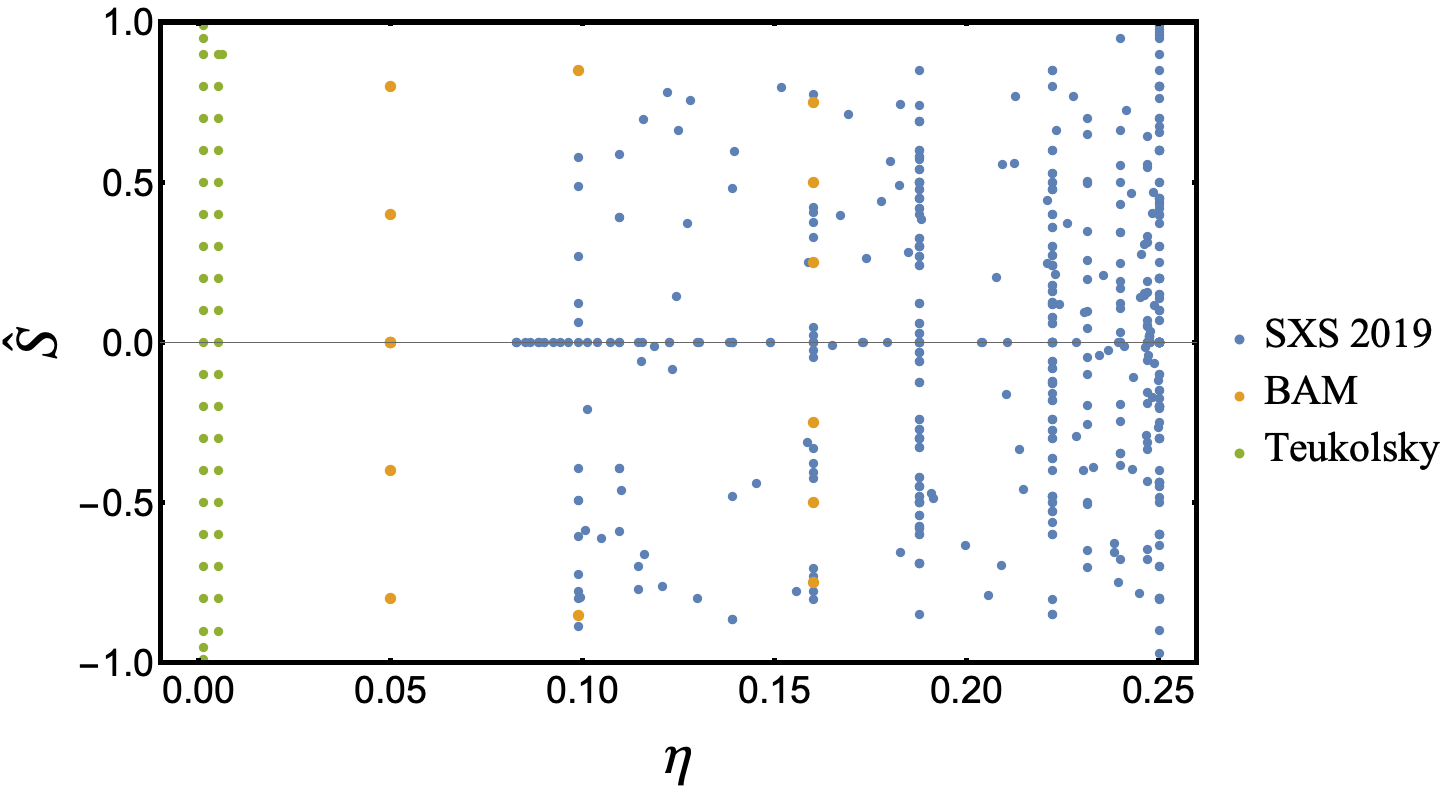}
    \caption{Parameter space coverage of the NR and Teukolsky dataset employed for the calibration.} 
    \label{fig:dataset}
\end{figure}

The data employed for the model calibration consist of 531 BBH nonprecessing quasi-circular NR simulations performed with the SpEC code and released in the SXS Collaboration Catalog of NR simulations \cite{Boyle_2019}, 15 simulations from the BAM code and 61 cases obtained from numerical solutions of the perturbative Teukolsky equation through the numerical code $\it{Teukode}$ \cite{Harms_2014}. The set of SpEC simulations offers a good coverage of the parameter space region below $q=10$, though the number of equal spin cases above $q=4$ is reduced. For this reason, we incorporate the BAM simulations at equal spin points placed at $q=4$ and $q=8$. We also employ BAM simulations at $q=18$ for different values of the primary spin, which help to connect the comparable masses region with the high mass ratio region of the Teukolsky waveforms, which are placed at $q=80$, $q=200$ and $q=1000$ for different spin values. However, we have treated the $q=18$ as equal spin cases, since the effective spin parameter employed is almost the same assuming that the secondary spin has the same value as the primary spin and since this helps the conditioning of the 2D fits in $\{\eta,\hat{S}\}$. Figure~\ref{fig:dataset} shows the parameter space coverage for $\eta$ and the effective spin $\hat{S}$.

For the $l=2$, $m=2$ inspiral frequency collocation points $\{\lambda^{\omega}_i\}$ that are placed at earlier times than the starting time of a particular NR simulation, the value has been taken from waveforms produced with the \texttt{SEOBNRv4} model at the same parameters as in the corresponding NR simulation. This procedure is completely equivalent to employing data from the EOB-NR hybridised waveforms that were employed in the calibration of \texttt{IMRPhenomXAS} and \texttt{IMRPhenomXHM}, since the waveforms were hybridised employing also the \texttt{SEOBNRv4} model.

For the SpEC simulations, we have selected the highest resolution level available from which we have used the rhOverM\_Asymptotic\_GeometricUnits\_CoM.h5 file, which contains the RZW strain with the center of mass correction as explained in \cite{Boyle_2019}. From this file, we select the $N=3$ extrapolated level for the inspiral and merger collocation points of the different modes, but we select the $N=2$ extrapolated level for the peak and ringdown quantities, since we have found that in general this level contains less noise for the peak and post-peak regions of the waveform modes.

For the BAM simulations and the Teukolsky waveforms, EOB-NR hybrid waveforms for each mode are employed. These hybrids, also used in the calibration of IMRPhenomXAS and IMRPhenomXHM, reproduce the $\psi_4$ Weyl scalar, and we obtain the strain through fixed-frequency integration \cite{Reisswig_2011}. The fixed-frequency integration technique is based on Fourier-transforming the $\psi_4$ waveform and performing the double time integration in the Fourier domain, which is equivalent to multiplying the waveform by a factor $-1/(4\pi^2 f^2)$, where $f$ is the Fourier frequency. However, to eliminate the effect of spectral leakage  and removing spurious non-linear drifts, $f$ is set to a value $f_0$ in the region outside the required frequency range of the final time domain strain waveform following \cite{Reisswig_2011}. As discussed in \cite{Reisswig_2011}, the cutoff frequency $f_0$ needs to be adjusted to the waveform. Too low values of $f_0$ lead to noisier time domain waveform, whereas too high values typically cause a loss of amplitude at late times.

\section{Performance of the phenomenological model}\label{sec:results}

In order to test and validate the construction and calibration of the model, as presented in the previous sections, here we will present a series of tests. First, we compare the model with the public LVCNR Catalog of NR simulations and with the surrogate model \texttt{NRHybSur3dq8}, which accurately reproduces and interpolates a hybridised version of the SXS NR Catalog, employing the usual mismatch quantity for assessment. Comparison with the phenomenological model \texttt{IMRPhenomXHM} is also presented, which allows to assess the validity of the model in more challenging regions of parameter space. And finally, a computation of the gravitational wave recoil, with a discussion of the accuracy implications of this quantity, will be compared to public NR fits for this quantity and with the results of other multimode models.

Mismatch is a common quantity in the gravitational wave modelling community to assess the disagreement between two gravitational waveforms. It is defined in terms of a noise-weighted inner product called overlap:
\begin{equation}
\label{eq:overlap}
    (h_1(t)|h_2(t))=4\Re\int_{f_\text{min}}^{f_\text{max}} df\dfrac{\tilde{h}_1(f)\tilde{h}^*_2(f)}{S_n(f)}
\end{equation}
where $\tilde{h}(f)$ is the Fourier transform of $h(t)$, $^*$ denotes complex conjugation and $S_n(f)$ is the power spectral density of a detector at a particular frequency $f$. From it, the faithfulness or match is obtained by maximising the normalised overlap over a global phase and time translation:
\begin{equation}
    \mathcal{M}\equiv\max_{\phi_0,\Delta t}\dfrac{(h_1(t)|h_2(t))}{\sqrt{|h_1(t)||h_2(t)|}},
\end{equation}
where the norm of a waveform $|h(t)|$ is defined in terms of the overlap product. Mismatch is the deviation of the match from unity:
\begin{equation}
\label{eq:mismatch}
    \mathcal{MM}\equiv 1-\mathcal{M}.
\end{equation}
Two completely equivalent waveform will have zero mismatch, while two completely different waveform will have a mismatch close to unity.

\subsection{Comparison with LVCNR Catalogue}\label{subsec:lvcmatches}

A first performance test of the model is the comparison with the dataset of NR simulations from the LVCNR Waveform Catalog \cite{schmidt2017numerical}, in particular with 110 simulations from the SXS Collaboration \cite{SXS:catalog} produced by the SpEC code \cite{SXS:code}. NR simulations are in general the most trustworthy source of information about the gravitational wave emission in the nonlinear highly dynamical regime of the binary merger, and provide information of this regime to the different calibrated waveform model families. Therefore, an important test for a NR calibrated waveform model is to show a good agreement with numerical relativity.

\begin{figure}[th!]
    \centering
        \includegraphics[width=0.48\textwidth]{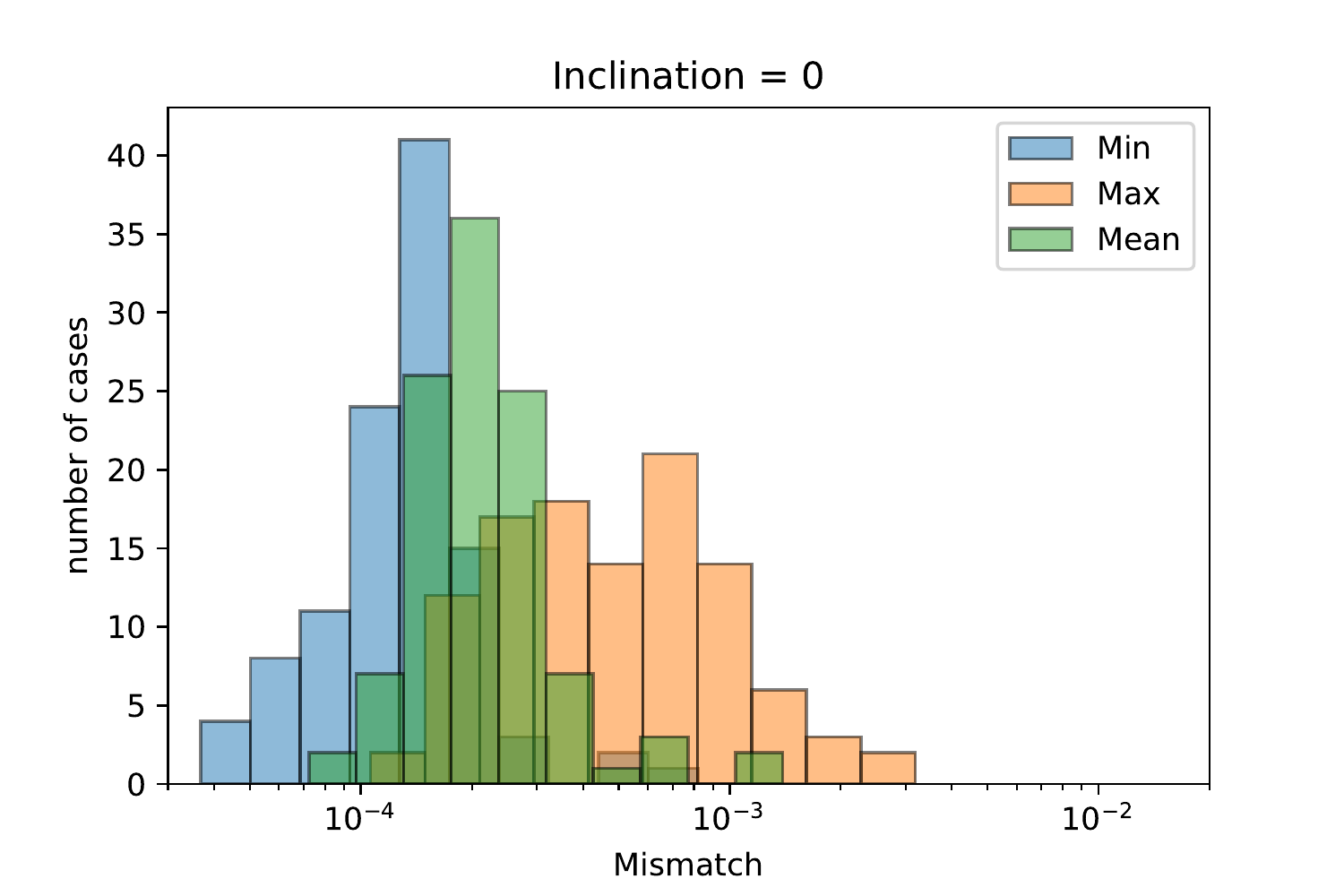}
        \includegraphics[width=0.48\textwidth]{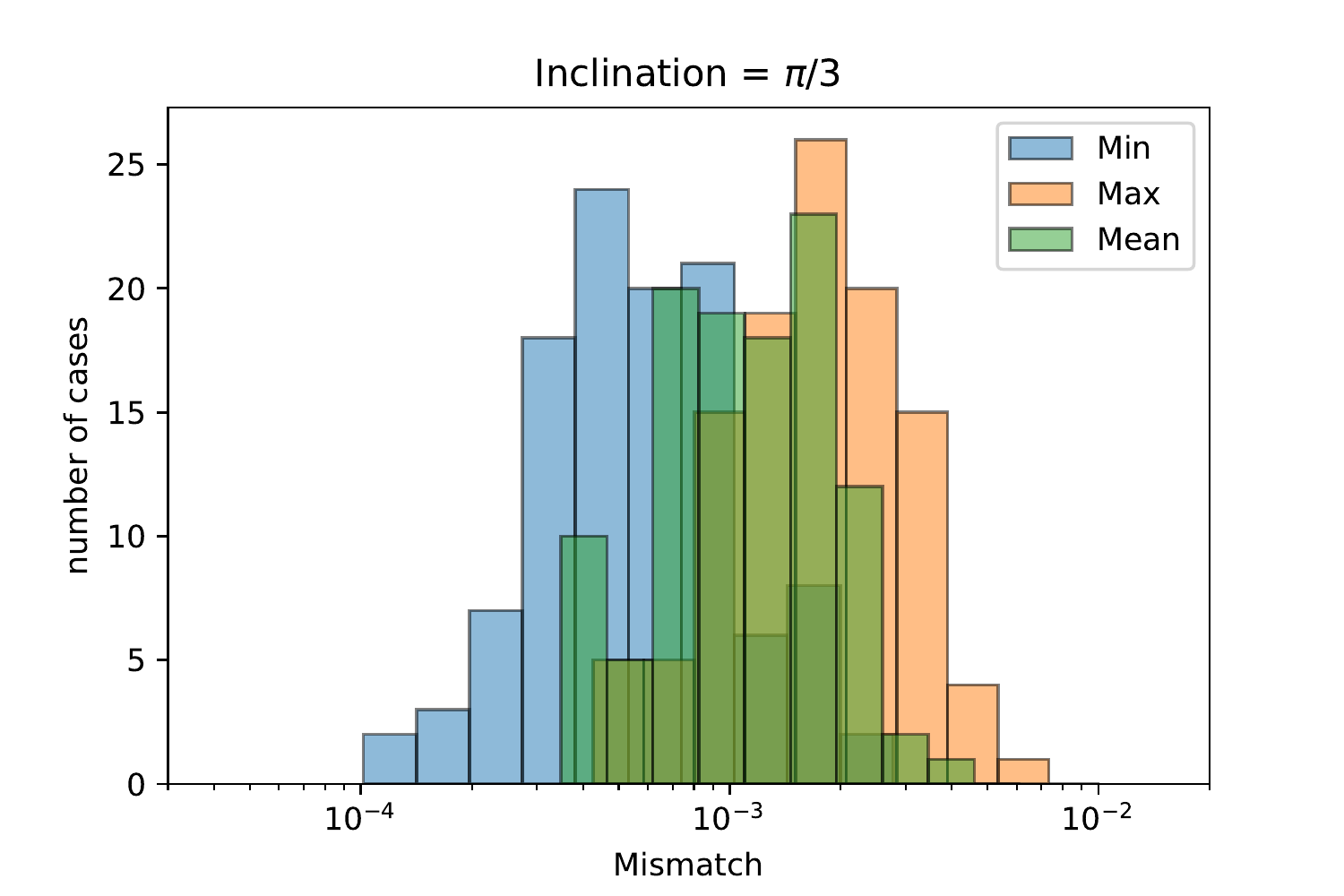}
        \includegraphics[width=0.48\textwidth]{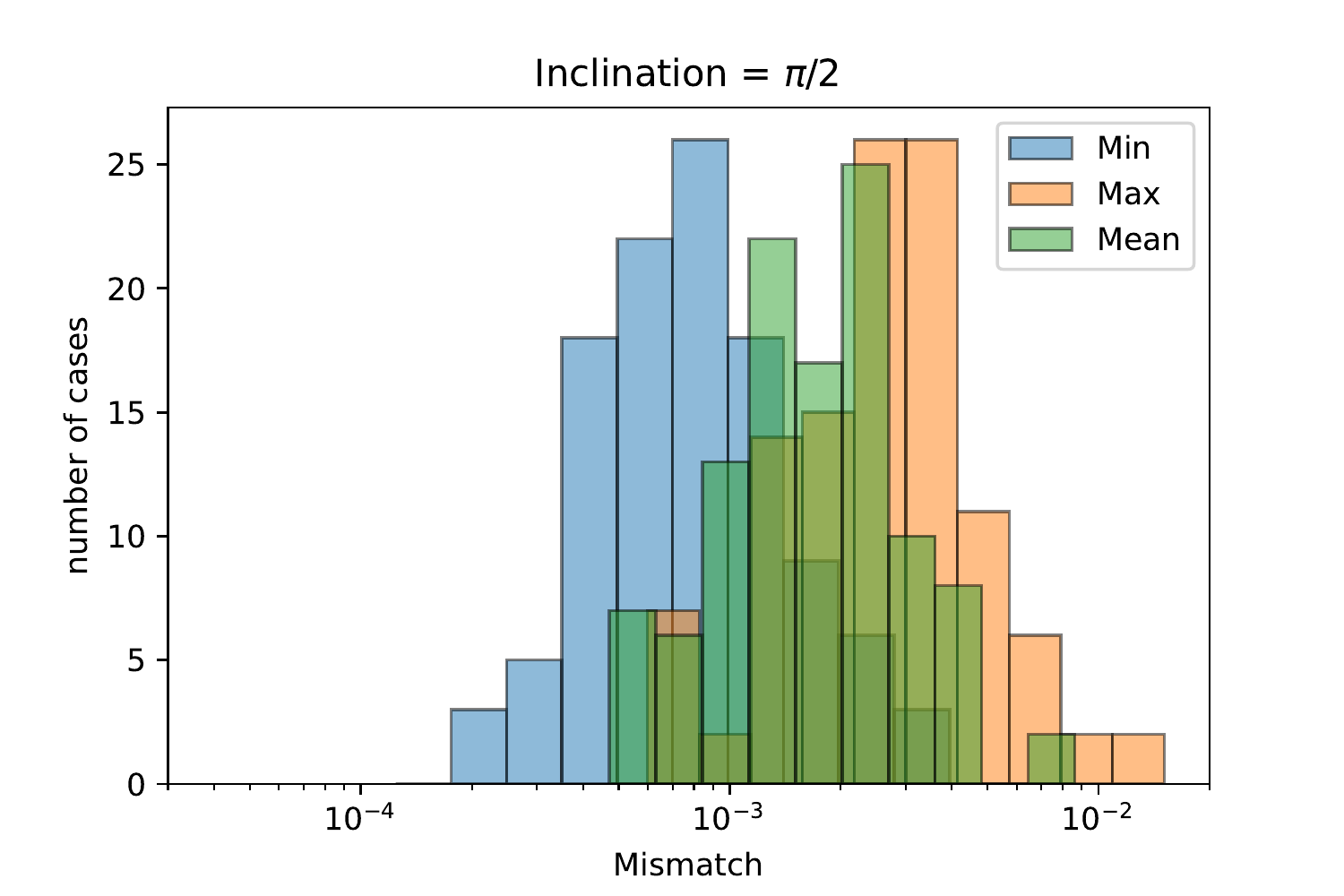}
    \caption{Mismatch distributions of IMRPhenomTHM polarisations with the LVCNR SXS Catalog. Histograms for the minimum, mean and maximum mismatch for each case are shown for the three different inclinations employed in the comparison. It can be seen that for face-on configuration and $\pi/3$ inclination, all cases are below $1\%$ mismatch, while for edge-on configurations ($\pi/2$ inclination) a few cases are above $1\%$ mismatch.}
\label{fig:mmpolarizsxs}
\end{figure}

\begin{figure}[th]
    \centering
        \includegraphics[width=0.24\textwidth]{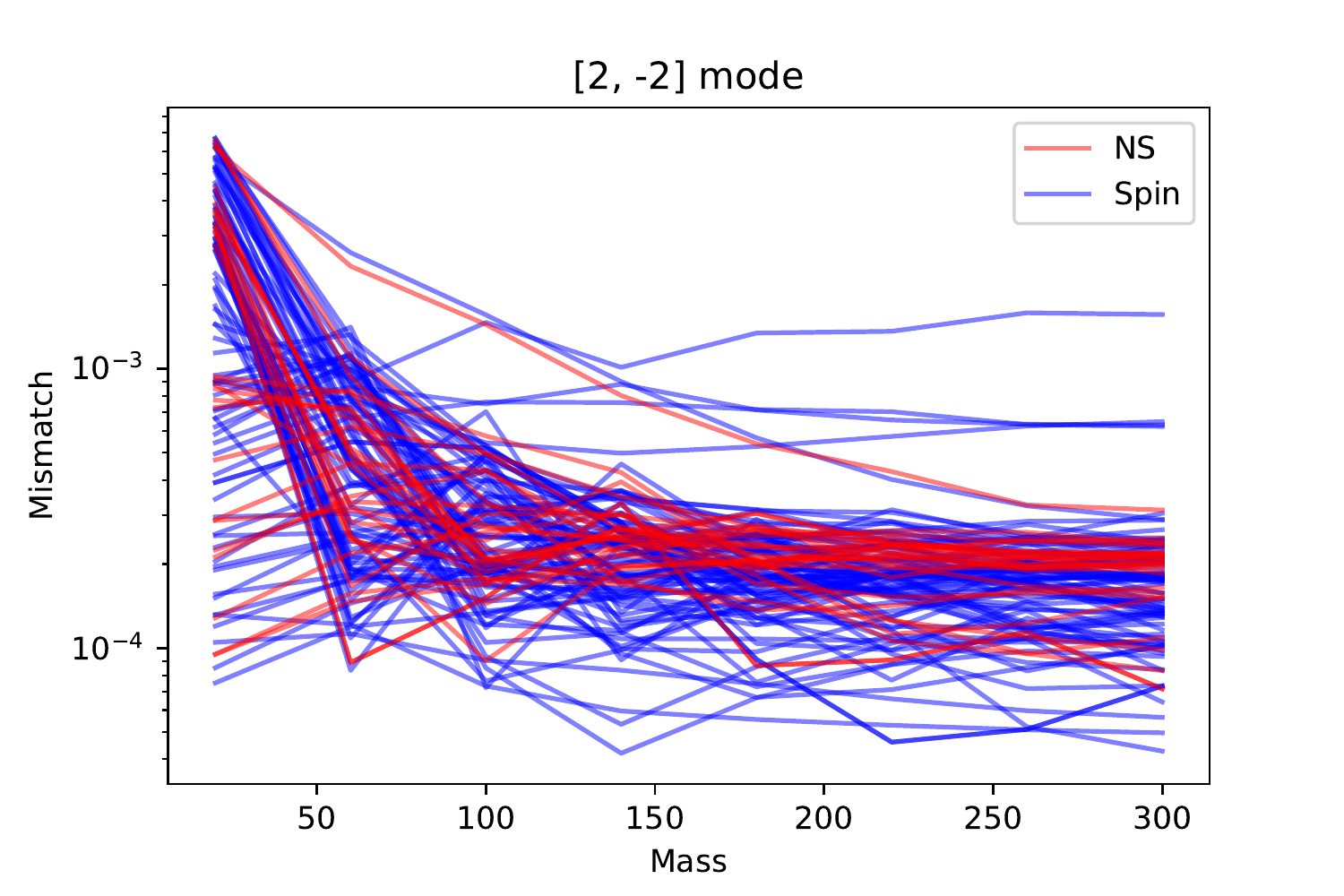}\includegraphics[width=0.24\textwidth]{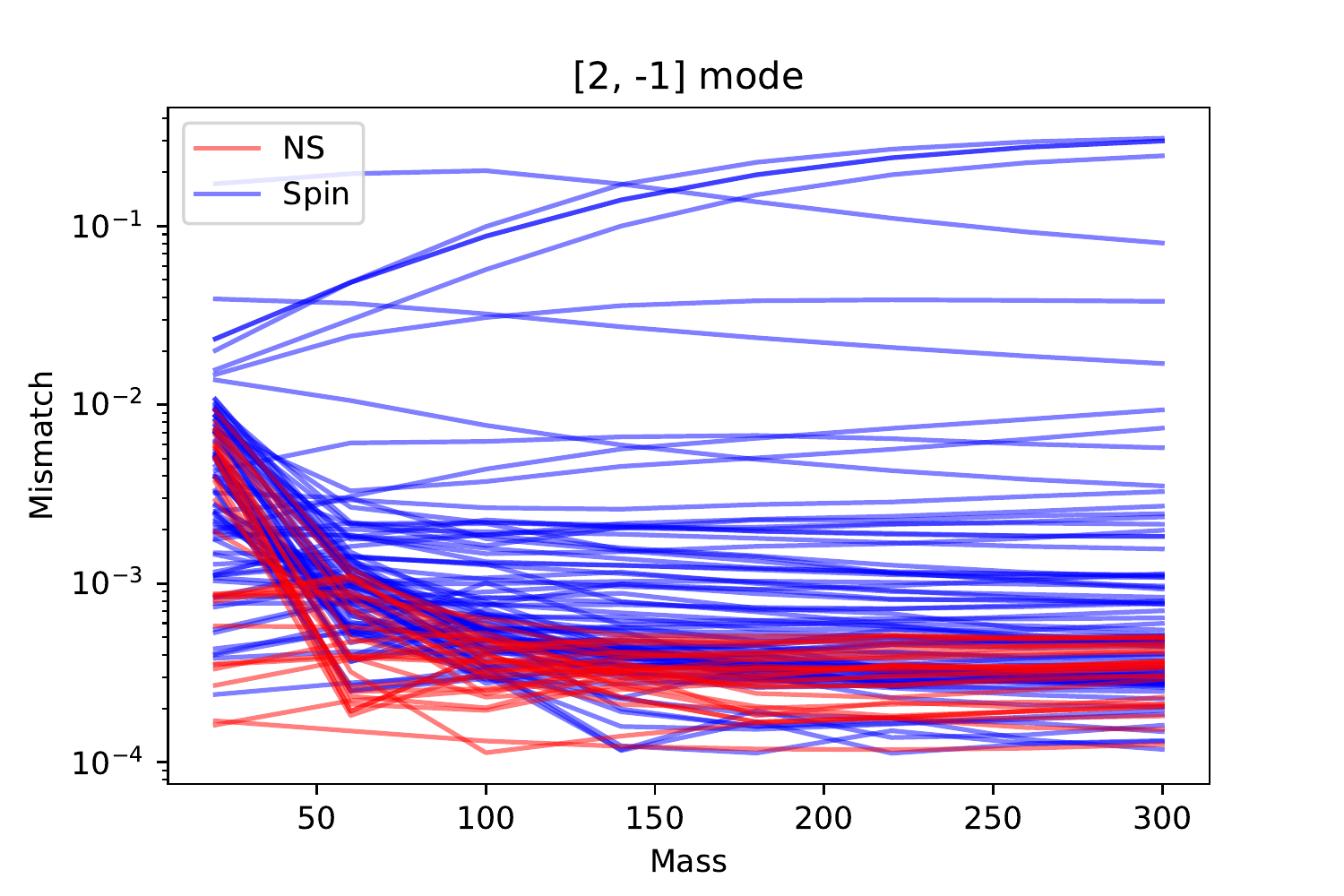}
        \includegraphics[width=0.24\textwidth]{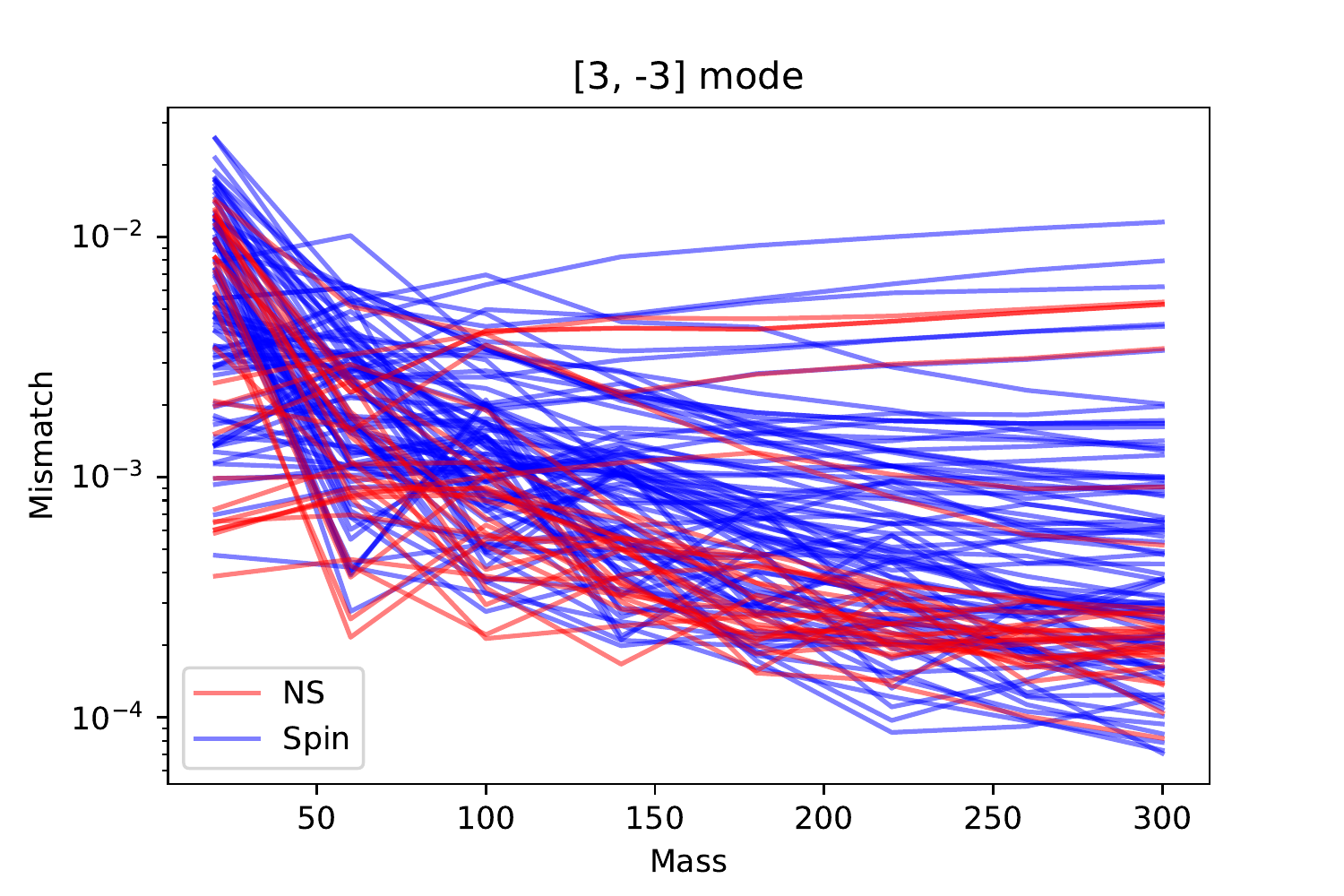}\includegraphics[width=0.24\textwidth]{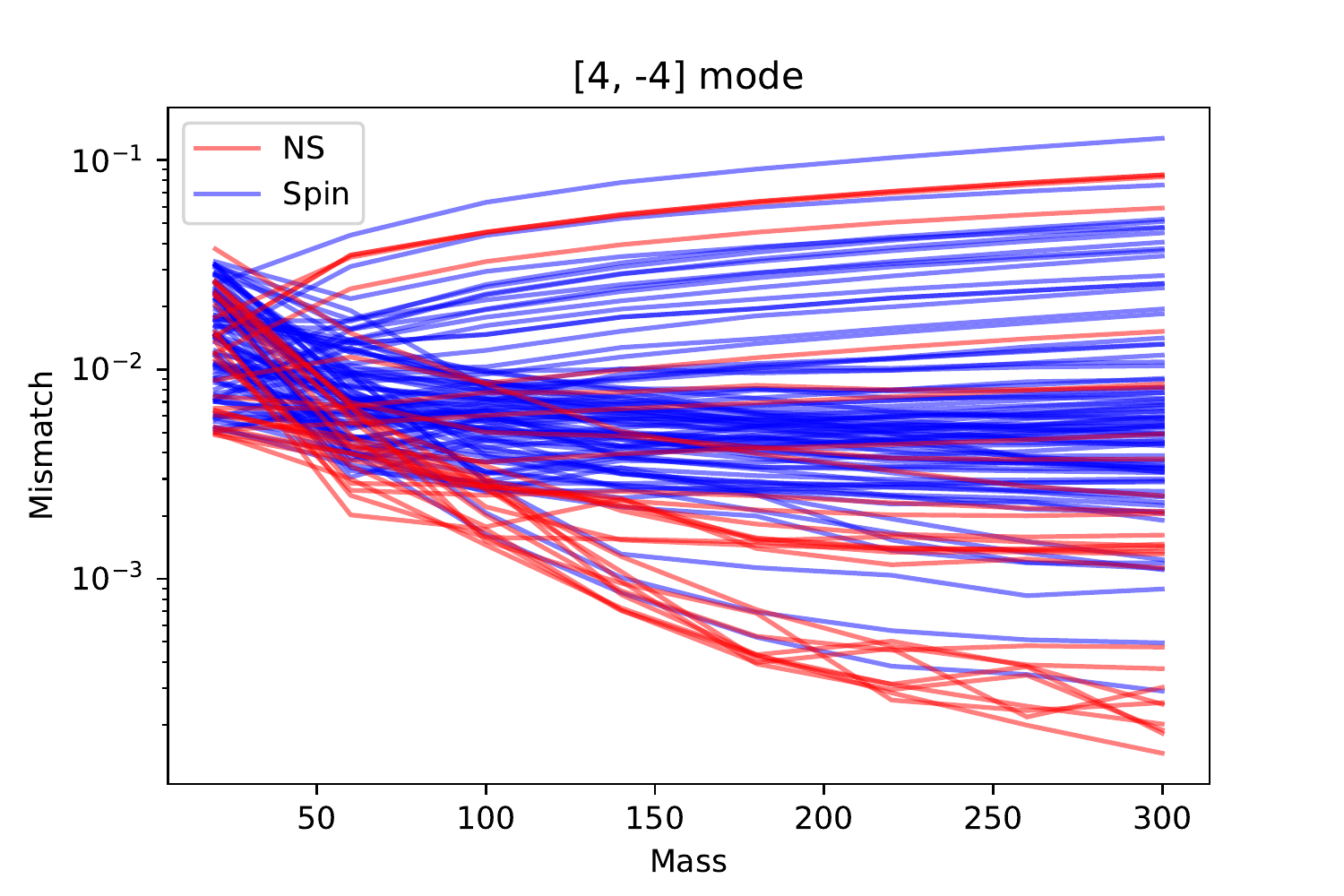}
        \includegraphics[width=0.24\textwidth]{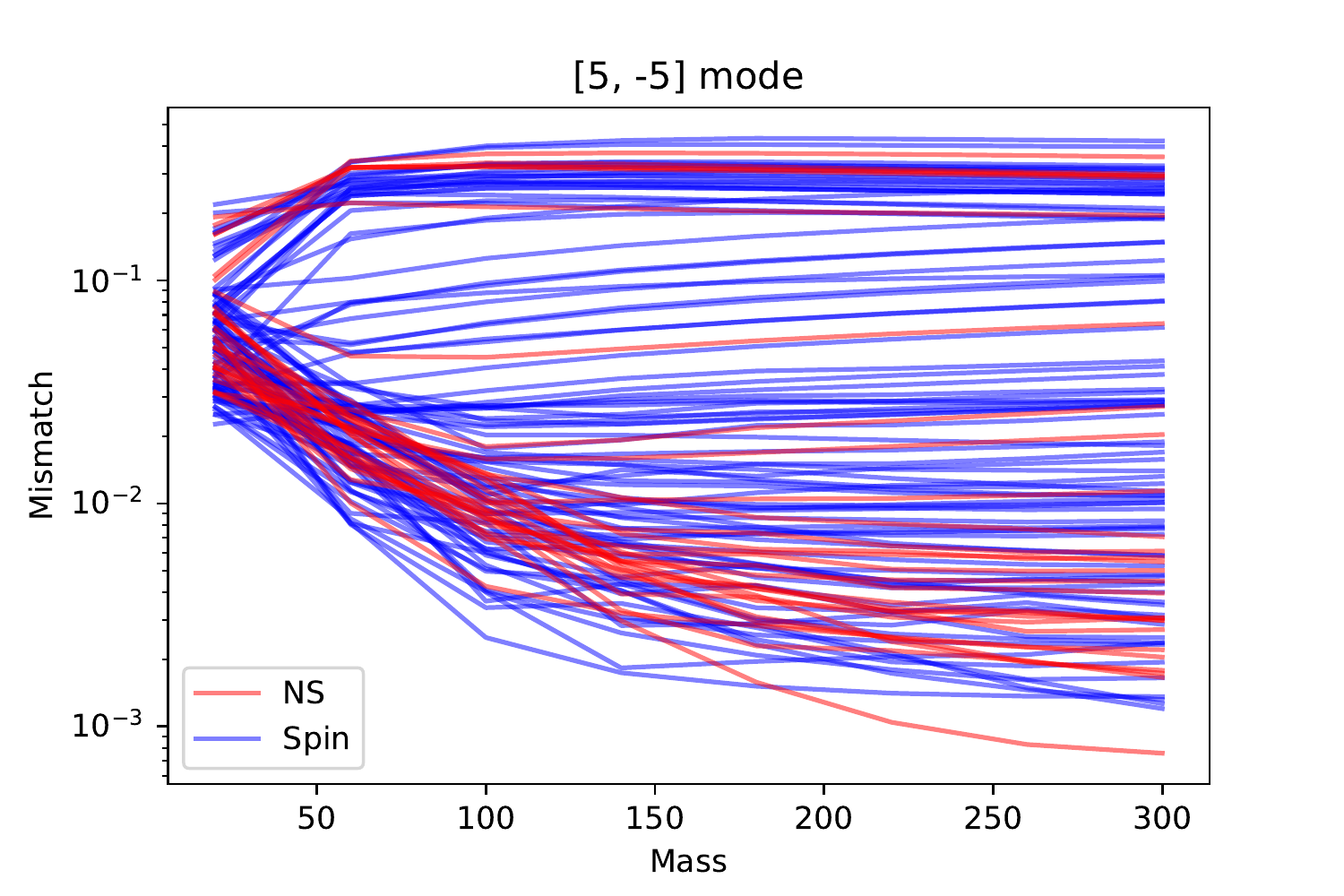}
    \caption{Single mode mismatches of IMRPhenomTHM with LVCNR SXS Catalog as a function of the total mass of the binary. Red: non-spinning configurations. Blue: spinning configurations. It can be seen that for the weaker modes $l=4$, $m=4$ and $l=5$ and $m=5$ a good portion of cases show poor agreement for all mass configurations, due to NR waveform quality.}
\label{fig:mmpmodessxs}
\end{figure}

We compute mismatches (Eq.~(\ref{eq:mismatch})) between polarisations obtained from the LVCNR Catalog simulations and the model presented in this work, for three different inclinations $\iota\in\{0,\pi/3,\pi/2\}$ and different total masses of the binary system, from $M=20M_{\odot}$ to $M=320M_{\odot}$ in steps of $40M_{\odot}$. Results can be seen in Fig.~\ref{fig:mmpolarizsxs}, where distributions of the minimum, mean and maximum mismatch for each case are shown for the three different inclinations. A degradation in the results can be observed with increasing inclination, since the subdominant modes contribution is becoming larger as the systems go from face-on configurations (zero inclination) to edge-on configurations (inclination $\pi/2$) where the dominant $l=2,m=2$ mode is turned off. Nevertheless, even for edge-on configurations, the maximum mismatches of only a few cases are above $1\%$ mismatch.

\begin{figure}[ht!]
    \centering
        \includegraphics[width=0.24\textwidth]{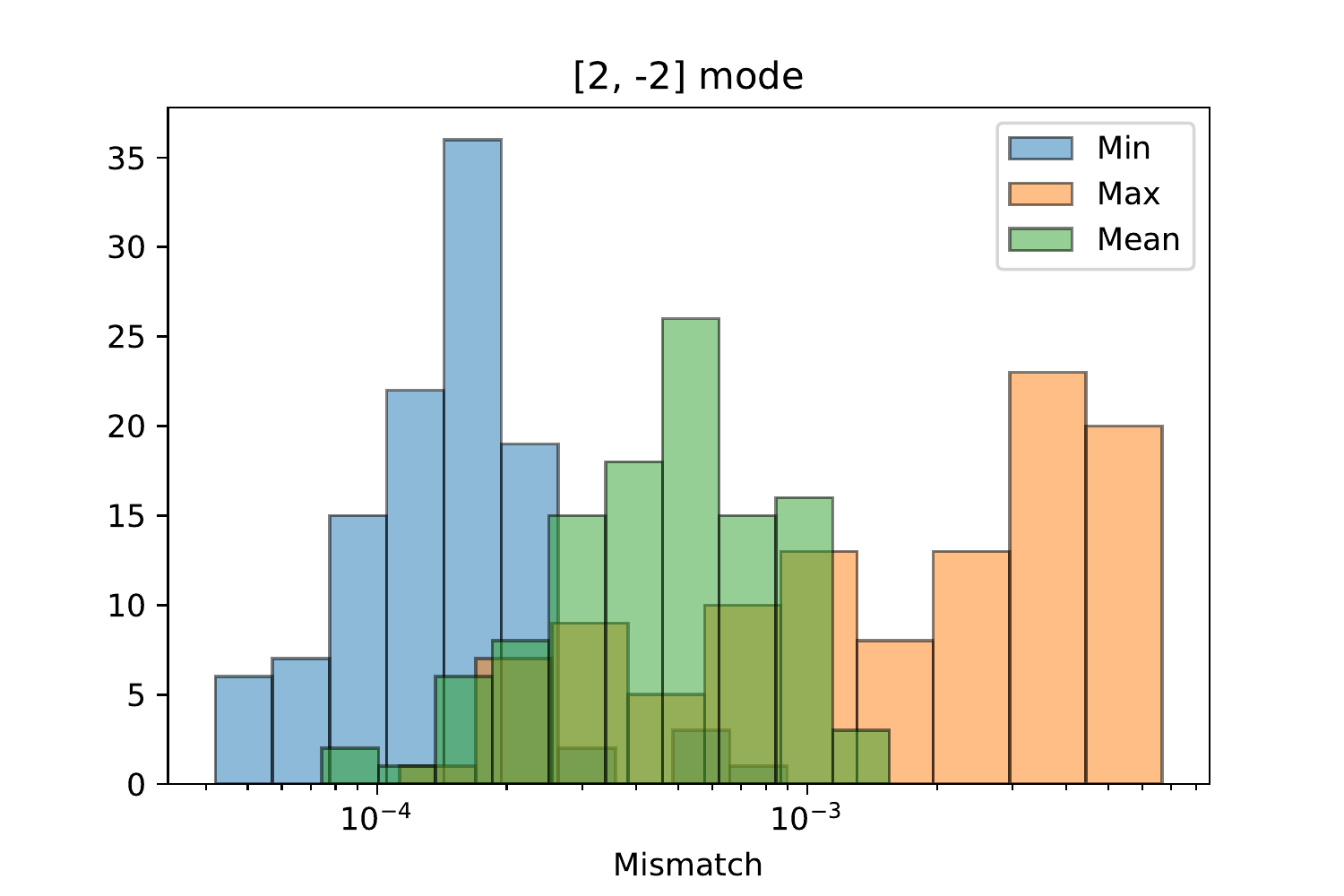}\includegraphics[width=0.24\textwidth]{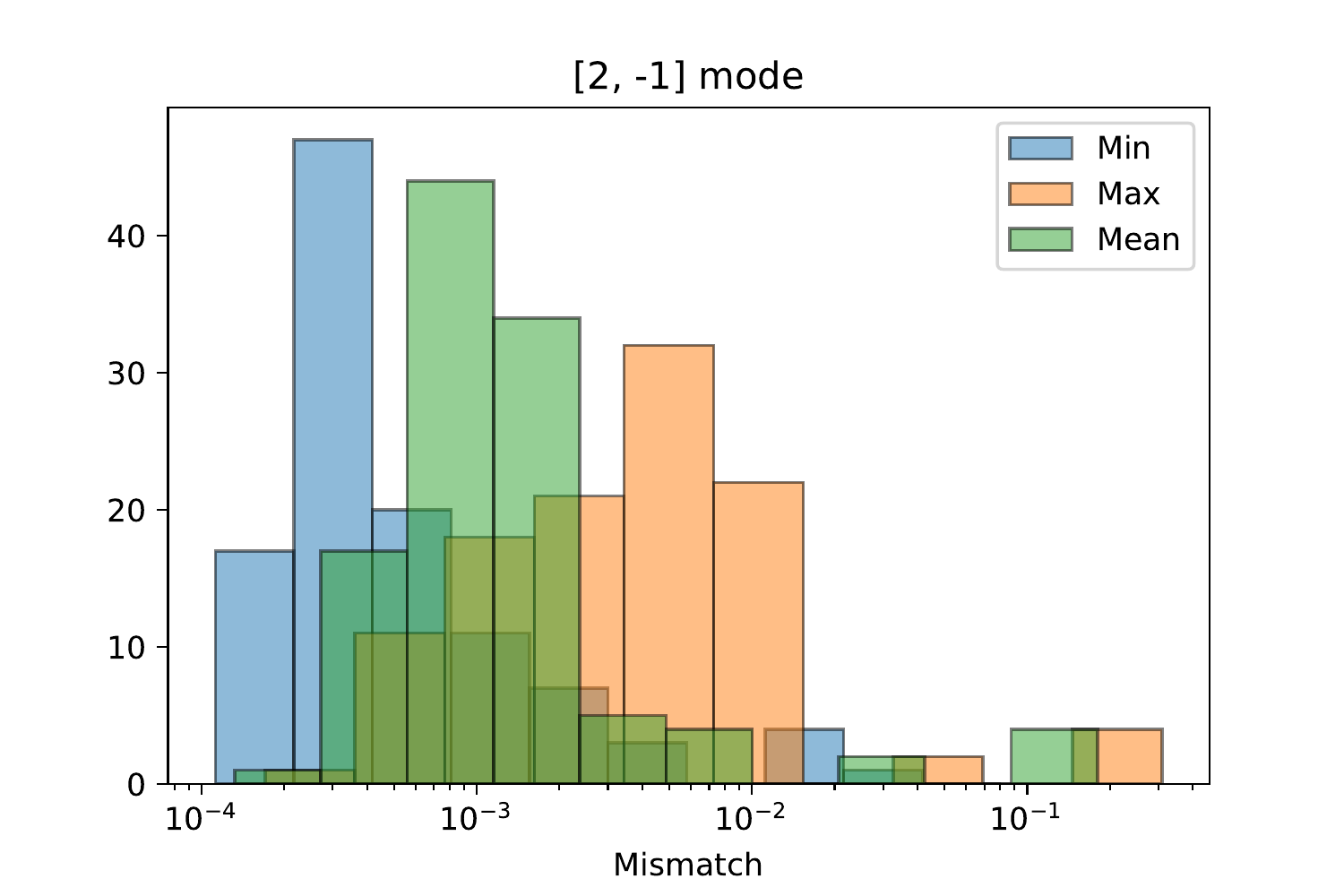}
        \includegraphics[width=0.24\textwidth]{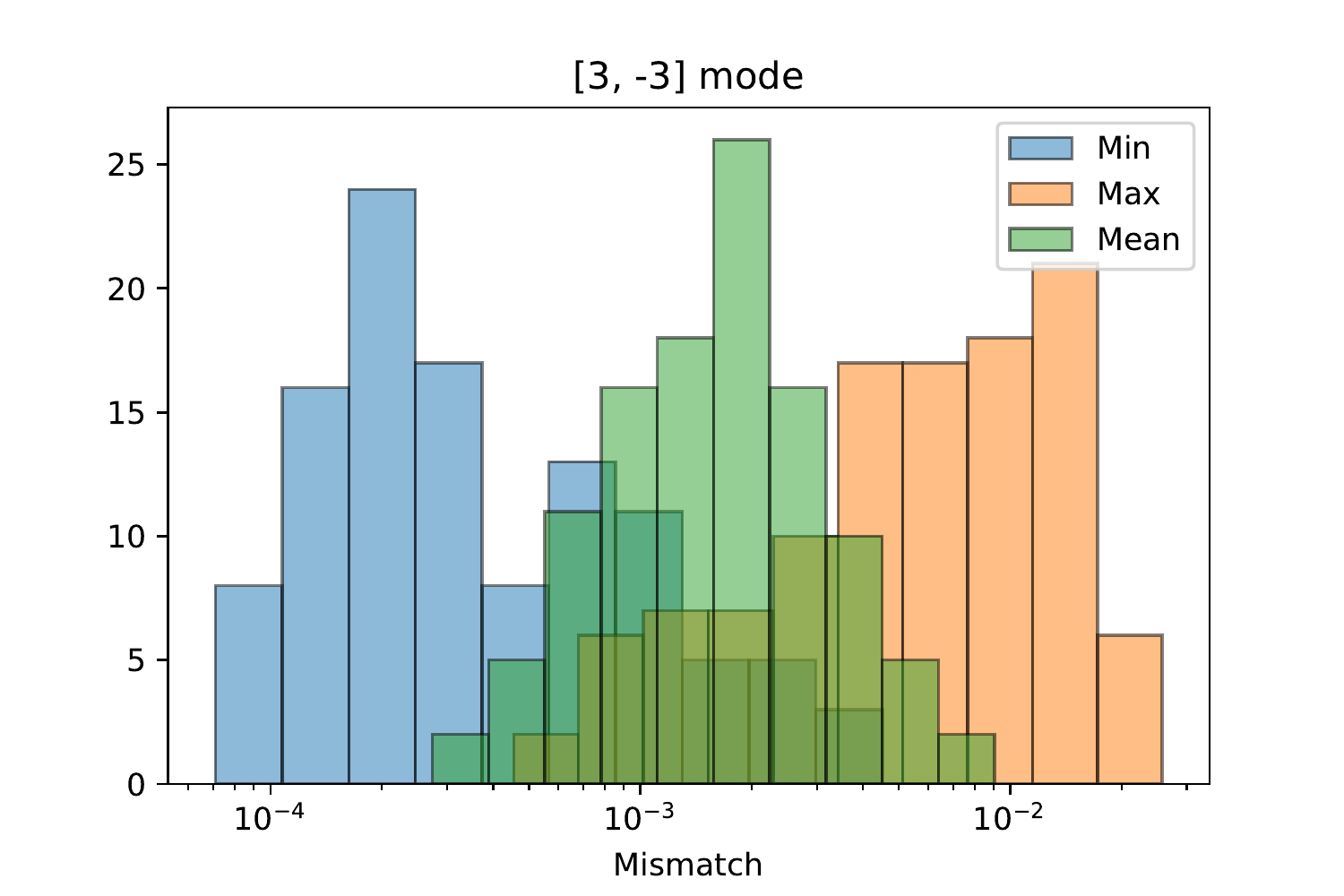}\includegraphics[width=0.24\textwidth]{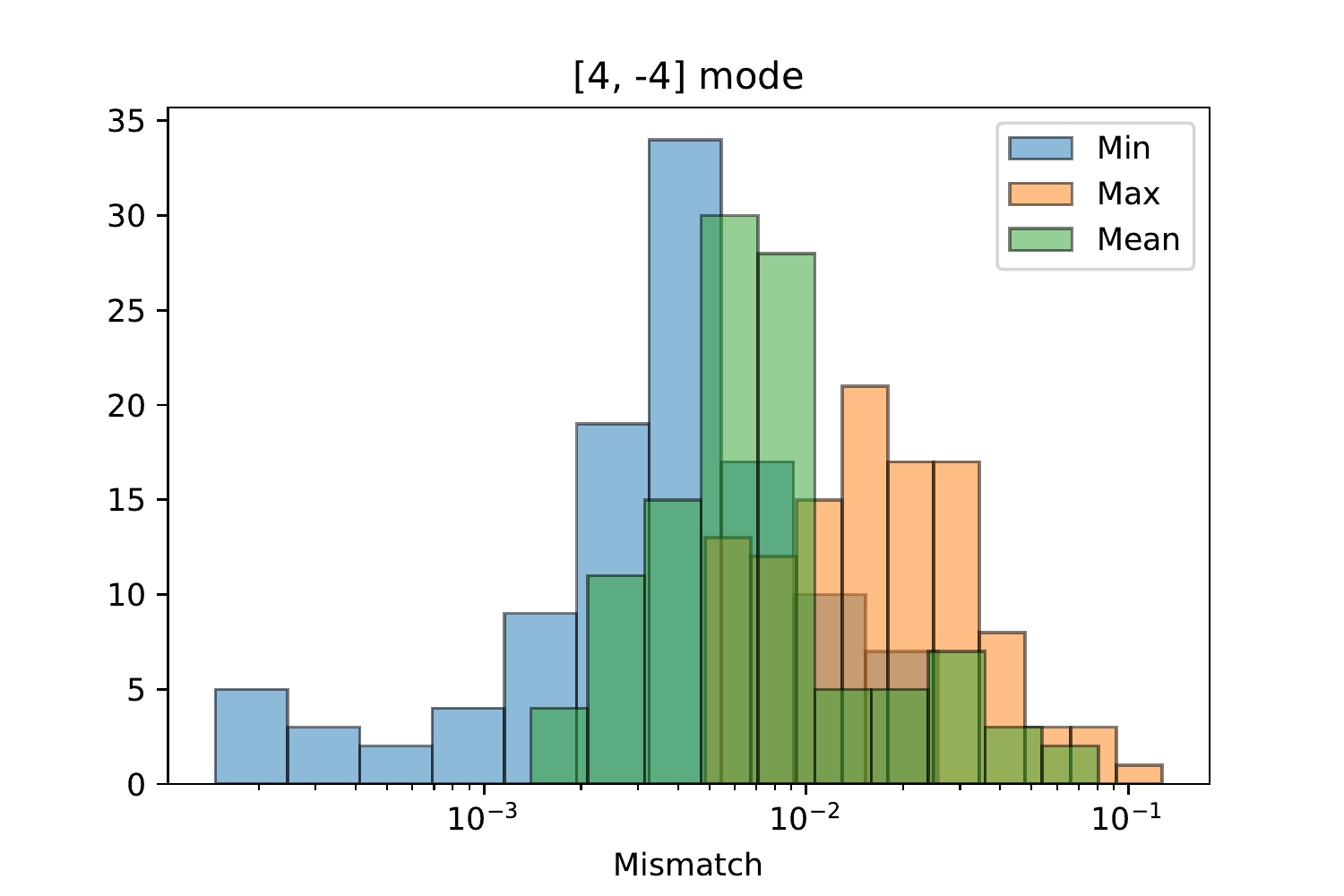}
        \includegraphics[width=0.24\textwidth]{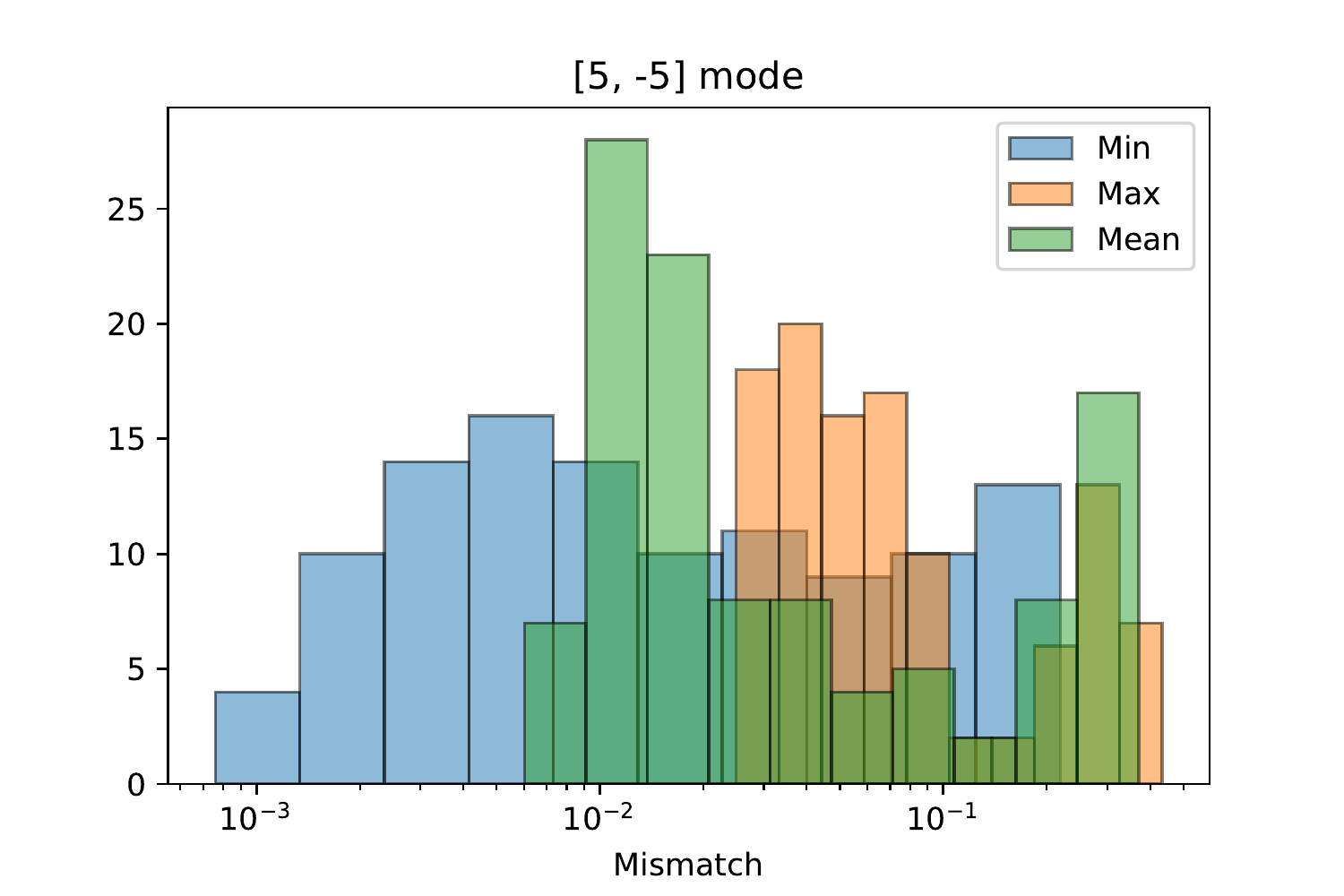}
    \caption{Single mode mismatch distributions of IMRPhenomTHM with LVCNR SXS Catalog. For each mode, the minimum, mean and maximum mismatch distributions are shown.}
\label{fig:mmpmodessxshist}
\end{figure}

\begin{figure}[h!]
    \centering
        \includegraphics[width=\columnwidth]{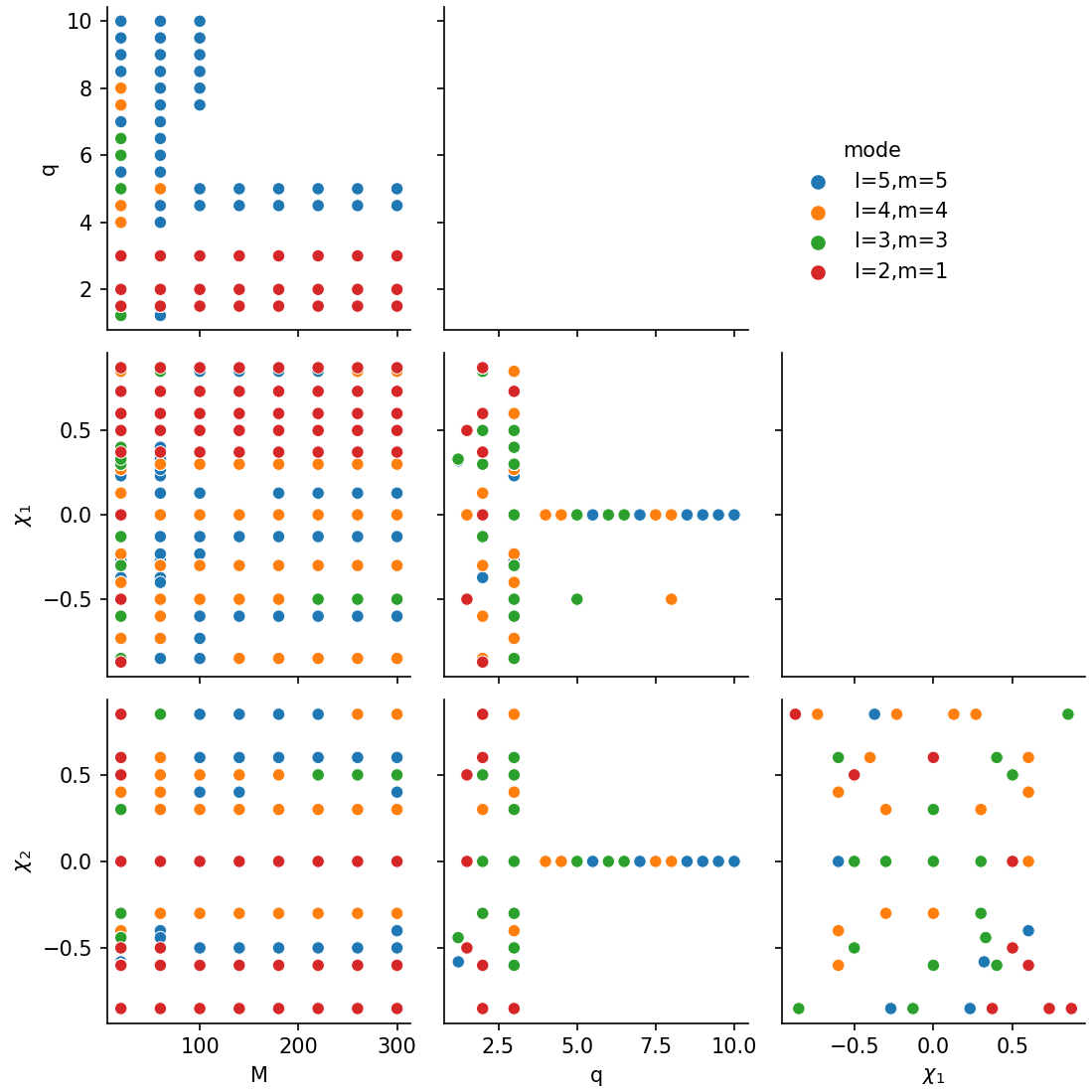}
    \caption{Parameter space distribution of mismatch greater than $1\%$ for the single mode comparison with the LVCNR Catalog. The dominant $l=2$, $m=2$ mode is included but not shown since no mismatch greater than $1\%$ was obtained.}
\label{fig:sxscorner}
\end{figure}

Mismatches for individual modes are also computed, for the same total mass list as in the previous comparison. (For individual modes, there is no need to compute at different inclinations, since this only scales the waveform amplitude for single mode content.) Results in terms of the total mass of the system are shown in Fig.~\ref{fig:mmpmodessxs} and mismatch distribution of minimum, mean and maximum mismatch for each mode is shown in Fig.~\ref{fig:mmpmodessxshist}. The parameter space distribution of mismatches greater than $1\%$ is shown in Fig.~\ref{fig:sxscorner}. A degradation of mismatch results can be observed when going from the dominant mode $l=2$, $m=2$ to the weakest mode $l=5$, $m=5$, which is consistent with the fact that for weaker modes, the quality of NR waveforms is worse since they are more difficult to extract from the global complex strain. 

\subsection{Comparison with other multimode models}\label{subsec:modelcomparison}

Although the dataset of NR simulations is the most trustworthy set of waveforms in the highly dynamical regime of binary mergers, it is a finite dataset, with less than a thousand simulations for non-precessing configurations, and they are also limited in length due to computational cost. As a more extensive test of the model, in this section we present comparisons with a NR-EOB hybrid surrogate model called \texttt{NRHybSur3dq8}, which has been constructed to infer waveforms of any length within the bounds of its training region, $q\leq9$, $|\chi_i|\leq0.91$, which essentially corresponds to the parameter space coverage of NR simulations. To test the model also outside the training boundaries of \texttt{NRHybSur3dq8}, in more challenging regions of parameter space, mismatch results with the Fourier domain phenomenological model \texttt{IMRPhenomXHM} are also presented, which allow to test the model up to extremal spin configurations, and for arbitrary mass ratio (although in the following results we have restringed to mass ratios below $q=20$).

With each model, single mode mismatches between the coincident modes in both models and full polarisations will be computed. Waveforms are generated for random parameter choices between some parameter bounds, employing the function \texttt{SimInspiralFD} from the LALSuite libraries \cite{lalsuite}, which provides the complex Fourier domain polarisations needed for the noise weighted product (\ref{eq:overlap}). Gaussian colored noise is modeled by the power spectral density estimation for the zero detuning high power configuration of the advanced LIGO detectors \cite{adligopsd}, selecting a sampling rate of $8192$ Hz (corresponding to a maximum frequency of $4096$ Hz) and with minimum frequencies for the match calculation and starting waveform generation frequencies listed in Table~\ref{tab:mismatch_settings}.

\begin{table}[h!]
    \centering
    \begin{tabular}{| c | c | c | }
\hline
 Targeted model & \texttt{NRHybSur3dq8} & IMRPhenomXHM \\ 
 \hline
 $M$ range & $[10,200]M_{\odot}$ & $[10,200]M_{\odot}$ \\
 \hline
 $q$ range & $[1,9]$ & $[1,20]$ \\  
 \hline
 $\chi_i$ range & $[-0.91,0.91]$ & $[-1,1]$\\
 \hline
 $f_\text{sampling}$ & \multicolumn{2}{c|}{$8192$ Hz}\\
 \hline
 $f_\text{min}$ & \multicolumn{2}{c|}{$20$ Hz}\\
 \hline
 $f_\text{start} $ (polarisations) & 8 Hz & 10 Hz\\
 \hline
 $f_\text{start}$ (single modes) & \multicolumn{2}{c|}{$(m/2)20$ Hz}\\
 \hline
  $S_n(f)$  & \multicolumn{2}{c|}{aLIGOzerodetHighpower}\\
 \hline
    \end{tabular}
 \caption{Setting for mismatch comparison.}
 \label{tab:mismatch_settings}
\end{table}

\begin{figure}[th!]
    \centering
        \includegraphics[width=0.48\textwidth]{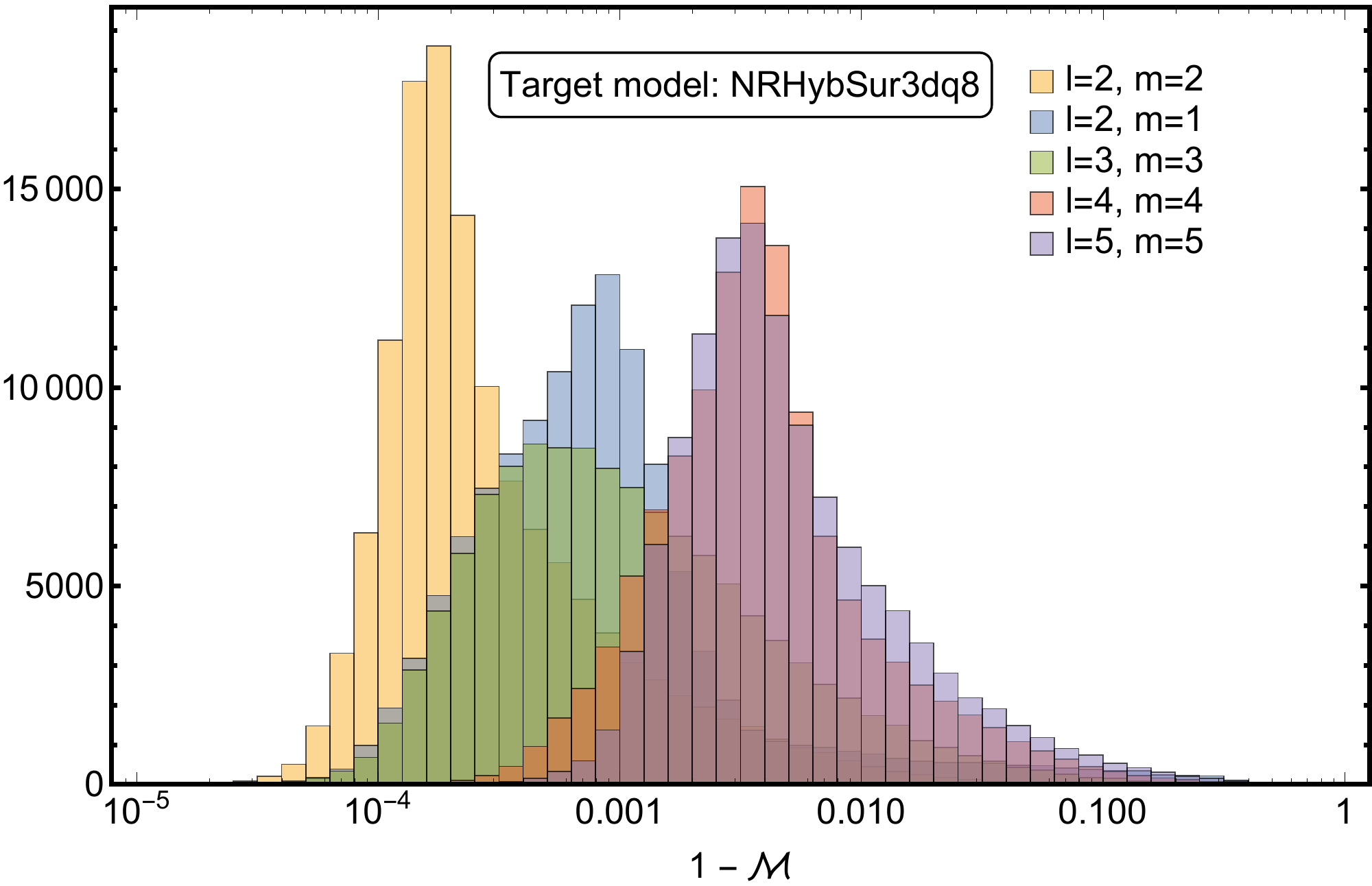}
        \includegraphics[width=0.48\textwidth]{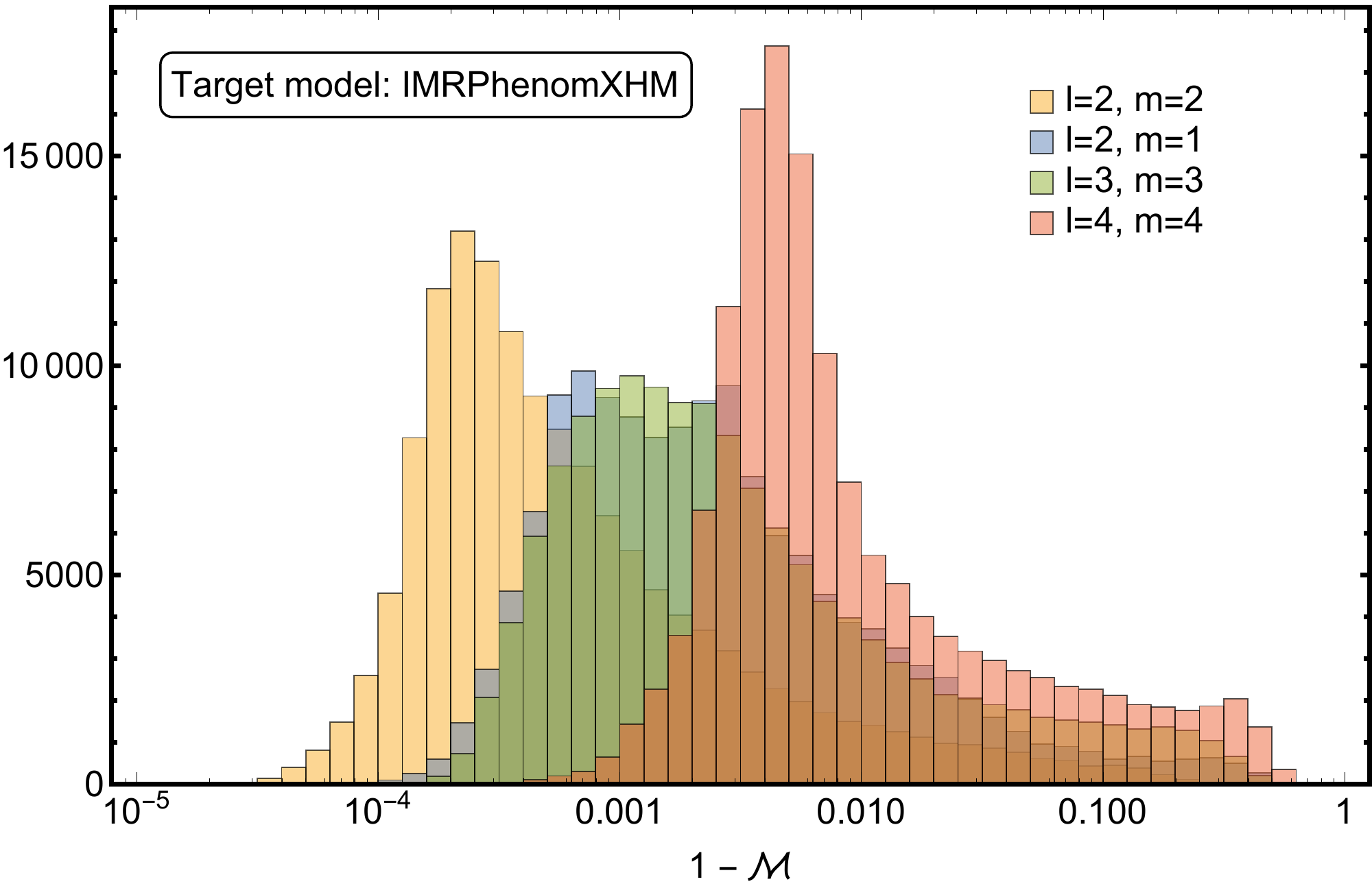}
    \caption{Mismatch distributions for individual modes for \texttt{IMRPhenomTHM} with the non-precessing state-of-the-art models \texttt{NRHybSur3dq8} (top) and \texttt{IMRPhenomXHM} (bottom). A correlation between model agreement and mode hierarchy can be seen in the results, with the dominant $l=2$, $m=2$ mode the most similar between the models.}
\label{fig:singlemodemismatch}
\end{figure}

In Fig.~\ref{fig:singlemodemismatch} we can see the distributions of mismatches for each mode with the corresponding mode of the model \texttt{NRHybSur3dq8} and \texttt{IMRPhenomXHM}. From both comparisons, we can see a hierarchy in the agreement of the modes, with the dominant $l=2$, $m=2$ mode having the best agreement with both models. The $l=2$, $m=1$ and $l=3$, $m=3$ modes also show good agreement, with a peak of their distributions close to $0.1\%$ mismatch. Finally, the weak $l=4$, $m=4$ and $l=5$, $m=5$ modes have the worst agreement, but still have the peak and most of their mismatch distributions below $1\%$. This can be understood in terms of the original NR data that has been employed in the construction of the three models, where the dominant $l=2$, $m=2$ mode is typically very well reproduced by simulations, but data quality degrades with increasing $l$, as the modes are weaker and it is more difficult to cleanly extract them from the simulations.

\begin{figure}[th!]
    \centering
        \includegraphics[width=0.48\textwidth]{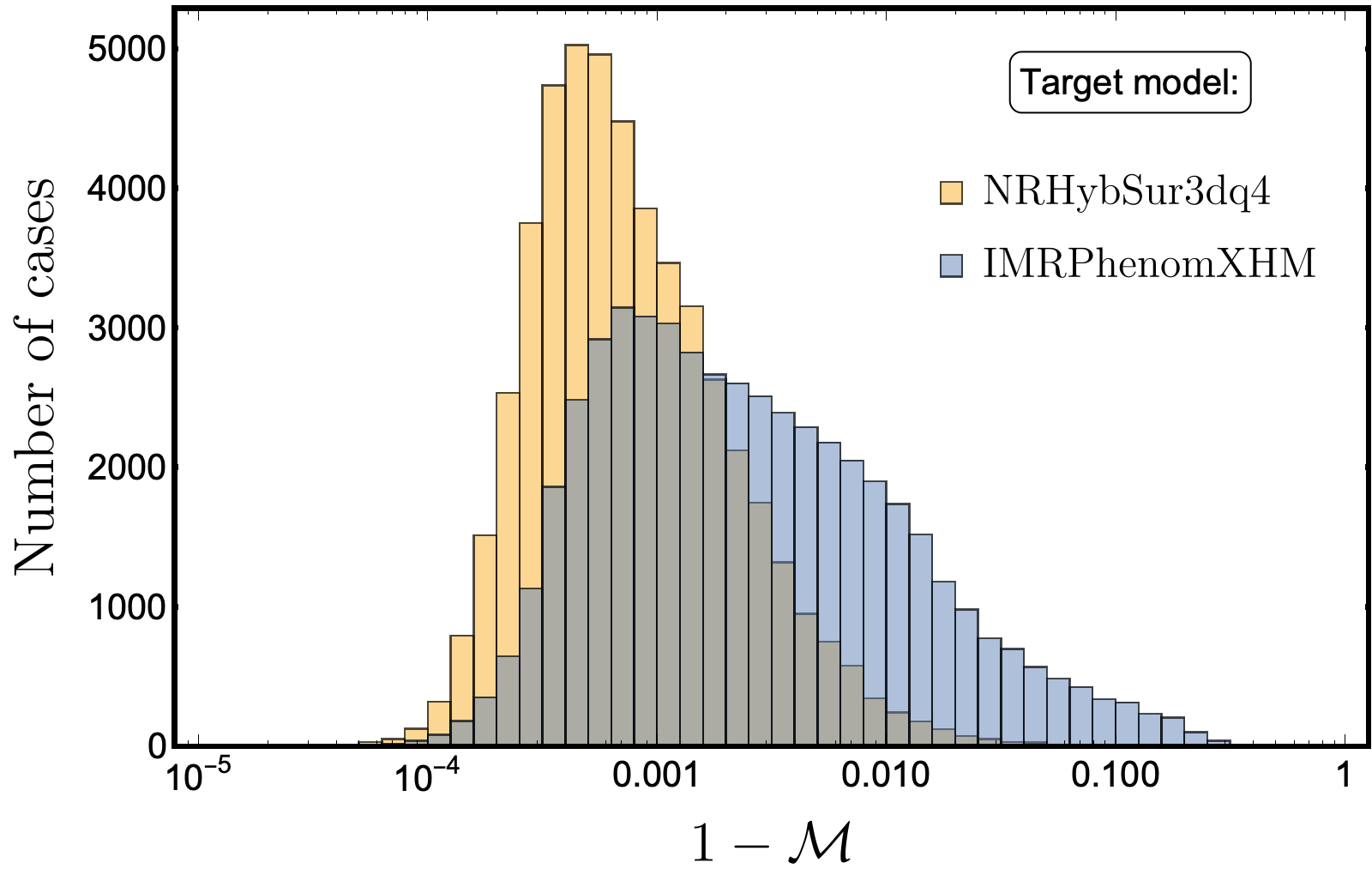}
        \includegraphics[width=0.48\textwidth]{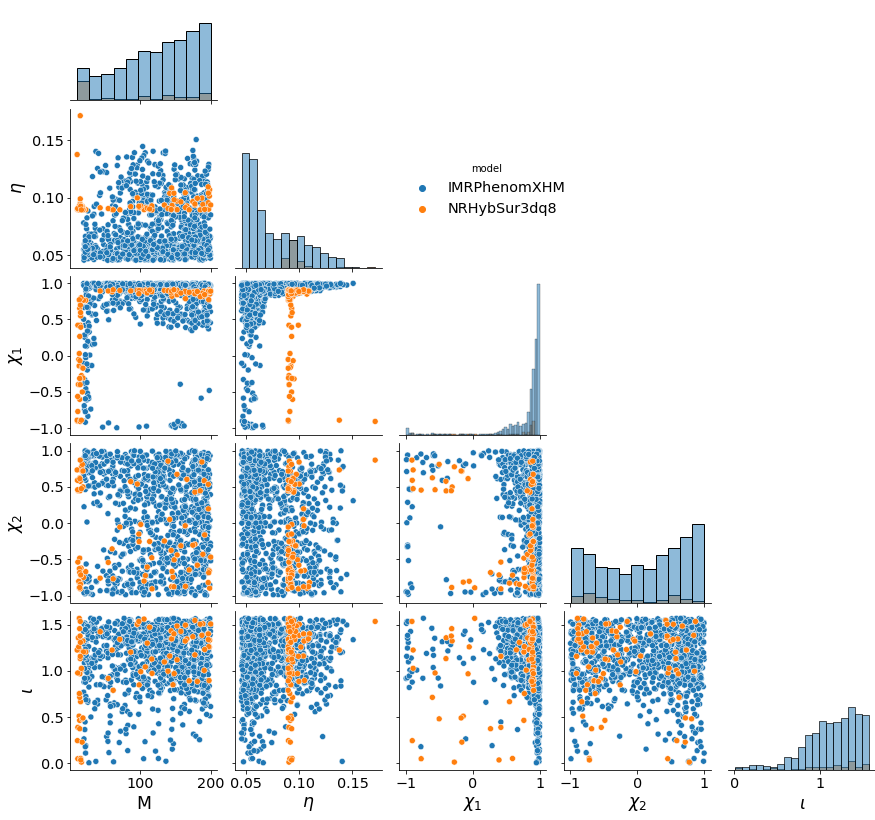}
    \caption{Mismatch distributions of \texttt{IMRPhenomTHM} with \texttt{NRHybSur3dq8} and \texttt{IMRPhenomXHM} for multimode polarisations. Each model contains its full available mode content. Top: histogram of the distributions for all cases. Bottom: parameter space locations of cases with higher mismatches ($>3\%$ for \texttt{NRHybSur3dq8} and $>10\%$ for \texttt{IMRPhenomXHM}).}
\label{fig:polarizmismatch}
\end{figure}

In Fig.~\ref{fig:polarizmismatch} we can see the distributions of mismatches for the full multimode polarisations compared with both models. We see that the peak of both distributions is close to $0.1\%$ mismatch, with the distribution of mismatches against \texttt{NRHybSur3dq8} being shifted to smaller values of mismatch and therefore better agreement found between this model and \texttt{NRHybSur3dq8}. Nevertheless, the parameter range for the comparison is not the same between both models. As we have explained, the validity region of the surrogate model \texttt{NRHybSur3dq8} coincides with the parameter space coverage of the bulk of NR simulations, and therefore calibrated models are expected to agree more in that region, while the parameter range of the comparison with \texttt{IMRPhenomXHM} has been extended to higher mass ratios (up to $q=20$) and extremal spins, precisely to compare the model outside the NR coverage region. In the bottom pannel of Fig.~\ref{fig:polarizmismatch} we can see the distribution of cases with a mismatch greater than $3\%$ for the \texttt{NRHybSur3dq8} comparison and greater than the  $10\%$ for the \texttt{IMRPhenomXHM} comparison (in order to capture the tail of both distributions). It can be seen that poor mismatch is typically located on the left boundary of the symmetric mass ratio (high unequal mass systems) and for large positive primary spin values, having a preference also towards edge-on configurations, where the multimode structure of the model is more relevant.

\subsection{Gravitational recoil}\label{subsec:recoil}

While mismatch studies provide a good measure for quantifying agreement between different waveforms and datasets across parameter space, some aspects relevant for the modelling of a multimode waveform are harder to extract from a single scalar quantity such as the mismatch. 
In particular, while individual modes can be compared intrinsically, i.e comparing their frequency and amplitude evolution, the relative orientation of the modes that compose the multimode structure of the polarisations cannot be tested with single mode mismatch studies, and in the polarisation mismatch studies this is typically shadowed by the quality of agreement of the dominant mode.

As an additional test for the multimode structure of the model, we provide comparisons for the gravitational recoil prediction. Gravitational recoil is an effect of asymmetric black hole binaries due to the anisotropic radiation of gravitational waves, which generates a net emission of angular momentum. Only multimode models of gravitational waves can capture this effect, since a single mode emission will not be anisotropic. Emitted angular momentum in the $\hat{\boldsymbol{n}}$ direction is given by:
\begin{equation}
    \dfrac{d\boldsymbol{P}}{dt}=\lim_{r\rightarrow\infty}\dfrac{r^2}{16\pi}\int d\Omega(\dot{h}_{+}+\dot{h}_{\times})\hat{\boldsymbol{n}}.
\end{equation}

With the spherical harmonic decomposition of the polarisations of Eq.~(\ref{eq:polariz_decomp}), the sky-sphere integral in the $\hat{\boldsymbol{n}}$ direction can be performed on the spherical harmonic basis. In terms of the parallel and perpendicular components of the emitted linear momentum with respect to the direction of the orbital angular momentum $\hat{\boldsymbol{L}}$, for non-precessing systems the parallel component is $P_{||}=0$ due to the equatorial symmetry of the system, and the perpendicular component (which lies on the orbital plane) is given in terms of the mode time derivatives:
\begin{equation}
\label{eq:recoil}
    \begin{split}
        \dfrac{d\boldsymbol{P}_{\perp}}{dt}&=\lim_{r\rightarrow\infty}\dfrac{r^2}{8\pi}\sum_{l,m}\dot{h}_{l,m}(a_{l,m}\dot{h}^*_{l,m+1}\\ &+ b_{l,-m}\dot{h}^*_{l-1,m-1} - b_{l+1,m+1}\dot{h}^*_{l+1,m+1}),
    \end{split}
\end{equation}
where the factors $a_{l,m}$ and $b_{l,m}$ are given by:
\begin{subequations}
    \begin{align}
        a_{l,m}&=\dfrac{\sqrt{(l-m)(l+m+1)}}{l(l+1)},\\
        b_{l,m}&=\frac{1}{2l}\sqrt{\dfrac{(l-2)(l+2)(l+m)(l+m-1)}{(2l-1)(2l+1)}}.
    \end{align}
\end{subequations}

\begin{figure}[th]
    \centering
        \includegraphics[width=0.48\textwidth]{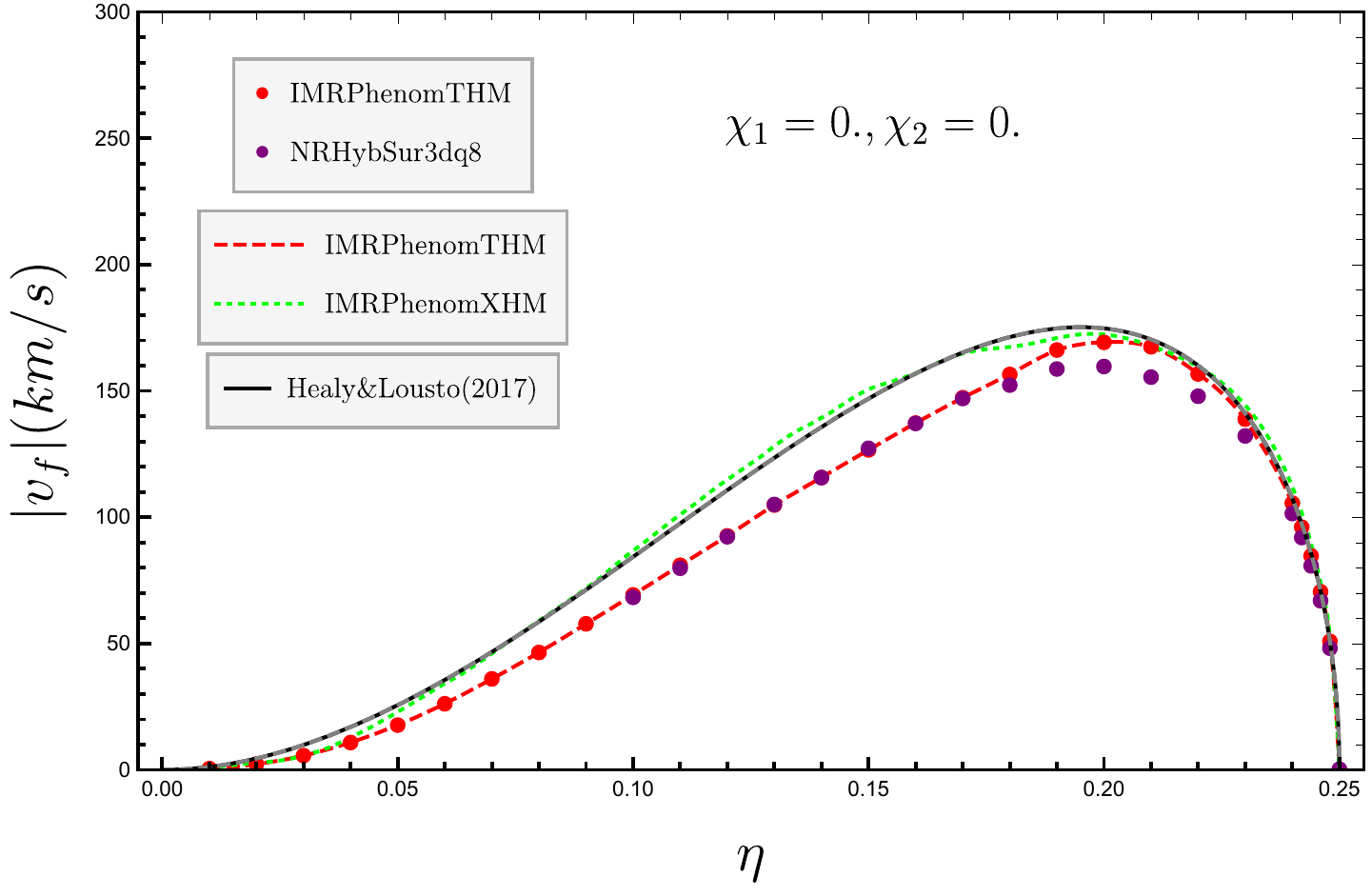}
        \includegraphics[width=0.48\textwidth]{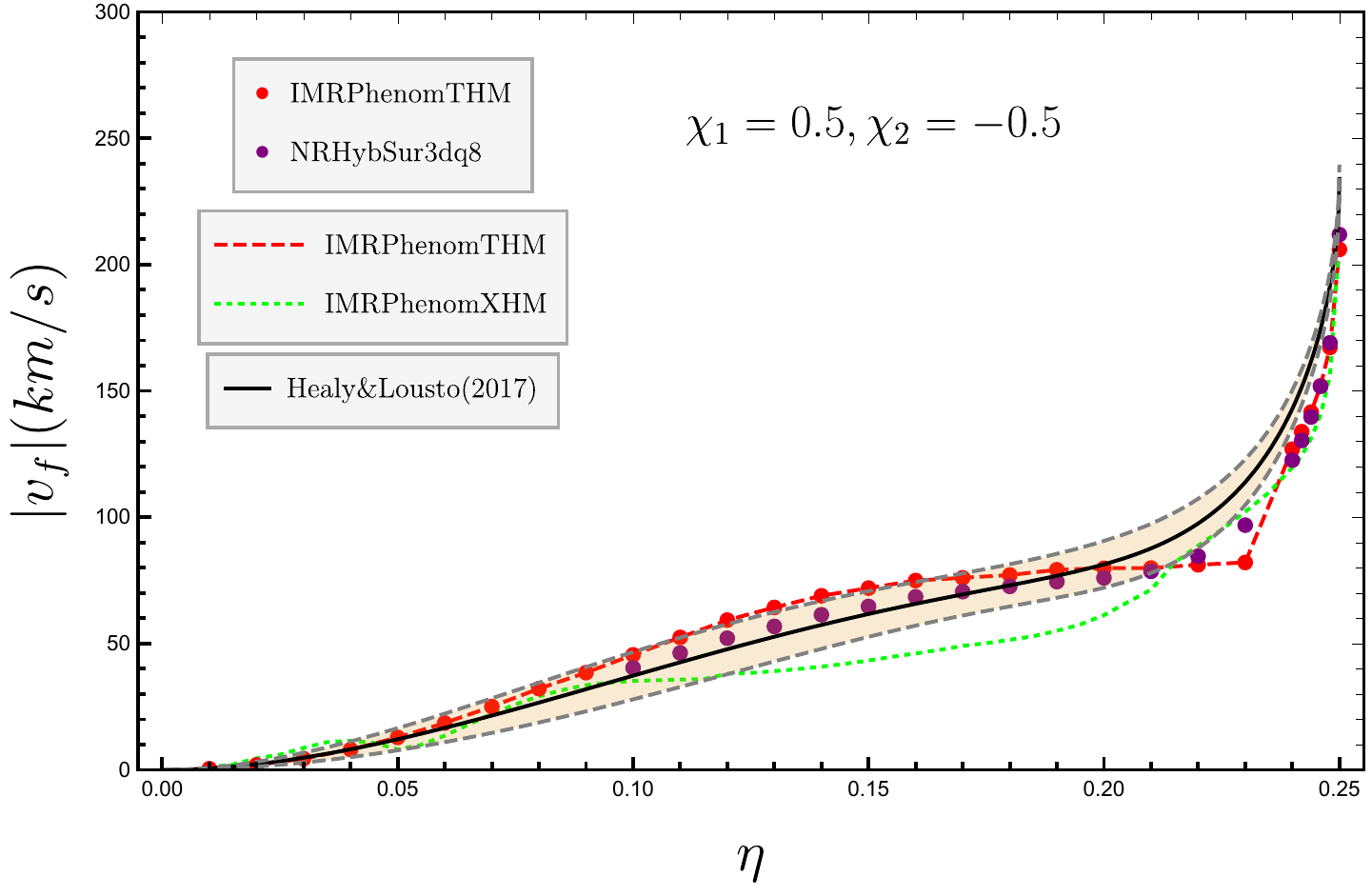}
        \includegraphics[width=0.48\textwidth]{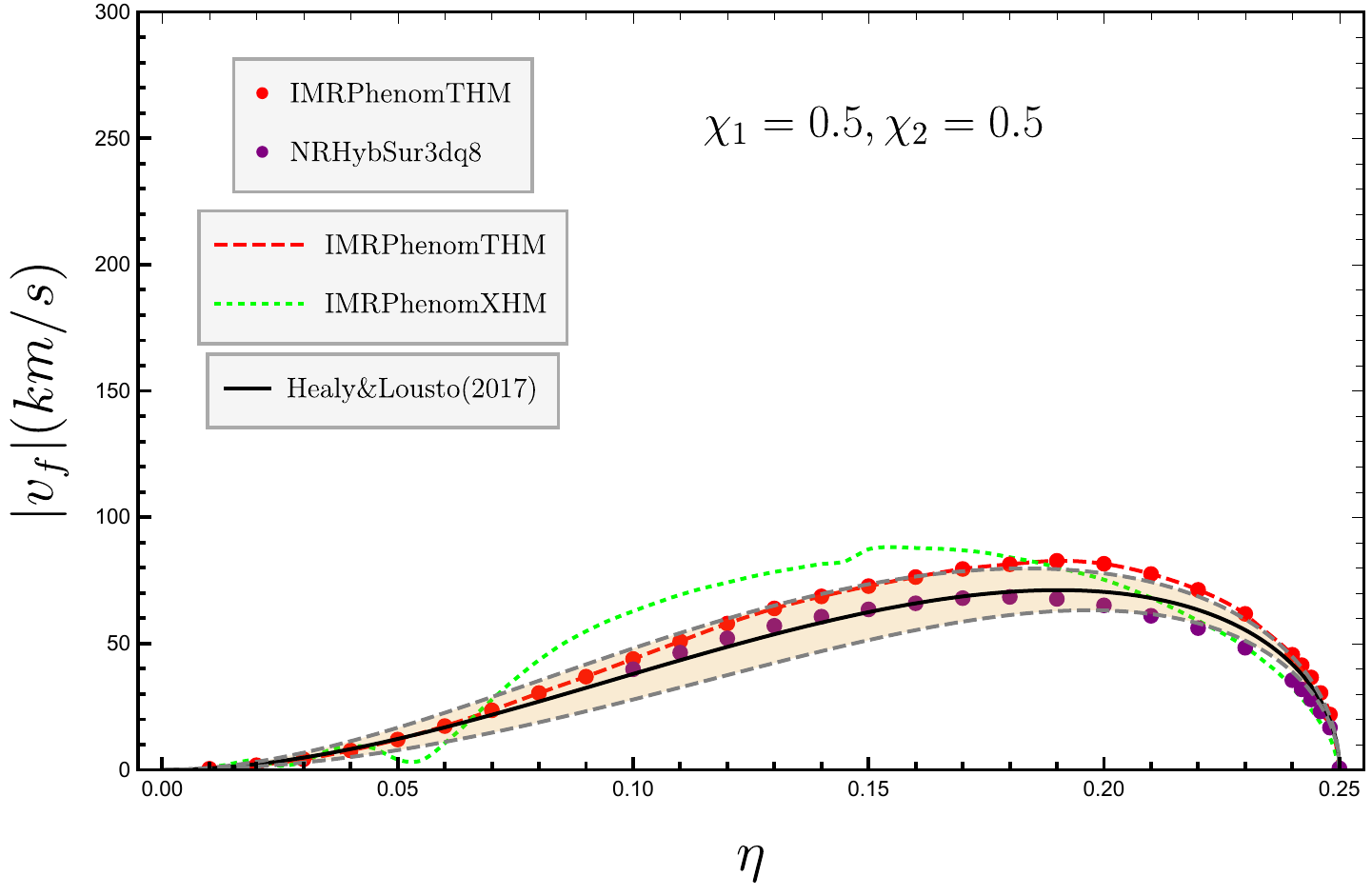}
    \caption{Comparison of final gravitational recoil kick velocity for different models and a NR fit of this quantity. Top: non-spinning configurations. Middle: Unequal spin configurations. Bottom: Equal spin configurations. Shaded regions correspond to the uncertainty of the NR fit.}
\label{fig:recoil}
\end{figure}

The final recoil velocity of the remnant will, due to linear momentum conservation, be the time integral over the whole binary history of the emitted linear momentum in the opposite direction:
\begin{equation}
\label{eq:kick}
    \boldsymbol{v}_f=-\int dt \dfrac{d\boldsymbol{P}_{\perp}}{dt},
\end{equation}
with $ \boldsymbol{v}_f$ being a complex number with the magnitude of the recoil kick and its direction in the orbital plane.

A prediction for the recoil velocity is a test well suited for the multimode structure of a model. Eq.~(\ref{eq:recoil}) can be seen as a sum of pairs of interacting modes with different mode number $m=m'\pm1$, and the integral of Eq.~(\ref{eq:kick}) can be split into different contributions from each pair of modes. To accurately predict the final recoil velocity, each of these contributing integrals has to be accurate, which is a test for the intrinsic frequency and amplitude of the modes; but also the total sum of the different terms has to be accurate, and this is a test for the different phase relations between the modes. 

Here we compute the magnitude $v_f=|\boldsymbol{v}_f|$ of the recoil velocity for three different models for different spin configurations as a function of the symmetric mass ratio of the system, and we compare with an existing NR fit of this quantity. Results for the three models and the Healy \& Lousto NR fit \cite{HealyLousto} prediction can be seen in Fig.~\ref{fig:recoil}. A similar trend to the other models and the NR fit can be seen to be reproduced by \texttt{IMRPhenomTHM}. For the spinning configurations, most of the computed values are inside the uncertainty interval of the NR fit, though for the non-spinning configurations the recoil is in general a bit underestimated. It can be seen, however, that in the underestimated region our model agrees quite well with \texttt{NRHybSur3dq8}. This difference with respect to the NR fit and the Fourier domain model \texttt{IMRPhenomXHM} could be due to the different mode content of the models.

\section{Performance in parameter estimation}\label{sec:PE}

One of the most important applications of gravitational wave models is the parameter estimation for detected gravitational wave events, to infer the physical properties of the source systems. In this section, we test the performance of the model in parameter estimation for simulated injected signals and for real gravitational wave detections, employing the Bayesian inference suite specialised in gravitational wave parameter estimation \texttt{Bilby} \cite{Ashton_2019}, in a parallelizable pipeline suited for HPC \texttt{parallelBilby} \cite{Smith_2020}, which performs a stochastic sampling of the likelihood function across parameter space employing the nested sampling algorithm \texttt{dynesty} \cite{Speagle_2020}.

\subsection{NR Injection recovery}\label{subsec:injection}

As a first test of the performance of the model in parameter estimation, we test the parameter recovery for two synthetic signals where a NR simulation from the LVCNR Catalog has been injected into Gaussian zero-mean noise. We select the SXS non-precessing simulations \texttt{SXS:BBH:0110}, as it was done for the validation of \texttt{IMRPhenomXHM} in \cite{garcaquirs2020imrphenomxhm}, and \texttt{SXS:BBH:0237}. In Table~\ref{tab:injection_settings} we list the intrinsic parameters of each simulation as well as the extrinsic parameters chosen for the synthetic signal, followed by the recovered value and its uncertainty. Parameter estimation has been performed with \texttt{parallelBilby} and \texttt{Dynesty} settings of \texttt{nlive=2048}, \texttt{walks=200}, \texttt{nact=5} and \texttt{maxmcmc=10000}.

\setlength{\extrarowheight}{7pt}
\begin{table}[h!]
    \centering
    \begin{tabular}{ c  | c  c  | c  c  }
\hline
\hline
\multirow{2}{*}{Parameters} & \multicolumn{2}{c|}{\texttt{SXS:BBH:0110}} &   \multicolumn{2}{c}{\texttt{SXS:BBH:0237}}   \\ \cline{2-5}
  &   Injected & Recovered  &   Injected  & Recovered \\ \cline{1-5}
 \hline
 \hline
$M_\mathrm{tot}/M_\odot$ &   $62.0$ & $61.28^{+5.32}_{-5.34}$ & $45.0$ & $46.38^{+3.93}_{-2.95}$ \\
 $q$ & $0.2 $ & $0.21^{+0.06}_{-0.04}$& $0.5$ & $0.43^{+0.17}_{-0.11}$\\
 $\chi_{\text{eff}}$ & $0.42 $ & $0.41^{+0.07}_{-0.07}$ & $ -0.2$ & $-0.2^{+0.13}_{-0.13}$ \\
 $\chi_{1z}$ & $0.5 $ & $0.5^{+0.07}_{-0.1}$ & $ -0.6$ & $-0.23^{+0.21}_{-0.26} $ \\
 $\chi_{2z}$ & $0.0$ & $-0.05^{+0.6}_{-0.56}$& $ 0.6$ & $-0.1^{+0.42}_{-0.46}$ \\
$\iota$ & $\pi/3$ & $0.9^{+0.22}_{-0.26}$ & $ \pi/3$ & $0.71^{+0.33}_{-0.37}$ \\
$\phi_{\text{ref}}$ & $0$ & $4.97^{+1.14}_{-4.77}$ & $0.8$ & $1.63^{4.48}_{-1.47}$ \\
 $\alpha$   & $1.4$ & $1.4^{+0.02}_{-0.02}$ & $3.81$ & $3.82^{+0.04}_{-0.04}$ \\
 $\delta$  & $-0.6$ & $-0.6^{+0.09}_{-0.1}$ & $0.63$ & $0.6^{+0.04}_{-0.05}$ \\
 \hline
 \hline
    \end{tabular}
 \caption{NR injection recovery. For each simulated signal, corresponding to the NR simulations \texttt{SXS:BBH:0110} and \texttt{SXS:BBH:0237}, the injected value and the recovered parameter with their uncertainties (90$\%$ confidence intervals) are listed. The listed parameters correspond to the total mass $M_\mathrm{tot}$ in solar masses $M_\odot$, mass ratio $q=m_2/m_1 < 1$, effective spin parameter $\chi_{\text{eff}}$, z-components of the dimensionless spin vector of each black hole ($\chi_{1z}$ and  $\chi_{2z}$), inclination $\iota$, coalescence phase $\phi_{\text{ref}}$ at the reference frequency, right ascension $\alpha$, and declination $\delta$.} 
 \label{tab:injection_settings}
\end{table}

\begin{figure*}[htpb!]
\includegraphics[width=\columnwidth]{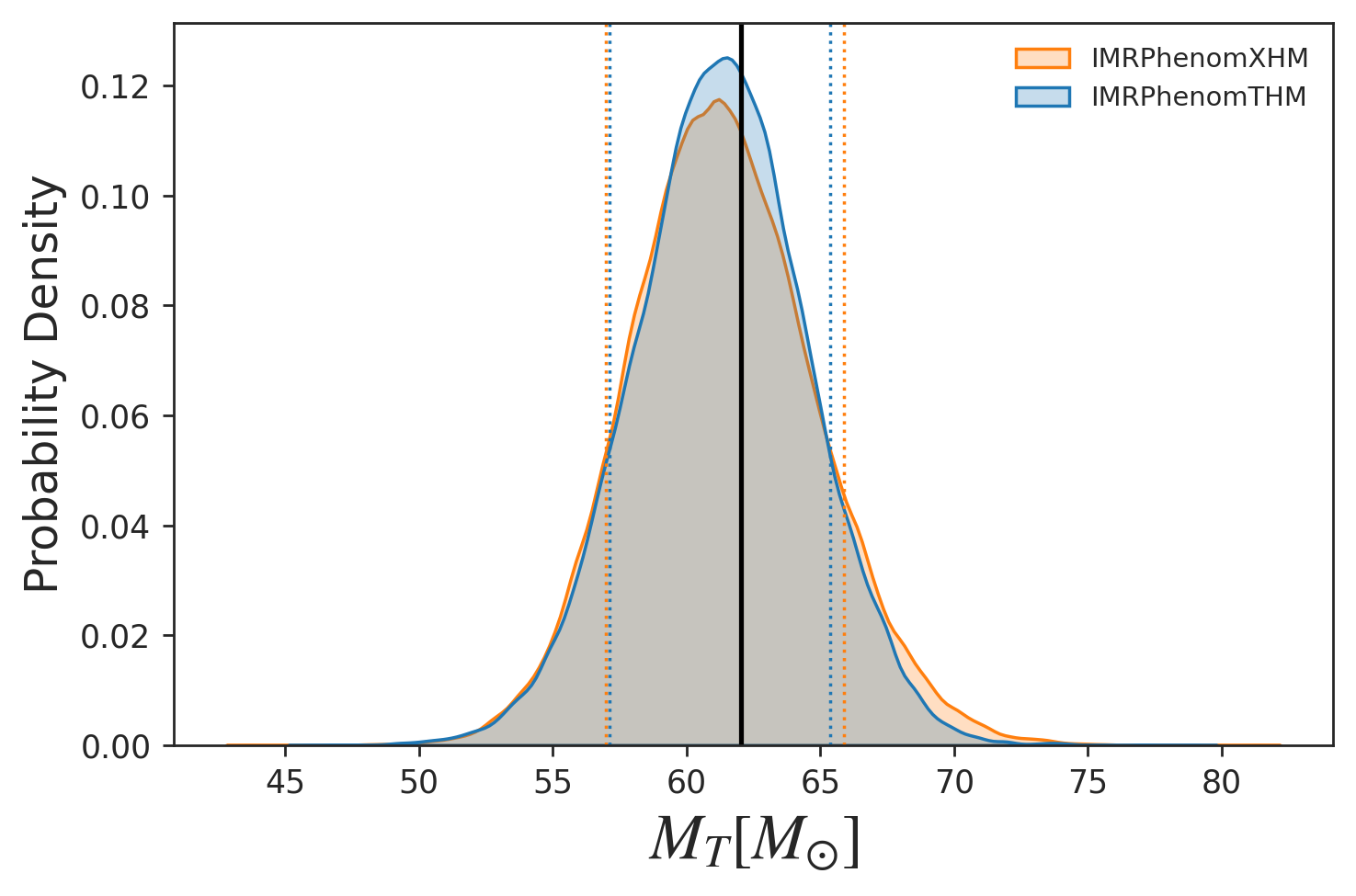}\includegraphics[width=\columnwidth]{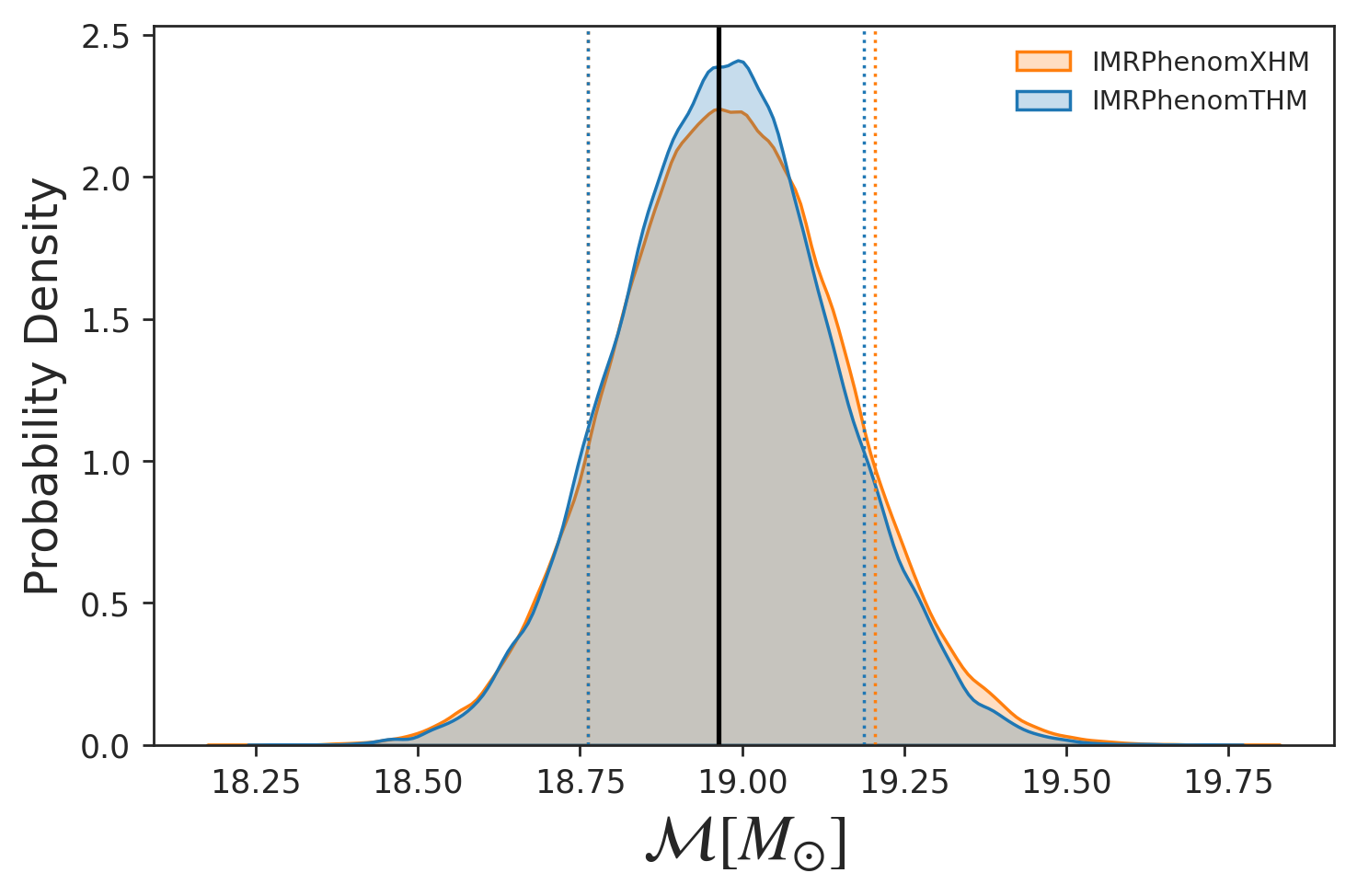}
\includegraphics[width=\columnwidth]{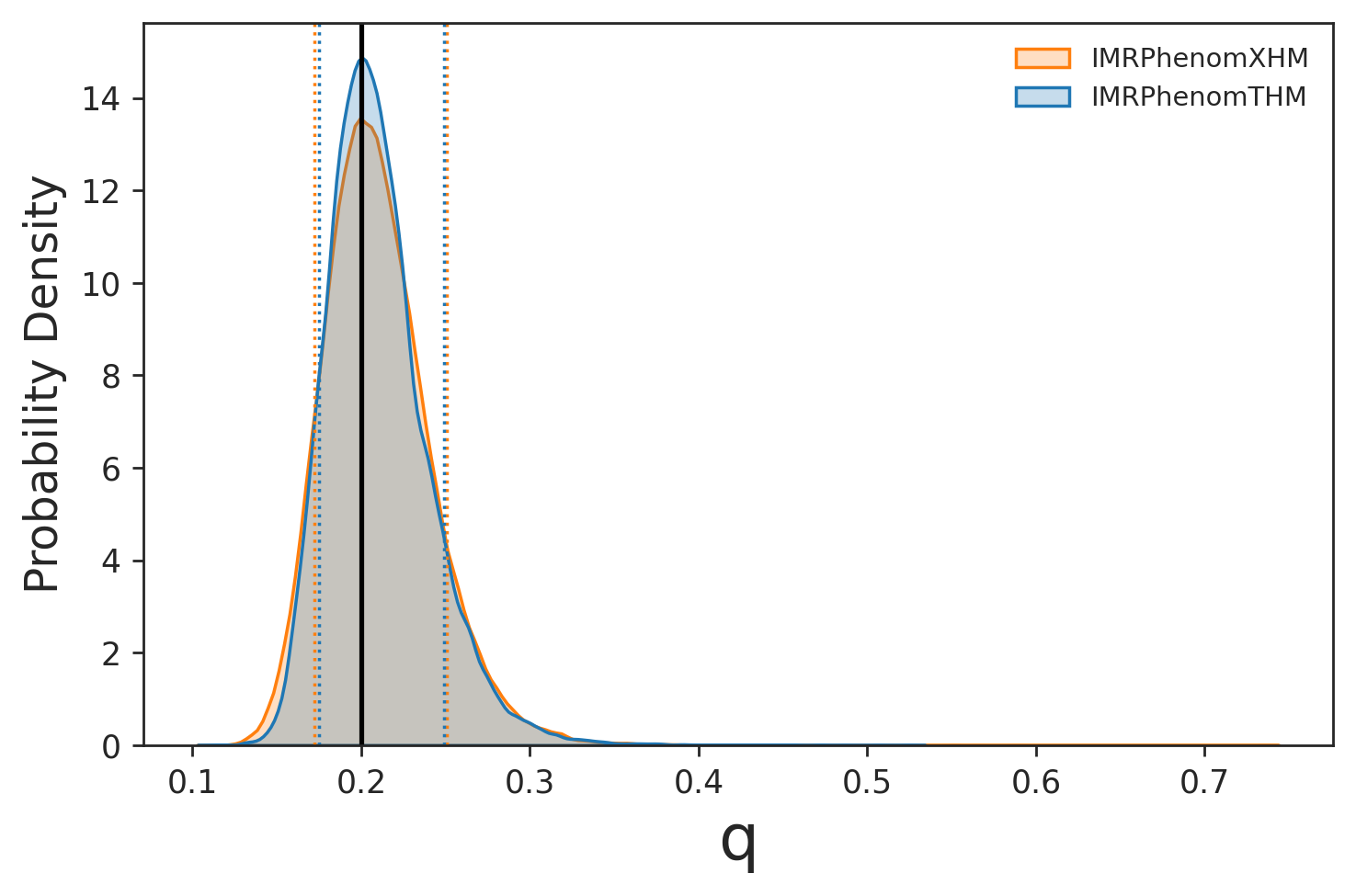}\includegraphics[width=\columnwidth]{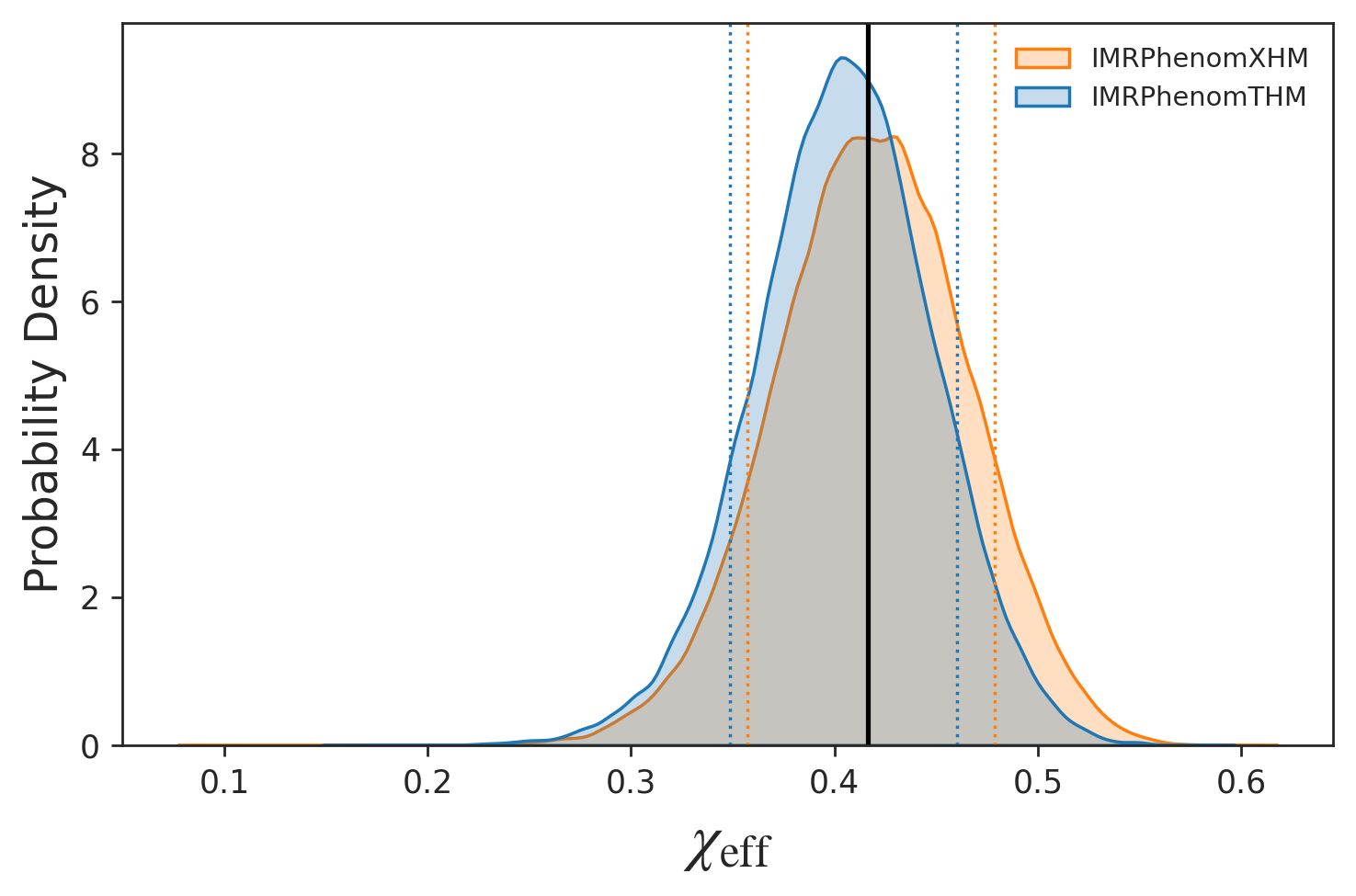}\\
 \caption{SXS:BBH:0110 recovered posteriors.}
\label{fig:sxs0110post}
\end{figure*}

\begin{figure*}[htpb!]
\includegraphics[width=\columnwidth]{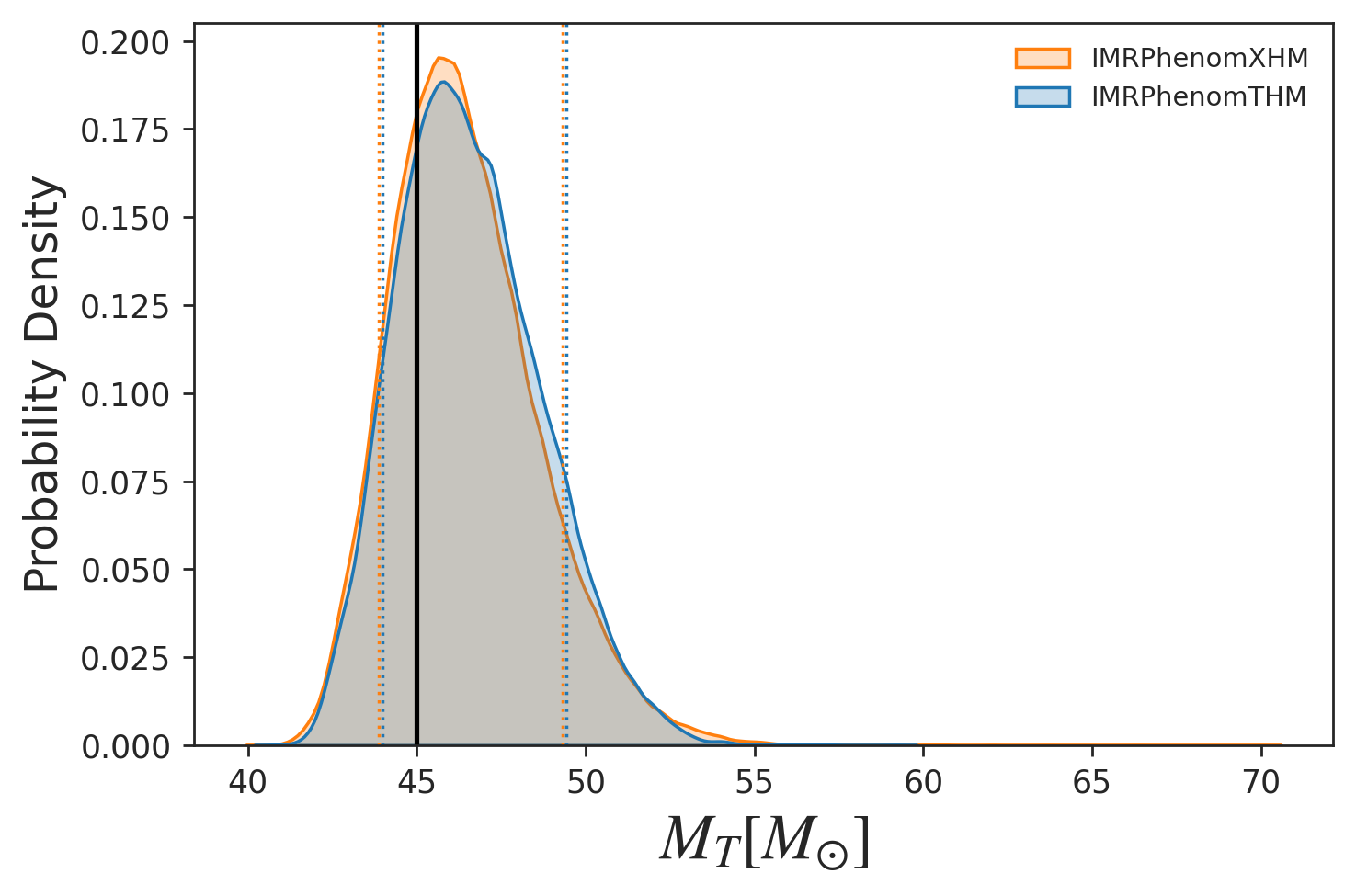}\includegraphics[width=\columnwidth]{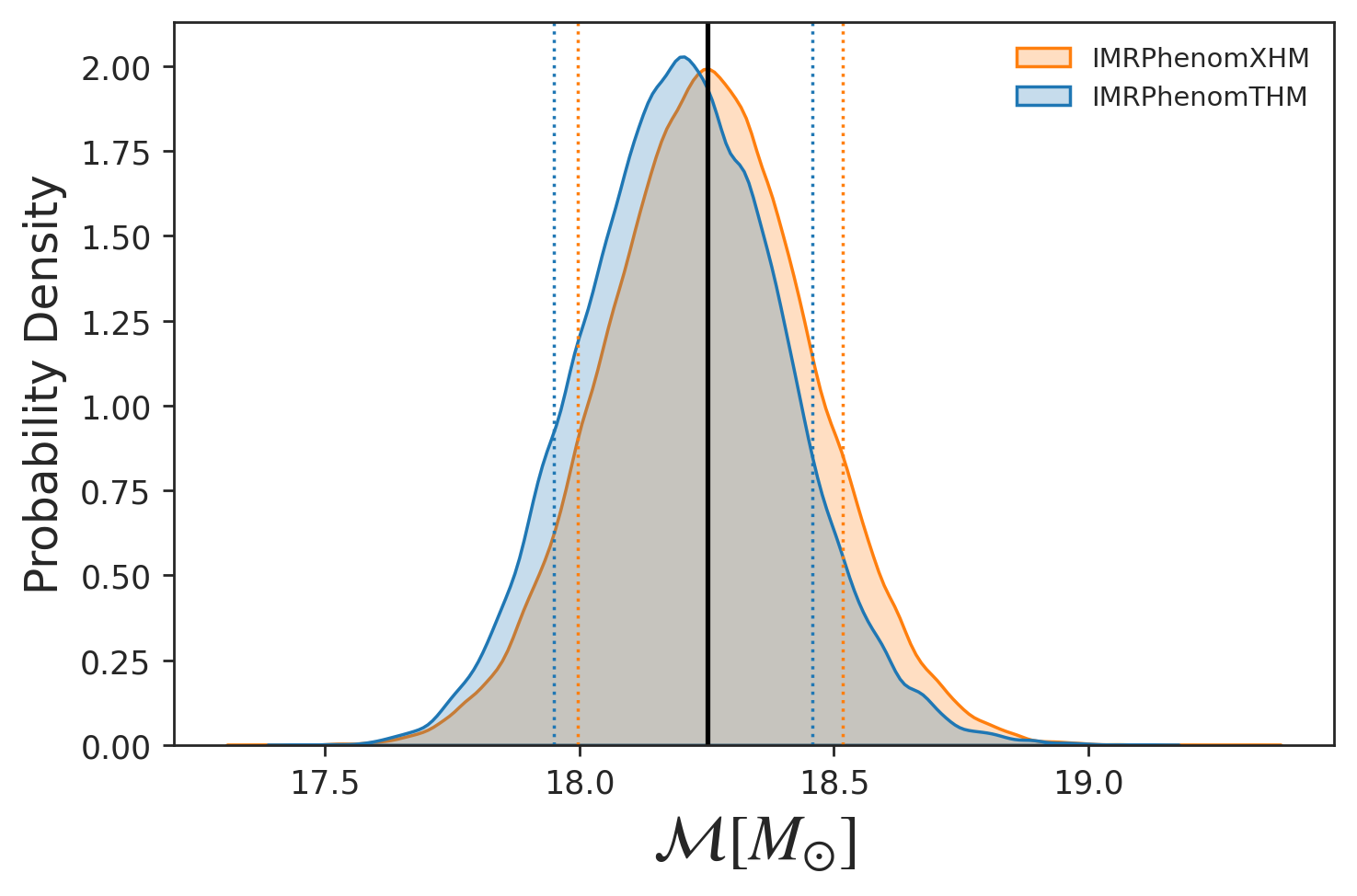}
\includegraphics[width=\columnwidth]{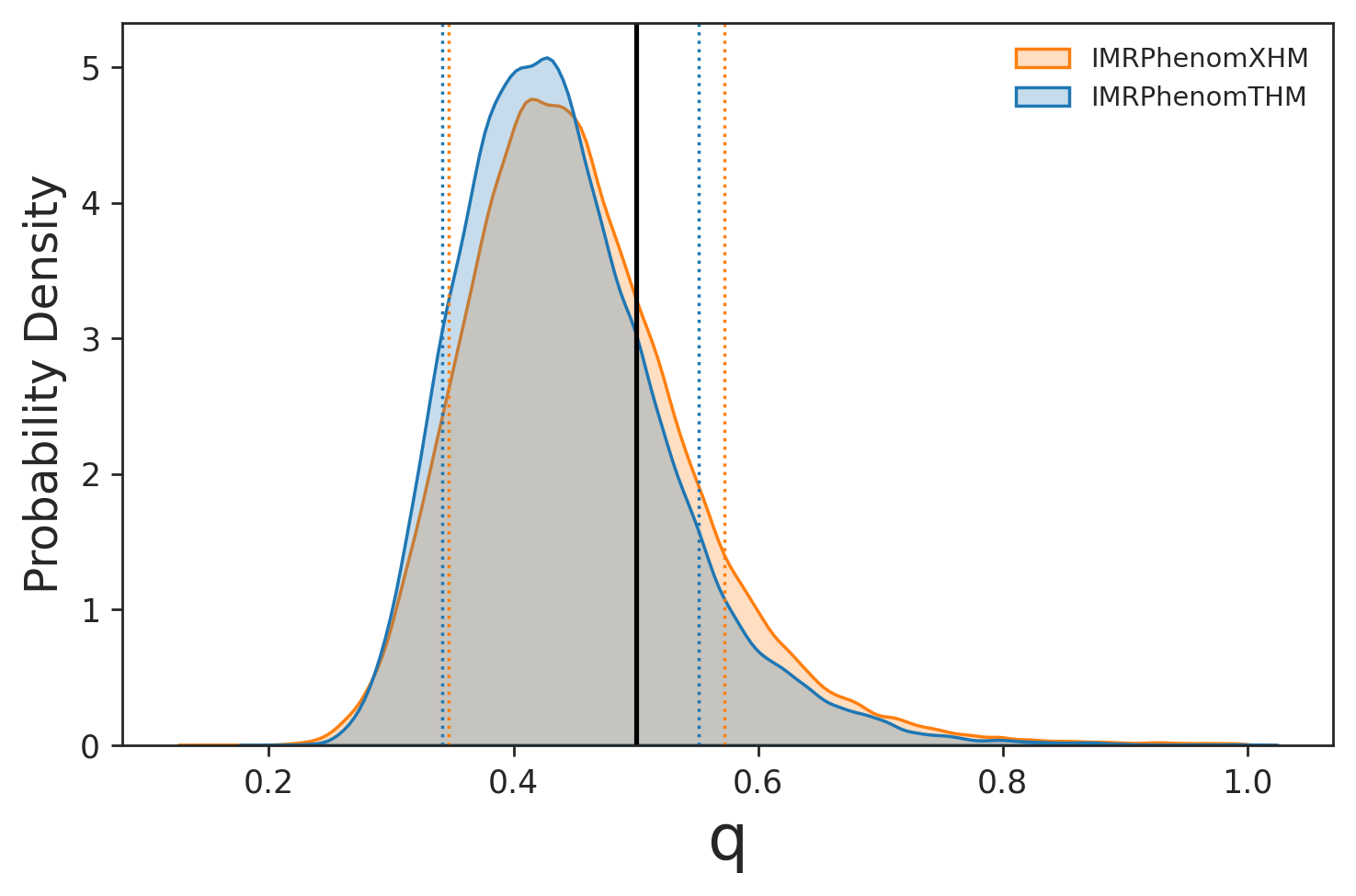}\includegraphics[width=\columnwidth]{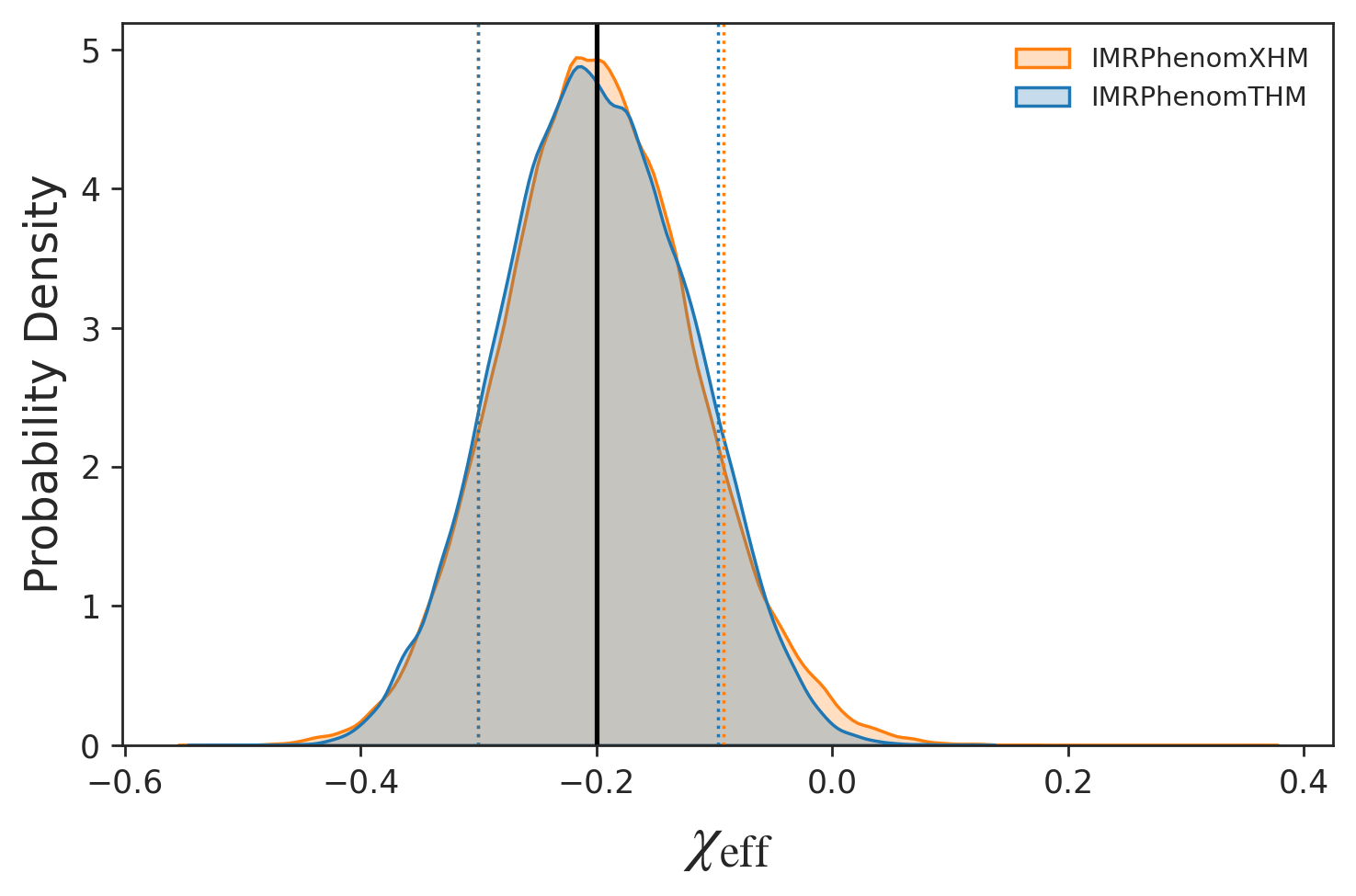}\\
 \caption{SXS:BBH:0237 recovered posteriors.}
\label{fig:sxs0237post}
\end{figure*}

As we can see from the results of Table~\ref{tab:injection_settings} and the posteriors shown in Fig.~\ref{fig:sxs0110post} and Fig.~\ref{fig:sxs0237post}, the model is able to reproduce quite well both the intrinsic and the extrinsic parameters from both simulated events. The main difference in intrinsic parameter recovery is seen for the mass ratio and individual spin components of the \texttt{SXS:BBH:0237} injection. For the mass ratio, as a cross-check, a injection recovery with the model \texttt{IMRPhenomXHM} was also performed, confirming a similar bias in the result, possibly due to a lack of mode content in the models with respect to the full NR waveform. But in any case the injected value is still inside the $90\%$ confidence interval (CI) of the recovered posterior. Regarding the individual spin components, although they are not well recovered for \texttt{SXS:BBH:0237}, the effective spin parameter $\chi_\mathrm{eff}$, which is the dominant spin parameter in the inspiral PN expressions, is very well recovered for both injections.

Regarding extrinsic parameter recovery, it can be seen from Table~(\ref{tab:injection_settings} that the sky location, parameterised as usual by right ascension and declination, is well recovered for both simulated signals, and the recovered inclination is inside the $90\%$ CI for both signals, with the result for \texttt{SXS:BBH:0237} slightly more biased. The orbital phase is generally not a very well recovered quantity, but the posterior distribution indicates a good recovery trend.

\subsection{Real events}\label{subsec:realevents}

After testing parameter estimation performance of the model on synthetic injected signals, a test on real gravitational wave data is performed. We have selected two events from the first GWTC-1 catalog~\cite{Abbott_2019} of detections from the first and second observing runs of the Advanced LIGO and Advanced Virgo detectors: GW150914, which was the first gravitational wave detection~\cite{Abbott:2016blz}, and GW170729, which has previously been studied with various higher-mode models due to its high mass \cite{Chatziioannou_2019}. We have employed the public 32 seconds strain data from GWOSC \cite{GWOSC} sampled at $4\text{kHz}$ around each event, with the public power spectral densities estimated for those segments of data \cite{gwtc1psd}, and the public calibration envelopes of the Hanford and Livingston detectors (for GW150914) and the Virgo detector (for GW170729) \cite{gwtc1calib}.

\begin{figure*}[htpb!]
\includegraphics[width=0.6\columnwidth]{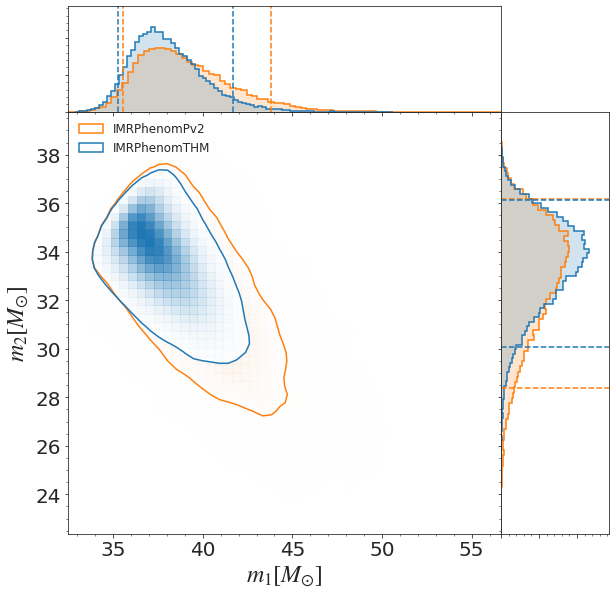}\includegraphics[width=0.6\columnwidth]{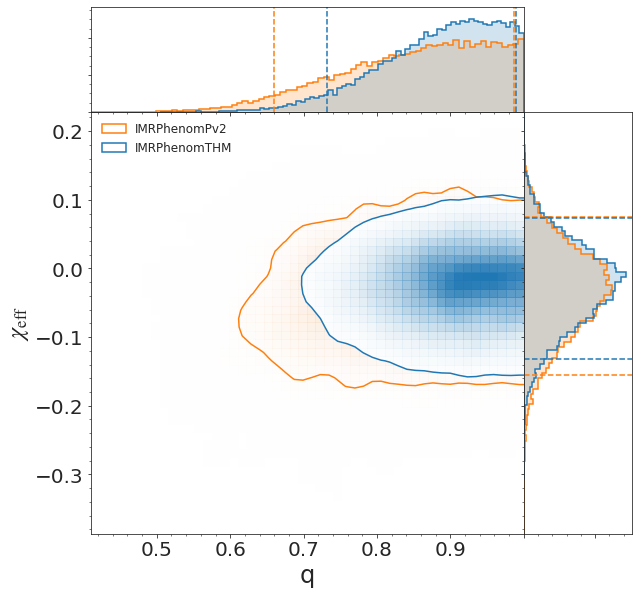}\includegraphics[width=0.6\columnwidth]{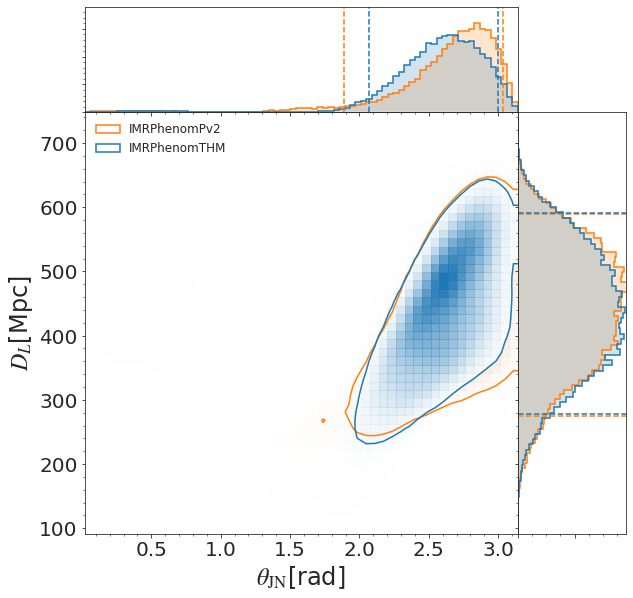}\\
 \caption{GW150914 recovered posteriors.}
\label{fig:150914corner}
\end{figure*}

\begin{figure*}[htpb!]
\includegraphics[width=0.6\columnwidth]{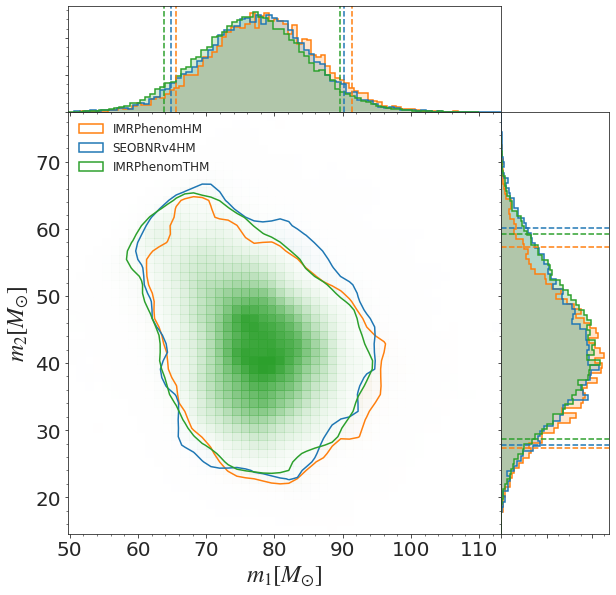}\includegraphics[width=0.6\columnwidth]{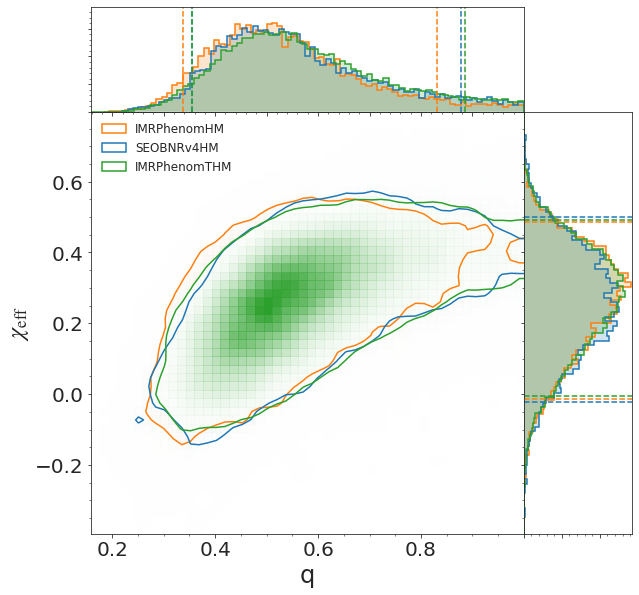}\includegraphics[width=0.6\columnwidth]{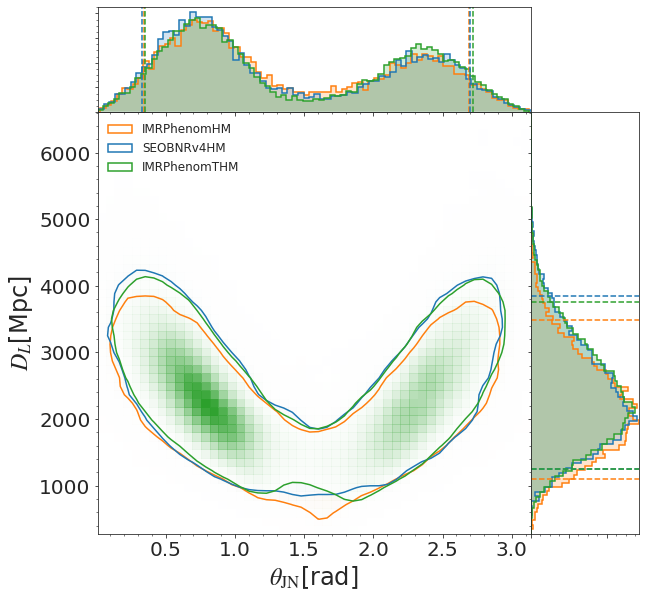}\\
 \caption{GW170729 recovered posteriors.}
\label{fig:170729corner}
\end{figure*}

Parameter estimation has been performed with the \texttt{parallelBilby} pipeline, with the following settings for the \texttt{Dynesty} sampler: \texttt{nlive=2048}, \texttt{walks=200}, \texttt{nact=10} and \texttt{maxmcmc=15000}. Results for GW150914 are compared in Fig.~\ref{fig:150914corner} with the published results from GWTC-1 \cite{Abbott_2019} obtained for this event with the phenomenological waveform model \texttt{IMRPhenomPv2}, which does not contain subdominant spherical harmonic modes information in its non-precessing core model, and allows for generic spin configurations through the twisting-up technique of the non-precessing model to describe precessing systems. Due to the different physical content of both models, complete agreement in the results is not expected, but results are consistent with the published ones. Results for GW170729 are shown in Fig.~\ref{fig:170729corner}, compared with the results of \cite{Chatziioannou_2019} where the non-precessing multimode models \texttt{IMRPhenomHM} and \texttt{SEOBNRv4HM} were employed, and a good consistency can be seen between the three models.

Apart from the results for these two real events, the model was employed in the re-analysis \cite{colleoni190412} of the O3a event GW190412 \cite{Abbott_2019}, which we analysed with the last generation of frequency domain phenomenological waveform models, the \texttt{IMRPhenomX} family and the dominant mode and multimode time domain model presented in this work. Excellent consistency was found in the results when comparing the last generation of calibrated multimode non-precessing waveform models for the different model families, \texttt{IMRPhenomXHM}, \texttt{SEOBNRv4HM}, \texttt{NRHybSur3dq8} and \texttt{IMRPhenomTHM}, while the previous generation of phenomenological waveform models, \texttt{IMRPhenomHM}, based on an analytical uncalibrated rescaling of the dominant mode model \texttt{IMRPhenomD}, showed higher discrepancies.

\subsection{Benchmarks}\label{subsec:benchmarks}
\begin{figure*}[thpb]
\includegraphics[width=\columnwidth]{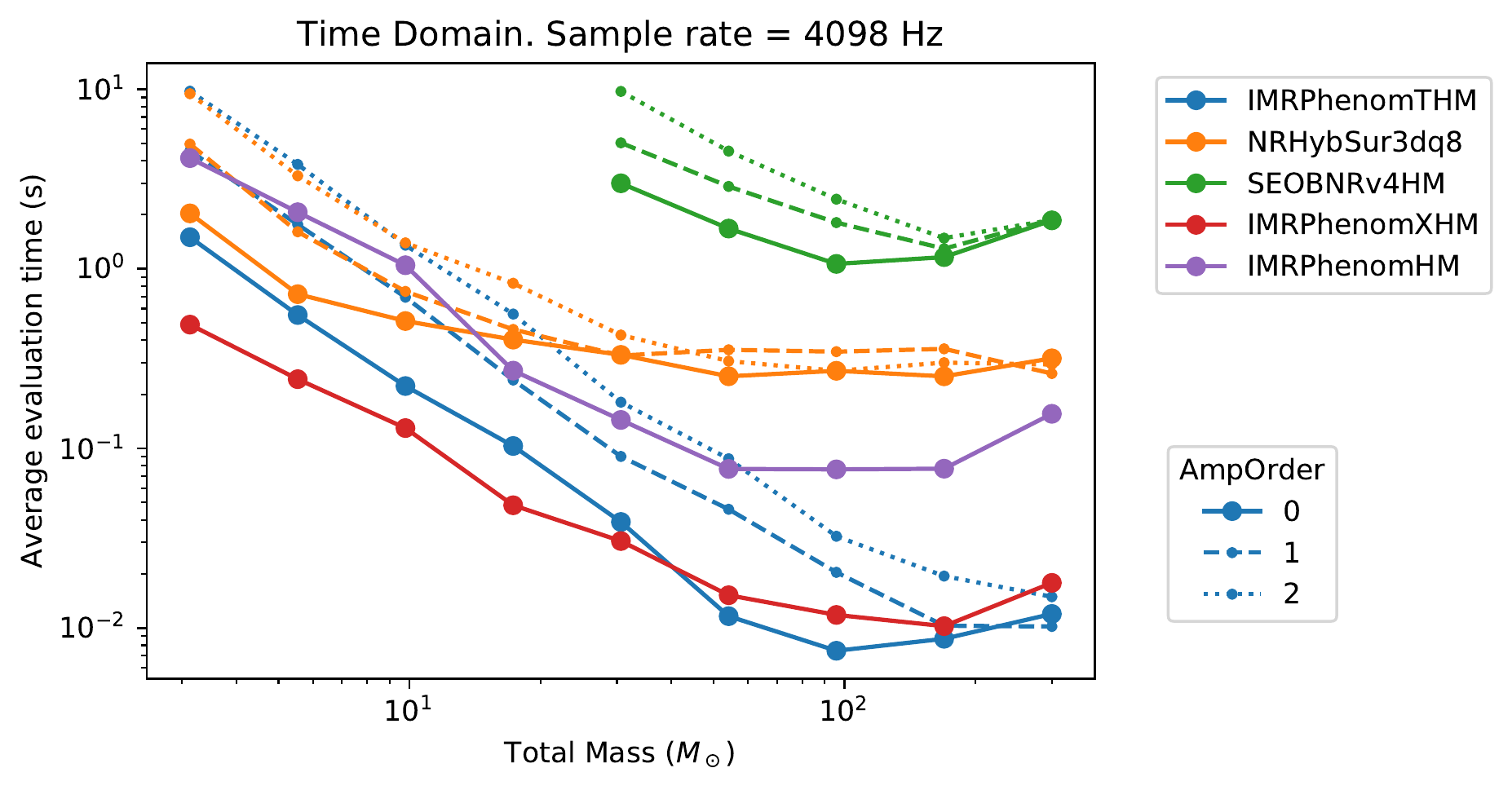}
\includegraphics[width=\columnwidth]{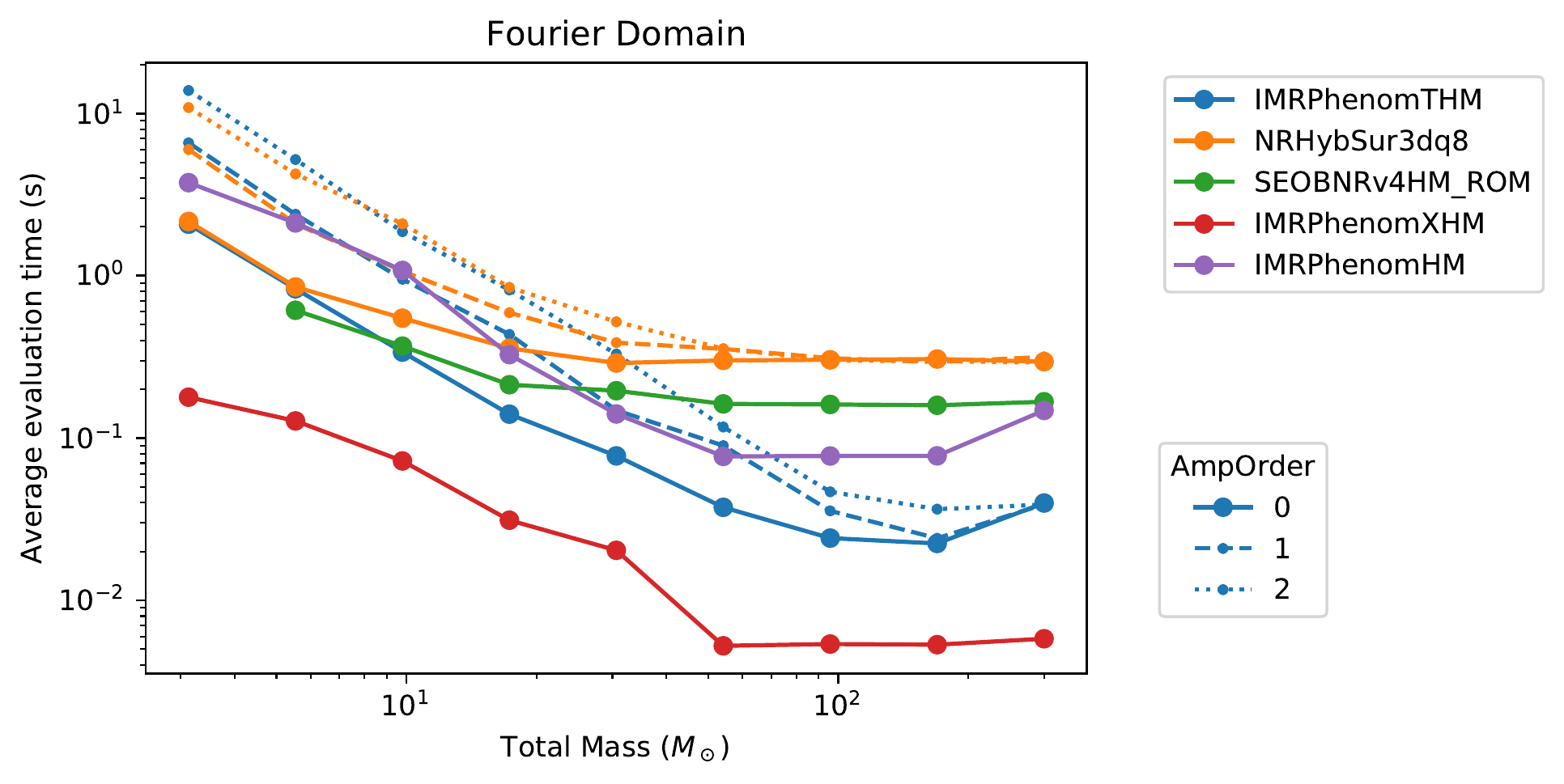}
 \caption{Average waveform evaluation time for different multimode non-precessing waveform models employing the \texttt{GenerateSimulation} interface of LALSimulation. Left: Time-domain waveform evaluation time as a function of the total mass of the system. Right: Frequency-domain waveform evaluation time as a function of the total mass of the system.}
 \label{fig:benchmarks}

\begin{center}
 \def\arraystretch{1.3 }
\begin{tabular}{  c  c  c  c  c  c c   c  }
\hline
\hline
$[M_{\min},M_{\max}]$ &  $\Delta T$ & \fontsize{8}{104}\selectfont IMRPhenomTHM & \fontsize{8}{104}\selectfont IMRPhenomXHM & \fontsize{8}{104}\selectfont  IMRPhenomHM & \fontsize{8}{104}\selectfont SEOBNRv4HM & \fontsize{8}{104}\selectfont SEOBNRv4HM$\_$ROM & \fontsize{8}{104}\selectfont NRHybSur3dq8  \\
\hline
\multirow{2}{*}{[50,100]} & 4\,s     &    9.5    &   5.0   &   67.2    &  1415.9 &   29.5      &      221.1     \\
                          & 8\,s     &    9.8    &   6.6   &   121.4   &  1430.2   &  35.2     &    230.2     \\
\hline
\multirow{2}{*}{[10,50]} & 4\,s     &    55.6    &   13.6   &   72.6    &   1630.4  &    38.7     &    307.1    \\
                         & 8\,s     &    69.1    &   20.6   &   166.0   &  1619.8   &  65.5      &    326.8       \\
\hline
\hline
 \end{tabular}
\end{center}
\captionof{table}{Mean likelihood evaluation time in milliseconds for several models including higher order modes for equal masses. The numbers represent averages over two different total mass ranges $[M_{\min},M_{\max}]=\{[10,50],[50,100]\}M_\odot$ and random spin orientations and magnitudes. The first column indicates the total mass range in which the models are evaluated and the second one specifies the data analysis segment length in seconds used for the calculations.}
\label{tab:tabBench}
\end{figure*}

In a similar way as we have done with previous models \cite{phenomxphm}, we estimate the speed of our new waveform model, \texttt{IMRPhenomTHM}, compared to other  non-precessing frequency- and time-domain waveform models including higher order modes: \texttt{IMRPhenomXHM} \cite{garcaquirs2020imrphenomxhm}, \texttt{IMRPhenomHM} \cite{phenomhm}, \texttt{SEOBNRv4HM} \cite{Cotesta_2018}, \texttt{SEOBNRv4HM$\_$ROM} \cite{Cotesta:2020qhw} and \texttt{NRHybSur3dq8} \cite{Varma_2019}. Thus, we show in Fig.~\ref{fig:benchmarks} benchmarking results for an arbitrarily chosen non-precessing case, $(q,\chi_{1z},\chi_{2z})=(3,0.5,-0.3)$, in a frequency range from 10 to 2048 Hz by comparing different waveform models. The timing test is performed with \texttt{GenerateSimulation} (included in the LALSimulation package), averaging over 100 repetitions.  In the left panel the dependency on the total mass is shown for a range between $(3,\:300)\: M_\odot$ for the waveforms in the time-domain generated with the \texttt{ChooseTDWaveform} function (included in LALSimulation) and a fixed sample rate of 4096 Hz. In the case of frequency-domain models the curves also include the time to perform an inverse Fourier transform. Furthermore, for the time-domain 
approximants one has to take into account the fact that in order to have all the higher order modes starting at the desired minimum frequency of the waveform for analysis purposes, one has to start the waveform generation earlier. This is caused by the fact that higher order modes with $m>2$ have a higher frequency content for the same array of times. This effect is encapsulated in the \texttt{amporder} parameter which changes the starting frequency of the waveform generation in the following way,
\begin{equation}
f_{\text{start}}=\frac{2 f_{\text{min}}}{\texttt{amporder}+2}.
\label{eq:eq999}
\end{equation}
Hence, in our timing test we loop also over different values of the  \texttt{amporder} parameter to evaluate its effect on the speed of the waveform generation. The right plot in Fig.~\ref{fig:benchmarks} shows the timing test performed in the Fourier domain by calling the distinct waveform models using the \texttt{SimInspiralFD} function (included in LALSimulation). In this plot the total mass is varied as in the previous panel, but the frequency resolution is computed internally to adapt to the length of the waveform. Moreover, in this plot we have replaced the time-domain \texttt{SEOBNRv4} model by its frequency-domain counterpart \texttt{SEOBNRv4HM$\_$ROM}.

From the left plot of Fig.~\ref{fig:benchmarks} one can conclude that \texttt{IMRPhenomTHM} performs faster than the other two time-domain waveform models for total masses above $30 M_\odot$, while for low total masses ($< 30 M_\odot$) the \texttt{SEOBNRv4HM} can not be generated and \texttt{IMRPhenomTHM} becomes comparable to \texttt{NRHybSur3dq8}. We note here also the important effect of the \texttt{amporder} parameter: as expected, the higher its value, the lower the starting frequency, and thus the generation of the time-domain model becomes slower. However, even with that we observe that \texttt{IMRPhenomTHM} becomes comparable to the frequency domain models for total masses $>30 M_\odot$, or even faster in the case of \texttt{IMRPhenomHM}. This is due to the fact that the symbolic phenomenological expressions constituting the core of \texttt{IMRPhenomTHM} make the model computationally efficient. Similar conclusions can be extracted from the right panel of Fig.~\ref{fig:benchmarks}, where one observes that while for total masses below  $30 M_\odot$ the model is slower than the frequency-domain \texttt{SEOBNRv4HM$\_$ROM} and \texttt{IMRPhenomXHM} models, for higher total masses it becomes comparable and even faster than \texttt{SEOBNRv4HM$\_$ROM}, but never reaching the speed of \texttt{IMRPhenomXHM}, which benefits from the multibanding technique \cite{Vinciguerra:2017ngf, Garcia-Quiros:2020qlt} to become the fastest higher order mode waveform model in the frequency domain.

We have also estimated the efficiency of our model \texttt{IMRPhenomTHM} compared to other models with higher order modes by computing their mean likelihood evaluation time in the LALInference Bayesian parameter estimation code \cite{Veitch:2014wba}.
This test has been performed choosing an equal-mass configuration with 100 different total masses in two different mass ranges: a low mass range with range $[M_{\min},M_{\max}]=[10,50]  M_\odot$, and high mass range  with  $[M_{\min},M_{\max}]=[50,100]  M_\odot$. Dimensionless spin magnitudes are distributed randomly between $[0,0.99]$ with a random isotropic distribution of spin vectors, and a reference frequency of $20$ Hz is chosen. 
Furthermore, two different segment lengths of $\Delta T= 4$\,s, 8\,s are studied as they are typical for the currently detected BBH GW signals \cite{Abbott_2019,Abbott:2020niy}. For the likelihood evaluations we employ the  Advanced LIGO zero detuned power spectral density \cite{adligopsd} as we also do for our match calculations in Sec.~\ref{sec:results} and NR injection studies in Sec.~\ref{subsec:injection}. For each total mass we perform 100 likelihood evaluations with randomly chosen spin configurations. In Table~\ref{tab:tabBench} the  average of these $10^4$ likelihood evaluations for each model is shown. 

The numbers in Table~\ref{tab:tabBench} are consistent with the results of Fig.~\ref{fig:benchmarks} as they show that \texttt{IMRPhenomTHM }is the most efficient time-domain model including higher order modes, and for higher total masses it becomes faster or comparable to some frequency domain models like \texttt{IMRPhenomHM} and  \texttt{SEOBNRv4HM$\_$ROM}. This shows that \texttt{IMRPhenomTHM} is perfectly suited for parameter estimation analysis of high mass GW events like the ones detected by the LIGO and Virgo collaboration \cite{Abbott_2019,Abbott:2020niy} and a valuable tool to complement and disentangle systematics coming from the use of frequency domain waveform models in parameter estimation studies of GW events.


\setlength{\extrarowheight}{0pt}

\section{Conclusions}\label{sec:conclusions}

In this work a new phenomenological model for non-precessing binary black hole signals has been presented, based on the construction of the \texttt{IMRPhenomT} model previously presented in \cite{estells2020imrphenomtp}. It incorporates additional spherical harmonic modes which are a requirement of current gravitational wave observations, and its construction is native in the time domain, differentiating it with respect to other phenomenological models from the current and past generations.

It has been shown that the model accurately reproduces the LVCNR Catalog of Numerical Relativity simulations, and that there is good agreement with other state of the art non-precessing multimode models like \texttt{NRHybSur3dq8} and  \texttt{IMRPhenomXHM}. The model also reproduces accurate results for the final kick velocity due to anisotropic gravitational wave emission, as compared to existing NR fits for this quantity and to the results of other multimode models, which is an exigent test of the correct global multimode structure of the waveforms. The model can be used to accurately estimate parameters from injected simulated signals and to produce consistent results for public gravitational wave detections, being also competitively fast for waveform generation and likelihood evaluation, despite its native time domain nature, which requires to perform a numerical Fast Fourier Transform in order to employ the model for parameter estimation. As a further proof of the reliability of the model, results from a preliminary version were employed in the re-analysis of the gravitational wave event GW190412, showing very good consistency with current generation calibrated non-precessing multimode models and showing likelihood evaluation times comparable to the fastest phenomenological model \texttt{IMRPhenomXHM}.

The model presented in this work, \texttt{IMRPhenomTHM}, complements the so-called ``generation X'' of phenomenological waveform models, which contains the fastest and most accurate family of frequency domain phenomenological models \texttt{IMRPhenomXAS}, \texttt{IMRPhenomXHM}, \texttt{IMRPhenomXP} and \texttt{IMRPhenomXPHM}. In a future work, an extension to model precessing systems, \texttt{IMRPhenomTPHM} will be presented, based on the features already introduced in \cite{estells2020imrphenomtp} for extending the dominant mode model \texttt{IMRPhenomT} into \texttt{IMRPhenomTP}. Having a complementary time domain phenomenological framework will help to disentangle possible modelling issues of the frequency domain framework, particulary elimitating the Stationary Phase Approximation from the precession description, and allowing to incorporate further methods of reproducing the precessing angular momenta dynamics. As a first step, the non-precessing model \texttt{IMRPhenomTHM} fulfills the requirements for the current state-of-the-art non-precessing waveform models.

\section*{Acknowledgements}
We thank the reviewers of the LIGO Scientific Collaboration, Maria Haney, Jonathan Thompson, Eleanor Hamilton and Jacob Lange for reviewing the \texttt{LALSuite} code implementation, and for carefully reading the manuscript and valuable feedback.
We thank Alessandro Nagar, Sebastiano Bernuzzi and Enno Harms for giving us access to $\it{Teukode}$ \cite{Harms_2014}, which was used to generate our extreme-mass-ratio waveforms. 

This work was supported by European Union FEDER funds, the Ministry of Science, 
Innovation and Universities and the Spanish Agencia Estatal de Investigación grants
PID2019-106416GB-I00/AEI/10.13039/501100011033,  
FPA2016-76821-P,     
RED2018-102661-T,    
RED2018-102573-E,    
FPA2017-90687-REDC,  
Vicepresidència i Conselleria d’Innovació, Recerca i Turisme, 
Conselleria d’Educació, i Universitats del Govern de les Illes Balears i Fons Social Europeu, 
Comunitat Autonoma de les Illes Balears through the Direcció General de Política Universitaria i Recerca with funds from the Tourist Stay Tax Law ITS 2017-006 (PRD2018/24),
Generalitat Valenciana (PROMETEO/2019/071),  
EU COST Actions CA18108, CA17137, CA16214, and CA16104, and
the Spanish Ministry of Education, Culture and Sport grants FPU15/03344 and FPU15/01319.
M.C.~acknowledges funding from the European Union's Horizon 2020 research and innovation programme, under the Marie Skłodowska-Curie grant agreement No. 751492.
D.K.~is supported by the Spanish Ministerio de Ciencia, Innovaci{\'o}n y
Universidades (ref.~BEAGAL 18/00148)
and cofinanced by the Universitat de les Illes Balears.
The authors thankfully acknowledge the computer resources at MareNostrum and the technical support provided by Barcelona Supercomputing Center (BSC) through Grants 
No. 
AECT-2020-2-0015,  
AECT-2020-1-0025,  
AECT-2019-3-0020,  
AECT-2019-2-0010,  
AECT-2019-2-0017,  
AECT-2019-1-0022,  
from the Red Española de Supercomputación (RES).

Authors also acknowledge the computational resources at the cluster CIT provided by LIGO Laboratory and supported by National Science Foundation Grants PHY-0757058 and PHY-0823459.
This research has made use of data obtained from the Gravitational Wave Open Science Center~\cite{GWOSC}, a service of LIGO Laboratory, the LIGO Scientific Collaboration and the Virgo Collaboration. LIGO is funded by the U.S. National Science Foundation. Virgo is funded by the French Centre National de Recherche Scientifique (CNRS), the Italian Istituto Nazionale della Fisica Nucleare (INFN) and the Dutch Nikhef, with contributions by Polish and Hungarian institutes.

\begin{widetext}

\appendix

\section{Post-Newtonian amplitudes}\label{appen:pnamp}
In the following we show the PN expressions used to construct the inspiral amplitudes as described in Sec. \ref{sec:InspiralAmplitude}.  The expressions for the different  modes in Eq.~\eqref{eq:pnamp} are based on 3PN order expressions from \cite{Blanchet_2008} with 2PN spin corrections from \cite{Buonanno_2013} and 1.5PN contributions from \cite{arun2008higherorder}. For the amplitude of the $(2,2)$ mode, we also include 3.5PN corrections from \cite{Faye_2012}. This is mostly equivalent to the PN amplitudes employed by the frequency domain model \texttt{IMRPhenomXHM} which are listed in Appendix E of \cite{garcaquirs2020imrphenomxhm}, except that the expressions listed there already contain the stationary phase approximation contributions needed for the frequency domain representation. The difference to \cite{garcaquirs2020imrphenomxhm} is that known 3PN contributions to the $l=3$, $m=3$ mode have not been included here since we find that a better performance with the calibrated higher order coefficients was achieved without including these 3PN terms. Here they are listed after factoring out the common prefactor $2\eta\sqrt{16/5}x(t)$:
\begin{equation}
\begin{split}
    \bar{H}_{22}^{PN}(x(t)) =& 1 + \Big( \frac{55 \eta }{42}-\frac{107}{42} \Big) x   + \Big[ -\frac{2 \delta  \chi _1}{3 \left(\frac{1-\delta }{2}+\frac{\delta +1}{2}\right)}+\frac{2 \delta  \chi _2}{3 \left(\frac{1-\delta }{2}+\frac{\delta +1}{2}\right)}+\frac{2 \eta  \chi _1}{3}+\frac{2 \eta  \chi
   _2}{3}-\frac{2 \chi _1}{3}-\frac{2 \chi _2}{3}+2 \pi \Big]  x^{3/2}\\
    & + \Big[    \frac{\delta  \chi _1^2}{2}-\frac{\delta  \chi _2^2}{2}+\frac{2047 \eta ^2}{1512}-\eta  \chi _1^2-\eta  \chi _2^2+2 \eta  \chi _1 \chi _2-\frac{1069 \eta }{216}+\frac{\chi _1^2}{2}+\frac{\chi
   _2^2}{2}-\frac{2173}{1512} \Big]x^2\\
    & + \Big( \frac{34 \pi  \eta }{21}-24 i \eta -\frac{107 \pi }{21} \Big)x^{5/2} - \Big[ -\frac{856 \gamma }{105}+\frac{114635 \eta ^3}{99792}-\frac{20261 \eta ^2}{2772}+\frac{41 \pi ^2 \eta }{96}-\frac{278185 \eta }{33264} \\
    & -\frac{428}{105} \log (16 x)+\frac{2 \pi ^2}{3}+\frac{428 i \pi
   }{105}+\frac{27027409}{646800} \Big]x^3 +  \Big( \frac{40 \pi  \eta ^2}{27}-\frac{4066 i \eta ^2}{945}-\frac{2495 \pi  \eta }{378}+\frac{14333 i \eta }{162}\\
   &-\frac{2173 \pi }{756} \Big)  x^{7/2},
\end{split}
\end{equation}

\begin{equation}
\begin{split}
    \bar{H}_{21}^{PN}(x(t)) =& \frac{1}{3} i \delta  \sqrt{x} -\frac{1}{2}i(m_1\chi_1 + m_2\chi_2)x - i\delta\Big(\frac{17}{84} - \frac{5\eta}{21}\Big)x^{3/2}\\
    & + \Big[\frac{1}{6}\delta(1+\log(16)) + i\Big(\delta\frac{\pi}{3} + \frac{11}{28}\delta\eta(\chi_1+\chi_2)-\frac{1}{6}(m_1\chi_1 + m_2\chi_2) + \frac{205}{84}(\chi_1-\chi_2)\Big)\Big]x^2\\
    & -i\Big(\frac{43\delta}{378}+\frac{509\delta\eta}{378}-\frac{79\delta\eta^2}{504}\Big)x^{5/2} - \Big[\frac{17\delta}{168}(1+\log(16)) - \delta\eta\Big(\frac{353}{84} + \frac{\log(2)}{7}\Big) + i\pi\delta\Big(\frac{\eta}{14}-\frac{17}{84}\Big)\Big]x^3,
\end{split}
\end{equation}

\begin{equation}
\begin{split}
    \bar{H}_{33}^{PN}(x(t)) =& -\frac{3}{4} i \sqrt{\frac{15}{14}} \delta  \sqrt{x}  + \Big(3 i \sqrt{\frac{15}{14}} \delta -\frac{3}{2} i \sqrt{\frac{15}{14}} \delta  \eta \Big)x^{3/2}\\
    & + \Big[  -\frac{3}{16} i \sqrt{\frac{3}{70}} \left(5 \chi _1 (\delta  (5 \eta -4)+19 \eta -4)+5 \chi _2 (5 \delta  \eta -4 \delta -19 \eta +4)+12 \delta  \left(-7 i+5 \pi \right. \right. \\
    & \left.  \left.  + 10 i \log \left(3/2 \right) \right)\right) \Big]x^2 + \Big( -\frac{887}{88} i \sqrt{\frac{3}{70}} \delta  \eta ^2+\frac{919}{22} i \sqrt{\frac{3}{70}} \delta  \eta -\frac{369}{88} i \sqrt{\frac{3}{70}} \delta \Big)x^{5/2},
\end{split}
\end{equation}

\begin{equation}
\begin{split}
    \bar{H}_{44}^{PN}(x(t)) =& \frac{8}{9} \sqrt{\frac{5}{7}} (3 \eta -1) x   -\frac{8}{9} \sqrt{\frac{5}{7}} \left(-\frac{175 \eta ^2}{22}+\frac{1273 \eta }{66}-\frac{593}{110}\right)  x^2\\
    & -\frac{8}{9} \sqrt{\frac{5}{7}} \Big[-12 \pi  \eta +i \left(\eta  \left(\frac{1193}{40}-24 \log (2)\right)-\frac{42}{5}+8 \log (2)\right)+4 \pi  \Big]x^{5/2} \\
    &   -\frac{8}{9} \sqrt{\frac{5}{7}} \left(-\frac{226097 \eta ^3}{17160}+\frac{146879 \eta ^2}{2340}-\frac{1088119 \eta }{28600}+\frac{1068671}{200200}\right) x^3,
\end{split}
\end{equation}

\begin{equation}
\begin{split}
    \bar{H}_{55}^{PN}(x(t)) =&    -\frac{3}{2} i \sqrt{\frac{15}{14}} \delta  (\eta -2)  x^{3/2}    -\frac{1}{88} i \sqrt{\frac{3}{70}} \delta  \left(887 \eta ^2-3676 \eta +369\right)  x^{5/2} \\
    &  +  \frac{\delta  \left(-3645 i \pi  (3 \eta -8)+\eta  \left(21870 \log \left(\frac{3}{2}\right)-96206\right)-5832 \left(10 \log \left(\frac{3}{2}\right)-7\right)\right)}{216 \sqrt{210}}    x^3.
\end{split}
\end{equation}

\end{widetext}

\bibliography{ligo.bib, pn.bib, ringdown.bib, wfmodels.bib, misc.bib, pe.bib}{}

\end{document}